\documentclass[11pt]{cernrep}
\usepackage{graphicx}
\usepackage{floatflt}
\usepackage{color}
\usepackage{colortbl}

\newcommand{\Pt}{{P_t}}
\newcommand{\Et}{{E_t}}
\newcommand{\dphi}{\Delta\phi}
\newcommand{\phigj}{\phi_{(\gamma,~jet)}}

\newcommand{\Ptgj}{$\Pt^{\gamma}$ and $\Pt^{Jet}$~~}
\newcommand{\ptgj}{$~\Pt^{\gamma}-\Pt^{Jet}~$}

\newcommand{\la}{\langle}
\newcommand{\ra}{\rangle}
\newcommand{\gpj}{~``$\gamma+jet$''~}
\newcommand{\gpp}{~``$\gamma+parton$''~}
\newcommand{\rrr}{\to} 

\newcommand{\pth}{\hat{p}_{\perp}^{\;min}}
\newcommand{\Db}{\Pt(O\!+\!\eta>5.0)}
\newcommand{\Ptg}{\Pt^{\gamma}}
\newcommand{\ptg}{$\Pt^{\gamma}$}
\newcommand{\Fptgj}{(\Pt^{\gamma}\!-\!\Pt^{Jet})/\Pt^{\gamma}}

\newcommand{\coltab}{0.69}

\newcommand{\aaa}{\hspace*{.39cm}}

\newcommand{\hmm}{\hspace*{-1.3mm}}

\newcommand{\Gvc}{\footnotesize{$(GeV/c)$} }

\newcommand{\gJ}{(\Pt^{\gamma}\!-\!\Pt^{J})/\Pt^{\gamma}}
\newcommand{\gpart}{(\Pt^{\gamma}\! - \! \Pt^{part})/\Pt^{\gamma}}
\newcommand{\Jpart}{(\Pt^{J}\! - \! \Pt^{part})/\Pt^{J}}

\newcommand{\sgmgj}{\sigma(Db[\gamma,J])}
\newcommand{\sgmgp}{\sigma(Db[\gamma,part])}

\newcommand{\lt}{\!<\!}\newcommand{\gt}{\!>\!}

\suppressfloats[!]

\def\baselinestretch{1.0}

\begin{document}


~\\[1cm]

%
\noindent                                                                      
\centerline{\bf\LARGE On the application of \gpj events }\\[4pt]
\centerline{\bf\LARGE for setting the absolute jet energy scale }\\[4pt]
\centerline{\bf\LARGE and determining the gluon distribution}\\[5pt]
\centerline{\bf\LARGE at the LHC .}
\thispagestyle{empty}
\thispagestyle{empty}

\vskip3cm

\centerline{\normalsize \sf D.V.~Bandurin, V.F.~Konoplyanikov, N.B.~Skachkov}
~\\[-5mm]

\centerline{\normalsize \it Joint Institute for Nuclear Research, Dubna, Russia}

\vskip6cm
\hspace*{6cm}{\bf Abstract}\\[19pt]
\noindent
We study the impact of new set of cuts, proposed in our 
previous works, on the improvement of accuracy of the jet energy calibration with 
``$p p\to\gamma+Jet+X$'' process at LHC.
Monte Carlo events produced by the PYTHIA 5.7 generator are used for this aim.
The selection criteria for \gpj event samples that would provide a good balance of
$\Pt^{\gamma}$ with $\Pt^{jet}$  and would allow to reduce the background are described.
The distributions of these events over \ptg ~and $\eta^{jet}$
are presented. The features of \gpj events in the barrel region of
the CMS detector ($|\eta^{jet}|\lt1.4$) are exposed.
The efficiency of the cuts used for background suppression is demonstrated.

It is shown that the samples of \gpj events, gained with the cuts for the jet energy calibration,
may have enough statistics for determining the gluon distribution inside a proton in the region 
of $x\geq2\cdot10^{-4}$ and of $Q^2$ by two orders higher than that studied at HERA.

\newpage

\tableofcontents
\thispagestyle{empty}

\newpage

\setcounter{page}{1}

\section{INTRODUCTION.} 

Setting an absolute energy scale for a jet, detected mostly by hadronic
and electromagnetic calorimeters (HCAL and ECAL), is an important task
for any of $pp$ and $p\bar{p}$  collider experiments (see e.g. [1--8]). 

The main goal of this work is to demonstrate the efficiency of the
selection criteria for ``$pp\to\gamma+Jet+X$'' events (we shall use 
in what follows the abreviation \gpj for them) that we proposed in 
\cite{9}--\cite{11} and which application may improve the precision
of the jet transverse momentum determination (i.e. of  $\Pt^{Jet}$)
based on assigning a photon $\Ptg$ to a signal produced by a jet.
This note summarises the results of our preliminary publications 
\cite{9}--\cite{11}
\footnote{The analogous work on application of methods developed in
\cite{9}--\cite{11} for a case of Tevatron energies and D0 detector 
geometry was recently fulfiled \cite{D0_Note}.
}
 and includes some modifications connected with the use of jetfinders 
\footnote{In contrast
with \cite{BKS_P1} -- \cite{BKS_P5}, here we use the geometry of
the CMS detector as given in CMSJET \cite{CMJ} and the corresponding
UA1 and UA2 jetfinders of this program (in addition to the LUCELL
jetfinder from PYTHIA) with their default values of parameters
(the only change is that we have increased the cone radius in the UA1
jetfinder from $R_{jet}=0.5$ to $R_{jet}=0.7$). The minijets or
clusters additional to the hard jet (of $\Pt^{jet}\geq 30~ GeV/c$)
are found by the program LUCELL in all events here.}.
Here we shall present the results of our analysis of \gpj events
generated by using PYTHIA 5.7 \cite{PYT}. The results  of
background suppression study in the framework of the GEANT
\cite{GEA} based detector simulation package CMSIM \cite{CMSIM}
are also included \cite{GMS},\cite{GMS_NN}. Further development based on the CMSIM
simulation of the detector response will be presented in our next
papers.

We consider here the case of the LHC luminosity 
$L=10^{33}~ cm^{-2}s^{-1}$. It will be shown below that this value is quite sufficient for selecting 
the event samples of a large enough volume even after an application of much more restrictive new
cuts as well as of new physical variables introduced in \cite{9}--\cite{11}.
Our aim is to select the samples of topologically clean
\gpj events with a good balance of $\Ptg$ and $\Pt^{Jet}$ and to use them for further 
modeling of the jet energy calibration procedure within packages based on the full GEANT simulation 
like CMSIM, for example. In this way one can estimate
a jet energy calibration accuracy that can be achieved with the proposed cuts in the experiment.

Section 2~ is a short introduction into the physics connected with the discussed problem.
General features of \gpj processes at LHC energy are presented here.
We review the possible sources of the $\Pt^{\gamma}$ and $\Pt^{Jet}$ disbalance and the ways of
selecting those events where this disbalance has a minimal value
on the particle level of simulation (we follow here the terminology of [1]).

In Section 3.1 the definitions  are given for the transverse momenta of
different physical objects that we have introduced in \cite{9}--\cite{BKS_P5}
as the quantities that have a meaning of a part of \gpj  event
and that we suppose to be important for studying the physics connected 
with a jet calibration
procedure. These values of transverse momenta 
enter into the $\Pt$-balance equation that reflects the
total $\Pt$ conservation law for the $pp$-collision event as a whole.

Section 3.2 describes the criteria we have chosen to select \gpj events
for the jet energy  calibration procedure. The ``cluster'' (or mini--jet)
 suppression criterion 
($\Pt^{clust}_{CUT}$) which was formulated in an evident form in our previous
publications \cite{9}--\cite{BKS_GLU} is used here.
\footnote{The analogous third jet cut threshold $E^{3}_{T}$ (varying from 20 to
8 GeV) was used in \cite{BERTRAM} for improving a single jet energy
resolution in di-jet events.}
(Its important impact on the  selection of events with a good balance of \Ptgj
will be illustrated in Sections 5--8.)
These clusters have a physical meaning of a part 
of another new experimentally measurable quantity,
introduced in \cite{9}--\cite{BKS_GLU} for the first time, namely,
the sum of $\vec{\Pt}$ of those particles that are {\it out} of the \gpj system
(denoted as $\Pt^{out}$) and are detectable
in the whole pseudorapidity $\eta$ region covered by the detector
($|\eta|\lt5.0$ for CMS). The vector and scalar forms of the total
$\Pt$ balance equation, used for the $pp$-event
as a whole, are given in Sections 3.1 and 3.2 respectively.

Another new thing is a use of a new physical object, proposed also in
\cite{9}--\cite{BKS_GLU} and named an ``isolated jet''. This jet is 
contained in the cone of radius $R=0.7$ in the $\eta-\phi$ 
space and it does not have any noticeable $\Pt$ activity in some ring around.
The width of this ring is taken to be of $\Delta R=0.26$ 
(or, approximately, of the width of 3 calorimeter towers).
In other words, we will select a class of events having  a total $\Pt$ activity 
inside the ring around this ``isolated jet'' within $3-8\%$ of jet $\Pt$.
(It will be shown in Sections 7, 8 and Appendices 2--5  that the number of events
with such a clean topological structure would not be small at LHC energy.)

Section 4 is devoted to the estimation of the size of
the non-detectable neutrino contribution to $\Pt^{Jet}$.
The correlation of the upper cut value, imposed onto $\Pt^{miss}$,
with the mean value of $\Pt$ of neutrinos belonging to 
the jet $\Pt$, i.e. $\la \Pt_{(\nu)}^{Jet}\ra$,
is considered. The detailed results of this section are presented
in the tables of Appendix 1. These tables also include 
the ratios of the ``gluonic events'' $qg\to q+\gamma$ 
containing the information about
the gluon distribution inside a proton. In the same tables
the expected number of events
(at $L_{int}= 3 ~fb^{-1}$) having charm ($c$) and beauty ($b$) quarks 
in the initial state of the gluonic subprocess are also given.

Since the jet energy calibration is rather a practical than an academic task,
in all the following sections we present the rates obtained with the cuts 
varying from strict to weak because their choice would be
a matter of step-by-step statistics collection during the data taking.

Section 5 includes the results of studying the dependence of the 
initial state radiation (ISR) $\Pt$-spectrum on the cut 
imposed on the clusters $\Pt$ ($\Pt^{clust}_{CUT}$) and on the angle 
between the transverse momenta vectors of a jet and a photon.
We also present the rates for three different types of \gpj events,
in which jet fits completely in one definite region of the hadronic
calorimeter: (1) in the Barrel (HB) with $|\eta|\lt1.4$ ; or (2) in the Endcap
 (HE) with $1.4<|\eta|\lt3.0$ or, (3) finally, 
in the Forward (HF) with $3.0\lt|\eta|\lt5.0$.

Starting with Section 6 our analysis is concentrated on  the ``$\gamma+1~jet$''
events having a jet entirely contained (on the particle level) of simulation
within the central calorimeter region. The dependence on $\Pt^{clust}_{CUT}$
of spectra of different physical variables
\footnote{mostly those that have a strong influence on the \ptgj balance in an event.}
(and among them of those appearing in the $\Pt$ balance equation of event as a whole),
as well as the dependence  on it of the spatial  distribution of $\Pt$ activity 
inside a jet, as well as outside it, are shown in Figs.~9--16.

The dependence of the number of events (for $L_{int}=3~fb^{-1}$) on
$\Pt^{clust}_{CUT}$  as well as the dependence on it of the fractional
disbalance $\Fptgj$  is studied in Section 7. The details of this study
are presented in the tables of  Appendices 2--5 that together with the
corresponding Figs.~17--23  can serve to justify the variables and cuts 
introduced in Section 3.  Figs.~18--23 as well as Tables 13--18 of 
Appendices 2 -- 5 demonstrate the influence of the jet isolation
criterion. The impact of $\Pt^{out}_{CUT}$ on the fractional $\Fptgj$
disbalance is shown in Figs.~24 and 25.


In Section 8 we present the estimation of the efficiency of background
suppression (that was one of the main guidelines to establish the 
selection rules proposed in Section 3) for different numerical values of cuts. 
Justification of some of these cuts
introduced in Section 3 for background suppression, based on the GEANT 
simulation with the CMSIM package, is given in \cite{GMS}.

The importance of the simultaneous use of the above-mentioned new parameters 
$\Pt^{clust}_{CUT}$ and $\Pt^{out}_{CUT}$ and also of the ``isolated jet''
 criterion for background 
suppression (as well as for improving the value of the \Ptgj balance)
is demonstrated in Tables 14--22 of Section 8, in Fig.33 and in the tables
of Appendix 6 that show  the dependence of selected events on 
$\Pt^{clust}_{CUT}$ and $\Pt^{out}_{CUT}$ for various $\Ptg$ intervals. 
The tables of Appendix 6 include  
the fractional disbalance values $\Fptgj$ that are found with an additional 
(as compared with tables of Appendix 2--5) account of the $\Pt^{out}$ cut.
In this sense the tables of Appendix 6 contain the final (as they include 
the background contribution)  and {\it first main} result of our study of the 
problem of  setting the absolute scale of  the jet energy at the particle
level defined by generation with PYTHIA.

In Section 9 we show the tables and some plots that demonstrate a possible 
influence of the intrinsic transverse parton momentum $k_t$ parameter variation
(including, as an illustration, some extreme $k_t$ values) on the \ptgj disbalance.


Section 10 contains the {\it second main} result of our  study of \gpj events 
at the LHC energy \cite{BKS_GLU} . Here we investigate the possibility of
using the same sample of the topologically clean \gpj events, obtained with 
the described cuts, for determining the gluon distribution in a proton
(see also \cite{MD1}, \cite{MD_}-- \cite{BALD02} ). The kinematic plot
presented here shows what a region of $x$ and $Q^2$ variables 
(namely: $2\cdot 10^{-4}\leq x\leq 1.0$ and 
$1.6\cdot 10^{3}\leq Q^2\leq8\cdot10^{4} ~(GeV/c)^2$)
can be covered at LHC energies, with a sufficient number of events for this
aim. The comparison with the kinematic regions covered by other experiments
where parton distributions were studied is also shown in the same plot (see Fig.~35).
It is also seen that at the LHC it would be possible to move to the values 
of $Q^2$ by about two orders higher than those that are reached at HERA.

About the Summary. We tried to write it in a way allowing a dedicated reader,
who is interested in result rather than in method, to pass directly to it
just after this sentence.

Since the results presented here were obtained with the PYTHIA simulation,
we are planning to carry out analogous estimations with another event generator
like HERWIG, for example, in subsequent papers.

\section{GENERALITIES OF THE \gpj PROCESS.}
\it\small
\hspace*{9mm}
The useful variables are introduced for studying the effects of the initial and final state radiation 
on the \Ptgj balance 
basing on the simulation in the framework of PYTHIA. Other effects of non-perturbative nature
like primordial parton $k_{~t}$ effect, parton-to-jet hadronization that may also
lead to \ptgj disbalance within the physical models used in PYTHIA are also discussed.
\rm\normalsize
\vskip4mm

\subsection{Leading order picture.}        

\setcounter{equation}{1}

The idea of setting of the absolute scale for a jet energy (and of 
hadronic calorimeter (HCAL) calibration) by means of the physical
process ``$pp(\bar{p})\rrr \gamma+jet+X$'' was realized many 
times in different experiments (see [1--8] and references therein).
It is based on the parton picture where two partons ($q\bar{q}$ or
$qg$), supposed to be moving in different colliding nucleons with
zero transverse momenta
(with respect to the beam line), produce a photon called the ``direct photon''.
This process is described by the leading order (LO) Feynman
diagrams shown in Fig.~1 (for the explanation of the numeration of lines see
Section 2.2) for the ``Compton-like'' subprocess (ISUB=29 in PYTHIA) \\[-20pt]
\begin{eqnarray}
\hspace*{6.04cm} qg\to q+\gamma \hspace*{7.5cm} (1a)
\nonumber
\end{eqnarray}
\vspace{-4mm}
and for the ``annihilation'' subprocess (ISUB=14)
~\\[-8pt]
\begin{eqnarray}
\hspace*{6.02cm} q\overline{q}\to g+\gamma.  \hspace*{7.4cm} (1b)
\nonumber
\end{eqnarray}

In a case when  initial partons have zero transverse momenta the 
$\Pt$ of the final state  ``$\gamma$+parton'' system produced in $2\to 2$
fundamental parton interactions (1) and (2) 
should be also equal to zero, i.e. the following $\Pt$ balance equation for 
photon and final parton should  take place \\[-15pt]
\begin{eqnarray}
\vec{\Pt}^{\gamma+part}=\vec{\Pt}^{\gamma}+\vec{\Pt}^{part} = 0.
\end{eqnarray}
Thus, one may expect that the transverse momentum
of the jet produced by the final state parton ($q$ or $g$), having 
$\vec{\Pt}^{part}=-\vec{\Pt}^{\gamma}$, will
be close in magnitude, with a reasonable precision, 
to the transverse momentum of the final state photon,
 i.e. $\vec{\Pt}^{Jet}\approx-\vec{\Pt}^{\gamma}$.

It allows the absolute jet energy scale to be determined 
(and the HCAL to be calibrated) in the experiments
with a well-calibrated electromagnetic calorimeter (ECAL). To put it simpler,
one can assign to the part of the jet transverse energy $E_t^{Jet}$ 
deposited in the HCAL the value of the difference between the values of
the transverse energy deposited in the ECAL in the photon direction
(i.e. $E_t^{\gamma}$) and  the transverse energy deposited in the ECAL 
in the jet direction. 
~\\[-6.1cm]
\begin{center}
\begin{figure}[h]
  \hspace{10mm} \includegraphics[width=13cm,height=8.3cm]{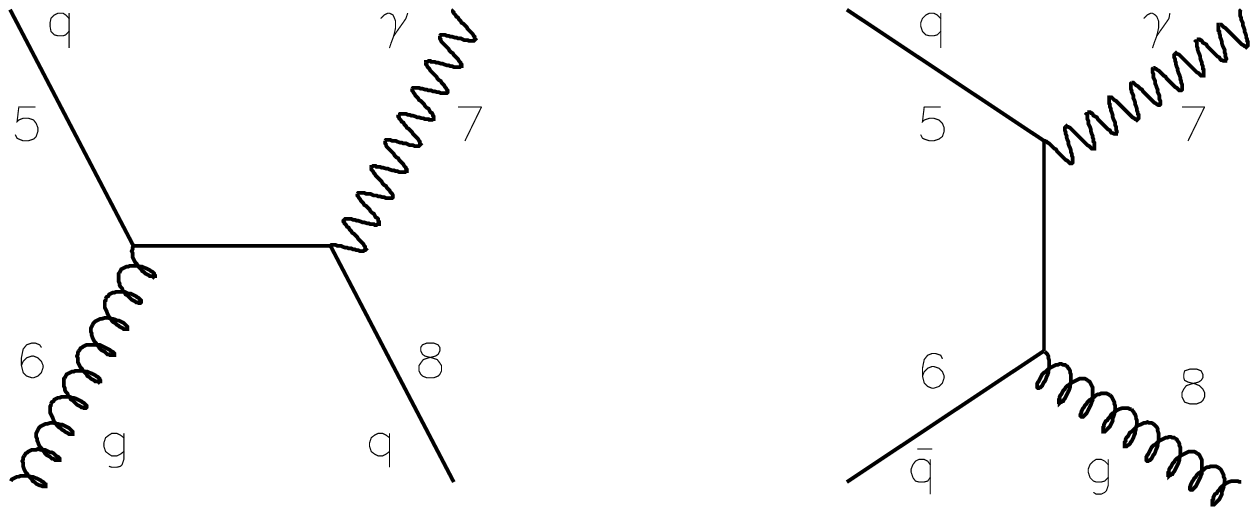} 
\vskip-11mm
\hspace*{42mm} {\small(a)} \hspace*{47mm} {\small(b)}
 \vskip-2mm
\caption{\hspace*{0.0cm} Some of the leading order Feynman diagrams for direct
photon production.} \label{fig1:LO}
\end{figure}
\vskip-10.5mm
\end{center}

\subsection{Initial state radiation.}                           

Since we believe in the perturbation theory, the leading
order (LO) picture described above is expected to be dominant and
to determine the main contribution to the cross section. The Next-to-Leading Order
(NLO) approximation (see some of the NLO diagrams in Figs.~2 and 4) introduces some
deviations from a rather straightforward LO-motivated idea of jet energy calibration.
A gluon ~radiated in the initial state (ISR), as it is seen from Fig.~2, 
can have its own non-zero transverse momentum
$\Pt^{gluon}\equiv \Pt^{ISR}\neq 0$ and thus the total 
transverse momenta of 2 partons that appear in the initial state of fundamental 
$2\rrr2$ QCD subprocesses (1a) and (1b) should not be equal to zero any more. As a 
result of the transverse momentum conservation there should arise a disbalance between 
the transverse momenta of a photon $\Pt^{\gamma}$ and of a parton $\Pt^{part}$ produced 
in the fundamental $2\rrr2$ process $5+6\to 7+8$ shown in Fig.~2 (and in Fig.~3)
and thus, finally, the disbalance between $\Ptg$ and $\Pt$ of a jet produced by 
this parton may appear.
~\\[-6cm]
\begin{center} \begin{figure}[h]
  \hspace{15mm} \includegraphics[width=13cm,height=8.3cm,angle=0]{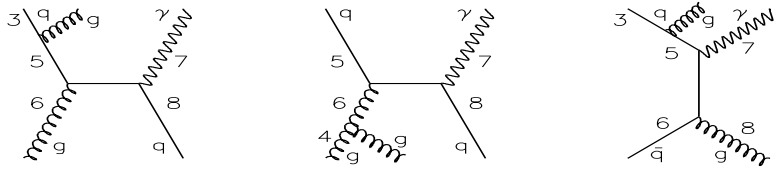}
  \vspace{-12mm}
  \caption{\hspace*{0.0cm} Some of Feynman diagrams of direct photon
production including gluon radiation in the initial state.}
    \label{fig2:NLO}
  \end{figure}
\end{center}
\vspace{-1.0cm}

Following \cite{9}--\cite{BKS_P5} and also \cite{D0_Note} we choose  the modulus of the vector sum of
the transverse momentum vectors $\vec{\Pt}^{5}$ and $\vec{\Pt}^{6}$
of the incoming  into $2\rrr 2$ fundamental QCD subprocesses $5+6\to 7+8$
partons (lines 5 and 6 in Fig.~2) and the sum of their modulus
as two quantitative measures \\[-5pt]
\vspace{-7mm}
\begin{eqnarray}
\Pt^{5+6}=|\vec{\Pt}^5+\vec{\Pt}^6|, \qquad \Pt{56}=|\Pt^5|+|\Pt^6|
\end{eqnarray}
to estimate the $\Pt$ disbalance caused by ISR
\footnote{The variable $\Pt^{5+6}$ was used in analysis in \cite{9}--\cite{BKS_P1}.}. 
The modulus of the vector sum \\[-5mm]
\begin{eqnarray}
\Pt^{\gamma+Jet}=|\vec{\Pt}^{\gamma}+\vec{\Pt}^{Jet}|
\end{eqnarray}
was also used as an estimator of the final state  $\Pt$
disbalance in the \gpj system in \cite{BKS_P1}--\cite{BKS_P5}.

The numerical notations in the Feynman diagrams  (shown in Figs. 1 and 2)
and in formula (3)  are chosen to be in correspondence with those
used in the PYTHIA event listing for description of the parton--parton subprocess
displayed schematically in Fig.~3. The ``ISR'' block describes the initial
state radiation process that can take place before the fundamental
hard $2\to 2$ process.
\begin{center}
  \begin{figure}[h]
  \vspace{-0.8cm}
   \hspace{3cm} \includegraphics[width=10cm,height=5cm,angle=0]{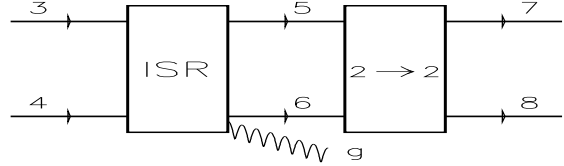}
  \vspace{-2.5cm}
 \caption{\hspace*{0.0cm} PYTHIA ``diagram'' of~ $2\to2$ process (5+6$\to$7+8)
following the block (3+4$\to$5+6) of initial state radiation (ISR), drawn here to illustrate the PYTHIA 
event listing information.}
    \label{fig4:PYT}
  \end{figure}
\end{center}

~\\[-2.3cm]

\subsection{Final state radiation.}                            
\begin{center}
\begin{figure}[htbp]
  \hspace{15mm} \includegraphics[width=13cm,height=8.4cm,angle=0]{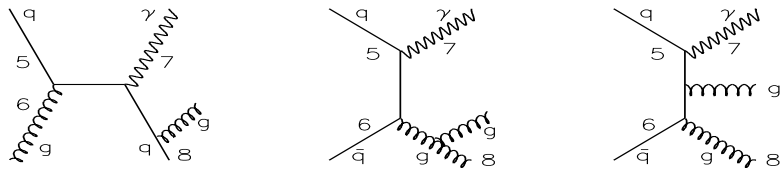}
  \vspace{-12mm}
  \caption{\hspace*{0.0cm} Some of Feynman diagrams of direct photon
production including gluon radiation in the final state.}
    \label{fig4:NLO}
  \end{figure}
\end{center}

Let us consider fundamental subprocesses in which there is
no initial state radiation but instead  final state radiation
(FSR) takes place. These subprocesses are described in the quantum
field theory by the NLO diagrams like those shown in Fig.~\ref{fig4:NLO}. It
is clear that appearance of an extra gluon leg in the final
state may lead to appearance of two (or more) jets or an
intense jet and a weaker jet (mini-jet or cluster) in an event as it happens 
in the case of ISR described above.
So, to suppress FSR (manifesting itself as some extra
jets or clusters) the same tools as for reducing ISR should be
used. But due to a usage in PYTHIA of the string model of fragmentation,
which has a non-perturbartive nature,  it is much more 
difficult to deduce (basing on the PYTHIA event listing information)
the variables (analogous to (3) and (4)) to describe the final state 
disbalance between $\Pt$ of a jet and $\Ptg$. That is why, keeping in
 mind a close analogy of the physical pictures of ISR and FSR (see
Figs.~\ref{fig2:NLO} and \ref{fig4:NLO}), we shall concentrate 
in the following sections on studing of the initial state radiation
supposing that the methods of its reduction should be also usful
for suppression of FSR. 
\\[-6cm]
\vspace{-1.0cm}

\subsection{Primordial parton $k_t$ effect.}                   

Now after considering the disbalance sources connected with the
perturbative corrections to the leading order diagrams let us
mention the physical effects of the non-perturbative nature. Thus,
a possible non-zero value of the intrinsic transverse parton
velocity inside a colliding proton may be another source of the
$\Pt^{\gamma}$ and $\Pt^{part}$ disbalance in the final state. 
This effect at the present stage of theoretical understanding of 
soft physics can be described mainly in a phenomenological way. Its
reasonable value is supposed to lead to the following limit on the value 
of  intrinsic transverse momentum $k_t \leq \,1.0 ~GeV/c$ of a parton. It 
should be noted that sometimes in the literature the total effect of ISR 
and of the intrinsic parton transverse momentum is denoted by a common
symbol ``$k_t$''. Here we follow the approach and the phenomenological  
model used in PYTHIA where these two sources of the \Ptgj disbalance, 
having different nature, perturbative and non-perturbative ones, 
can be  switched on separately by different keys (MSTP(61) for ISR and 
PARP(91), PARP(93), MSTP(91) for intrinsic parton transverse momentum 
$k_t$). In what follows we shall keep the value of $k_t$ mainly to be 
fixed  by the PYTHIA default value $\langle k_t \rangle=0.44~ GeV/c$. 
The dependence of the disbalance between $\Pt^{\gamma}$ and $\Pt^{Jet}$
on a possible variation of $k_t$ will be discussed in detail in Section 9.
The general conclusion from there is that any variation of $k_t$ within
reasonable boundaries (as well as slightly beyond them) does not produce a
large effect in the case when the initial state radiation is
switched on. The latter makes a dominant contribution.

 \subsection{Parton-to-jet hadronization.}                      

Another non-perturbative effect that leads to the \ptgj disbalance
is connected with hadronization (or fragmentation into hadrons) of the 
parton produced in the fundamental $2\to 2$ subprocess into a jet. The
hadronization of the parton into a jet is described in PYTHIA 
within the Lund string fragmentation model. The mean values of 
the fractional $\Pt^{Jet}-\Pt^{parton}$ disbalance will be presented 
in the tables of Appendices 2 -- 5 for three different jetfinders. 
As it will be shown in Section 7 (see also tables of Appendices 2--5)
the hadronization effect contribution into \ptgj disbalance may be 
approximately of the same size as that of ISR.





\section{CHOICE OF MEASURABLE PHYSICAL VARIABLES FOR THE \gpj
PROCESS AND THE CUTS FOR BACKGROUND REDUCTION.}               

\it\small
\hspace*{9mm}
The classification of different physical objects that participate in \gpj events and that may give
a noticeable contribution into the total $\Pt$-balance in the event as a whole is done.

Two new physical observables, namely, $\Pt$ of a cluster and $\Pt$ of all detectable particles beyond 
\gpj system, as well as the definion of isolated jet, proposed for studying \ptgj disbalance in 
\cite{9}--\cite{BKS_P5}, are discussed.

The selection cuts that would be imposed onto the physical observables of \gpj events are presented.

The  $\Pt$-balance equation for the event as a whole is written in scalar form that allow to express
the \ptgj disbalance in terms of the considered physical variables.
\rm\normalsize
\vskip4mm

Apart from (1a) and (1b), other QCD subprocesses with large cross sections, 
by orders of magnitude larger than the
cross sections of (1a) and (1b), can also lead to high $\Pt$ photons
and jets in final state. So, we face the problem of selecting
signal \gpj events from a large QCD background.
Here we shall discuss the choice of physical variables that would be
useful, under some cuts on their values, for separation of the
desirable processes with direct photon (``$\gamma^{dir}$'') 
from the background events. The possible ``$\gamma^{dir}-$candidate''
may originate from the $\pi^0,~\eta,~\omega$ and $K^0_s$ meson decays 
\cite{GMS}, \cite{GMS_NN} or may be caused by a bremsstrahlung photon 
or by an electron (see Section 8).

We take the CMS ECAL size to be limited by
$|\eta|\!\! \leq \!\! 2.61$ and the HCAL to consist of the Barrel (HB), Endcap
(HE) and Forward (HF) parts and to be limited by $|\eta| \leq 5.0$,
 where $\eta = -ln~(tan~(\theta/2))$ is a pseudorapidity
defined in terms of a polar angle $\theta$ counted from the beam line. In the plane
transverse to the beam line the azimuthal angle $\phi$ defines the
directions of $\vec{\Pt}^{Jet}$ and $\vec{\Pt}^{\gamma}$. \\

\subsection{Measurable physical observables and the $\Pt$ vector balance equation.}

In $p p\to \gamma + Jet + X$ events we are going to study
the main physical object will be a high $\Pt$ jet to be detected
in the $|\eta|\lt5.0$ region and a direct photon registered by the
ECAL up to $|\eta|\lt2.61$. In these events there will be a set of
particles mainly caused by beam remnants, i.e. by spectator parton
fragments, that are flying mostly in the direction of a non-instrumented
volume ($|\eta|>5.0$) in the detector. Let us denote the total transverse
momentum of these non-observable particles ($i$) as \\[-4mm]
\begin{equation}
\sum\limits_{i
\in |\eta| \gt5.0} \vec{\Pt}^i \equiv \vec{\Pt}^{|\eta| \gt5.0}.  \label{eq:sel1}
\end{equation} 

Among the particles  with $|\eta|\lt5.0$ there may also be neutrinos.
We shall denote their total momentum as
\begin{equation}
\sum\limits_{i \in |\eta|\lt5.0} \vec{\Pt}_{(\nu)}^i
 \equiv \vec{\Pt}_{(\nu)}.
\label{eq:sel2}
\end{equation}
\vspace{-2mm}

\noindent
The sum of the transverse momenta of these two kinds of non-detectable particles
will be denoted as $\Pt^{miss}$
\footnote{This value is a part of true missing $\Pt$ in an experiment that includes the detector
effects (see [1, 2]).}:
\vspace{-3mm}
\begin{eqnarray}
 \vec{\Pt}^{miss} = \vec{\Pt}_{(\nu)} + \vec{\Pt}^{|\eta|>5.0}.
\end{eqnarray}

\vspace{-2mm}

A high-energy jet may also contain neutrinos that may carry
part of the total jet energy and of $\Pt^{Jet}$. The average
values of these neutrino parts can be estimated from simulation.

We shall separate from the total jet transverse momentum $\vec{\Pt}^{Jet}$  
the part that can be measured in the detector, i.e. in
the ECAL+HCAL calorimeter system and in the muon system. Let us denote
this detectable part as $\vec{\Pt}^{jet}$ (small ``j''!).
 So, we shall
present the total jet transverse momentum $\vec{\Pt}^{Jet}$ as a sum of three
parts:

1. $\vec{\Pt}^{Jet}_{(\nu)}$, containing the
contribution of neutrinos that belong to the jet, i.e.
a non-detectable part of a jet $\Pt$ ($i$ - neutrino):
\vspace{-4mm}
\begin{eqnarray}
 \vec{\Pt}^{Jet}_{(\nu)} = \sum\limits_{i \in Jet} \vec{\Pt}_{(\nu)}^i.
\end{eqnarray}
~\\[-20pt]

2. $\vec{\Pt}^{Jet}_{(\mu)}$, containing the
contribution of jet muons to $\vec{\Pt}^{Jet}$ ($i$ - muon):
\vspace{-2mm}
\begin{eqnarray}
 \vec{\Pt}^{Jet}_{(\mu)} = \sum\limits_{i \in Jet} \vec{\Pt}_{(\mu)}^i.
\end{eqnarray}
~\\[-20pt]

These muons make a weak signal in the calorimeter
but their energy can be measured, in principle, in the muon or muon+tracker systems (in the
region of $|\eta|\lt2.4$ in the case of CMS geometry).
Due to the absence of the muon system and the tracker
beyond the $|\eta|\lt2.4$ region,
there exists a part of $\Pt^{Jet}$ caused by muons with $|\eta|>2.4$.
We denote this part as $\Pt^{Jet}_{(\mu,|\eta|>2.4)}$. It
can be considered, in some sense, as the analogue of $\Pt^{Jet}_{(\nu)}$ 
since the only trace of its presence would be weak MIP signals in calorimeter towers.

As for both points 1 and 2 above, let us say in advance that
the estimation of the average values of the neutrino and muon
contributions to $\Pt^{Jet}$ (see Section 4
and Tables 1--8 of Appendix 1) has shown
that they are quite small: about $0.35\%$--$0.50\%$ of $\la\Pt^{Jet}\ra_{all}$ 
is due to neutrinos and about $0.25\%$ of $\la\Pt^{Jet}\ra_{all}$ to muons,
where ``$all$'' means averaging over
all events including those without neutrinos in jets.
So, they together may cause  approximately about $0.6\%$ of \Ptgj disbalance
if muon signal is lost.

3. And finally, as we have mentioned before,
we use $\vec{\Pt}^{jet}$ to denote the part of $\vec{\Pt}^{Jet}$ which
includes all detectable particles of the jet
\footnote{We shall consider the issue of charged particles
contribution with small $\Pt$ 
into the total jet $\Pt$ while discussing the results of the full GEANT
 simulation (with account of the magnetic field effect) in our forthcoming
papers.} , i.e. the sum of $\Pt$ of jet particles that may produce a signal
in the calorimeter and muon system (calo=ECAL+HCAL signal)
\vspace{-2mm}
\begin{eqnarray}
 \vec{\Pt}^{jet} =  \vec{\Pt}^{Jet}_{(calo)} +\vec{\Pt}^{Jet}_{(\mu)} ,
\quad |\eta^{\mu}|\lt2.4.
\end{eqnarray}

Thus, in a  general case we can write for any $\eta$ values:
\vspace{-1.5mm}
\begin{eqnarray}
 \vec{\Pt}^{Jet}
=\vec{\Pt}^{jet}
+\vec{\Pt}^{Jet}_{(\nu)}
+\vec{\Pt}^{Jet}_{(\mu,|\eta^{\mu}|>2.4)}.
\end{eqnarray}

In the case of $p p\to \gamma + Jet + X$ events
the particles detected in the $|\eta|\lt5.0$ region may originate
from the fundamental subprocesses (1a) and (1b) corresponding to 
LO diagrams shown in Fig.~1, as well as from the processes corresponding 
to NLO diagrams (like those in Figs.~2, 4 that include ISR and FSR),
and also from the ``underlying'' event [1], of course.

As was already mentioned in Section 2,
the final states of fundamental subprocesses (1a) and (1b) may contain
additional jets due to the ISR and final state radiation (FSR) caused by the
higher order QCD corrections to the LO Feynman diagrams shown in Fig.~2 and 
Fig.~4. To understand and then to realize the jet energy calibration procedure,
we need to use the event generator to find the criteria for selection
of events with a good balance of $\vec{\Pt}^{\gamma}$ with the
$\vec{\Pt}^{jet}$ part measurable in the detector. It means that to make 
a reasonable simulation of the calibration procedure,
we need to have a selected sample of generated events having a small
$\Pt^{miss}$ (see Section 4) contribution and use it as a model.
We also have to find a way to select 
events without additional jets or with jets suppressed down to the level of mini-jets or clusters
having very small $\Pt$.

So, for any event we separate the particles in the $|\eta|\lt5.0$ region
into two subsystems. The first one consists of the particles belonging to the 
``$\gamma +Jet$'' system (here ``$Jet$'' denotes the jet with the highest
$\Pt \geq 30 ~GeV/c$) having the total transverse momentum $\vec{\Pt}^{\gamma +Jet}$ 
(large ``Jet'', see (11)). The second subsystem involves all other ($O$) particles
beyond the ``$\gamma +Jet$'' system in the  region, covered by the detector, 
i.e. $|\eta|\lt5.0$.
Let us mention that the value of $\vec{\Pt}^{\gamma +Jet}$ 
may be different from the value of observable:\\[-15pt]
\begin{eqnarray}
\vec{\Pt}^{\gamma +jet} =
\vec{\Pt}^{\gamma} + \vec{\Pt}^{jet} \quad {\rm (small ``jet'')},
\end{eqnarray}
~\\[-10pt]
\noindent
in the case of non-detectable particles presence in a jet.
The total transverse momentum of this
$O$-system is  denoted as $\Pt^{O}$ and it is a sum of
$\Pt$  of additional mini-jets (or clusters) and $\Pt$ of
single hadrons, photons and leptons in the $|\eta| \lt 5.0$ region. Since a part of
neutrinos are also present among these leptons, 
the difference of $\vec{\Pt}_{(\nu)}$ and $\vec{\Pt}^{Jet}_{(\nu)}$
gives us the transverse momentum \\[-12pt]
\begin{eqnarray}
 \vec{\Pt}^{O}_{(\nu)} = \vec{\Pt}_{(\nu)} - \vec{\Pt}^{Jet}_{(\nu)} \quad
|\eta^{\nu}|\lt5.0,
\end{eqnarray}

\noindent
carried out by the neutrinos that do not belong to the jet but are
contained in the $|\eta|\lt5.0$ region.

We denote by $\vec{\Pt}^{out}$ a part of $\vec{\Pt}^O$ that can be measured,
in principle, in the detector.
Thus, $\vec{\Pt}^{out}$ is a sum of $\Pt$ of other mini-jets or, generally,
clusters (with $\Pt^{clust}$ smaller than $\Pt^{Jet}$) and $\Pt$ of single
hadrons ($h$), photons ($\gamma$) and electrons ($e$) with $|\eta| \lt 5.0$
and muons ($\mu$) with $|\eta^\mu| \lt 2.4$ that are out of the \gpj system.
For simplicity these mini-jets and clusters will be called ``clusters''
\footnote{As was already mentioned in  Introduction, these clusters are found
by the LUCELL jetfinder with the same value of the cone radius as 
for jets: $R^{clust}=R^{jet}=0.7$.}.
So, for our \gpj events $\vec{\Pt}^{out}$ is the following sum 
(all $\{h,~\gamma,~e,~\mu \} \not\in$ Jet):\\[-12pt]
\begin{eqnarray}
 \vec{\Pt}^{out} =
\vec{\Pt}^{clust}
+\vec{\Pt}^{sing}_{(h)}
+\vec{\Pt}^{nondir}_{(\gamma)}
+\vec{\Pt}^{}_{(e)}+\vec{\Pt}^{O}_{(\mu, |\eta^\mu|\lt2.4)}; \quad  |\eta|\lt5.0.
\end{eqnarray}

\noindent
And thus, finally, we have:\\[-18pt]
\begin{eqnarray}
 \vec{\Pt}^{O} =
\vec{\Pt}^{out}+\vec{\Pt}^{O}_{(\nu)}+\vec{\Pt}^{O}_{(\mu, |\eta^\mu|>2.4)}.
\end{eqnarray}

\noindent
With these notations we come to the following vector form \cite{BKS_P1} of
the $\Pt$- conservation law for the ``$\gamma + Jet$'' event
(where $\gamma$ is a direct photon) as a whole (supposing that the jet and the photon are contained
in the corresponding detectable regions):\\[-14pt]
\begin{eqnarray}
\vec{\Pt}^{\gamma} +
\vec{\Pt}^{Jet} +
\vec{\Pt}^{O}+
\vec{\Pt}^{|\eta|>5.0} = 0
\end{eqnarray}

\noindent
with last three terms defined correspondingly by (11), (15) and (5) respectively.

\subsection{Definition of selection cuts for physical variables and
the scalar form of the $\Pt$ balance equation.}

\noindent
1. We shall select the events with one jet and one ``$\gamma^{dir}$-candidate''
(in what follows we shall designate it as $\gamma$ and call the
``photon'' for brevity and only in Section 8, devoted to the backgrounds, 
we shall denote $\gamma^{dir}$-candidate by $\tilde{\gamma}$) with\\[-11pt]
\begin{equation}
\Pt^{\gamma} \geq 40~ GeV/c~\quad {\rm and} \quad \Pt^{Jet}\geq 30 \;GeV/c.
\label{eq:sc1}
\end{equation}
The ECAL signal can be considered as a candidate for a direct photon
if it fits inside the 5$\times$5 ECAL crystal cell window having a cell
with the highest $\Pt$ $\gamma/e$ in the center (\cite{CMS_EC}).

For most of our applications in Sections 4, 5 and 6 mainly the PYTHIA
jetfinding algorithm LUCELL will be used.
The jet cone radius R in the $\eta-\phi$ space, counted from the
``jet initiator cell (ic)'', is
taken to be $R_{ic}=((\Delta\eta)^2+(\Delta\phi)^2)^{1/2}=0.7$.
Below in Section 6 we shall also consider the jet radius counted
from the center of gravity (gc) of the jet, i.e. $R_{gc}$.
Comparison with the UA1 and UA2 jetfinding algorithms (taken from the CMSJET
program of fast simulation \cite{CMJ}) is presented in Sections 6 and 7.

\noindent
2. To suppress the contribution of background processes, i.e. to select mostly the events with ``isolated''
direct photons and to discard the events with fake ``photons'' (that
may originate as $\gamma^{dir}$-candidates from meson decays, for instance), we restrict:

a) the value of the scalar sum of $\Pt$ of hadrons and other particles surrounding
a ``photon'' within a cone of $R^{\gamma}_{isol}=( (\Delta\eta)^2+(\Delta\phi)^2)^{1/2}=0.7$
(``absolute isolation cut'')
\footnote{We have found that $S/B$ ratio with $R^{\gamma}_{isol}\!=\!0.7$ is in about 1.5 times better
than with $R^{\gamma}_{isol}\!=\!0.4$ what is accompanied by only $10\%$ of additional loss of the number of
signal events.}
\\[-7pt]
\begin{equation}
\sum\limits_{i \in R} \Pt^i \equiv \Pt^{isol} \leq \Pt_{CUT}^{isol};
\label{eq:sc2}
\end{equation}
\vspace{-2.6mm}

b) the value of a fraction (``fractional isolation cut'')\\[-7pt]
\begin{equation}
\sum\limits_{i \in R} \Pt^i/\Pt^{\gamma} \equiv \epsilon^{\gamma} \leq
\epsilon^{\gamma}_{CUT}.
\label{eq:sc3}
\end{equation}

\noindent
3. To be consistent with the application condition of the NLO
formulae, one should avoid an infrared dangerous region and take care of
$\Pt$ population in the region close to a $\gamma^{dir}$-candidate
(see \cite{Fri}, \cite{Cat}). In accordance with \cite{Fri} and \cite{Cat}, 
we also restrict the scalar sum of $\Pt$ of particles
 around a ``photon'' within a cone of a smaller radius $R^{\gamma}_{singl}=0.2$.

Due to this cut,\\[-4mm]
\begin{equation}
\sum\limits_{i \in R^{\gamma}_{singl}} \Pt^i \equiv \Pt^{singl} \leq 2~ GeV/c
~~~~~(i\neq ~\gamma^{dir}),
\label{eq:sc4}
\end{equation}
an ``isolated" photon with high $\Pt$ also becomes a ``single'' one within
an area of 8 calorimeter towers (of size 0.087$\times$0.087  according to CMS geometry)
which surround the tower fired by it, i.e. a tower with the highest $\Pt$ 
(an analog of the 3$\times$3 tower window algorithm).

\noindent
4. We accept only the events having no charged tracks (particles)
with $\Pt>1~GeV/c$ within the $R=0.4$ cone around the $\gamma^{dir}$-candidate.

\noindent
5. We also consider the structure of every event with the photon
candidate at a more precise level of the 5$\times$5 crystal cell window 
with a cell size of 0.0175$\times$0.0175. 
To suppress the background events with photons resulting from
$\pi^0$, $\eta$, $\omega$
and $K_S^0$ meson decays, we require the absence of a high $\Pt$ hadron
in the tower containing the $\gamma^{dir}$-candidate:\\[-10pt]
\begin{equation}
\Pt^{hadr} \leq 5~ GeV/c.
\label{eq:sc5}
\end{equation}

\noindent
At the PYTHIA level of simulation this cut may effectively take into account 
the imposing of an upper cut on the HCAL signal in the towers behind
the ECAL tower fired by the direct photon (see Section 8 for details).
We can not reduce this value down to, for example, 2-3 $GeV/c$, because
a hadron with $\Pt$ below 2-3 $GeV/c$ deposits with high probability most of
its energy in ECAL and
may not reveal itself in HCAL. The value 5 $GeV/c$ is chosen with account of possible
loss of hadron energy in ECAL (see  \cite{GMS}).

\noindent   
6. We select the events with the vector $\vec{\Pt}^{Jet}$ being ``back-to-back'' to
the vector $\vec{\Pt}^{\gamma}$ (in the plane transverse to the beam line)
within $\dphi$ defined by the equation:\\[-12pt]
\begin{equation}
\phigj=180^\circ \pm \Delta\phi,
\label{eq:sc7}
\end{equation}
where $\phigj$ is the angle
between the \Ptgj vectors: 
$\vec{\Pt}^{\gamma}\vec{\Pt}^{Jet}=\Pt^{\gamma}\Pt^{Jet}\cdot cos(\phigj)$,
$\Pt^{\gamma}=|\vec{\Pt}^{\gamma}|,~~\Pt^{Jet}=|\vec{\Pt}^{Jet}|$.
The cases $\Delta\phi \leq 15^\circ, 10^\circ, 5^\circ$ (see Figs. 25,27,29 of
Section 8) are considered in this paper
($5^\circ$ is, approximately, one CMS HCAL tower  size in $\phi$).

\noindent
7. The initial and final state radiations (ISR and FSR) manifest themselves most clearly
as some final state mini-jets or clusters activity.
To suppress it, we impose a new cut condition that was not formulated in
an evident form in previous experiments: we choose the \gpj events
that do not have any other
jet-like or cluster high $\Pt$ activity  by selecting the events with the values of
$\Pt^{clust}$ (the cluster cone $R_{clust}(\eta,\phi)=0.7$), being lower than some threshold
$\Pt^{clust}_{CUT}$ value \cite{BERTRAM}, i.e. we select the events with\\[-10pt]
\begin{equation}
\Pt^{clust} \leq \Pt^{clust}_{CUT}
\label{eq:sc8}
\end{equation}
($\Pt^{clust}_{CUT}=15, 10, 5 ~GeV/c$ are most effective as will be shown in Sections 6--8).
Here, in contrast to \cite{BKS_P1}--\cite{BKS_P5}, the clusters are found
by one and the same jetfinder LUCELL while three different jetfinders UA1, UA2 and LUCELL
are used to find the ``leading jet'' (i.e. with $\Pt^{Jet}\geq 30~ GeV/c$) in the event.

\noindent
8. Now we pass to another new quantity (proposed also for the first time in \cite{BKS_P1}--\cite{BKS_P5}) 
that can be measured at the experiment.
We limit the value of the modulus of the vector sum of $\vec{\Pt}$ of all
particles, except those of the \gpj system, that fit into the region
 $|\eta|\lt5.0$ covered by the ECAL and HCAL, i.e., we limit the 
signal in the cells ``beyond the jet and photon'' region,
i.e. $i\not\in Jet,\gamma-dir$, by the following cut:
\begin{equation}
\left|\sum_{i\not\in Jet,\gamma-dir}\!\!\!\vec{\Pt}^i\right| \equiv \Pt^{out} \leq \Pt^{out}_{CUT},
~~|\eta^i|\lt5.0.
\label{eq:sc9}
\end{equation}

\noindent
The importance of $\Pt^{out}_{CUT}$ and $\Pt^{clust}_{CUT}$
for selection of events with a good balance of \Ptgj and for
the background reduction will be demonstrated in Sections 7 and 8.

Below the set of selection cuts 1 -- 8 will be referred to as
``Selection 1''. The last two of them, 7 and 8, are new criteria \cite{BKS_P1}
not used in previous experiments. 

9. In addition to them one more new object, introduced in 
\cite{BKS_P1} -- \cite{BKS_P5} and named an ``isolated jet'',  will be
 used in our analysis,i.e. we shall require the presence of a ``clean 
enough'' (in the sense of a limited $\Pt$ activity) region inside 
the ring of $\Delta R=0.26$ width (or approximately of a size of 
three calorimeter towers) around the jet.  Following this picture,
we restrict the ratio of the scalar sum
of transverse momenta of particles belonging to this ring, i.e.\\[-5pt]
\begin{equation}
\Pt^{ring}/\Pt^{jet} \equiv \epsilon^{jet}\leq\epsilon^{jet}_0, \quad {\rm where ~~~~ }
\Pt^{ring}=\!\!\!\!\sum\limits_{\footnotesize i \in 0.7\lt R \lt1.0} \!\!\!\!|\vec{\Pt}^i|.
\label{eq:sc10}
\end{equation}
~\\[-4pt]
($\epsilon^{jet}_0$ is chosen to be $3-8\%$, see Sections 7 and 8).

The set of cuts 1 -- 9 will be called in what follows ``Selection 2''.

\noindent
10. In  the following we shall consider also ``Selection 3'' where we shall 
keep only those events in which one and the same jet  is found
 simultaneously by every of three jetfinders used here: UA1, UA2 and LUCELL 
(i.e. up to a good accuracy having the same values of 
$\Pt^{Jet}, ~R^{jet}_{gc}$ and $\Delta\phi$).
For these jets (and also clusters in the same event) we require 
the following conditions:\\[-5mm]
\begin{eqnarray}
\Pt^{Jet}>30~GeV/c, \qquad \Pt^{clust}\lt \Pt^{clust}_{CUT},\qquad
\dphi\lt15^\circ (10^\circ,~5^\circ), \qquad \epsilon^{jet} \leq 3-8\%
\end{eqnarray}

The exact values of the cut parameters $\Pt^{isol}_{CUT}$,
$\epsilon^{\gamma}_{CUT}$, $\epsilon^{jet}$, $\Pt^{clust}_{CUT}$, $\Pt^{out}_{CUT}$
 will be specified below, since they may be
different, for instance, for various $\Pt^{\gamma}$ intervals
(being looser for higher  $\Pt^{\gamma}$).

\noindent 
11. As we have already mentioned in Section 3.1, one can
expect reasonable results of the jet energy calibration procedure
modeling and subsequent practical realization
 only if one uses a set of selected events with small $\Pt^{miss}$. So, we also use
the following cut:\\[-17pt]
\begin{eqnarray}
\Pt^{miss}~\leq \Pt^{miss}_{CUT}.
\label{eq:sc11}
\end{eqnarray}
For this reason we shall study in the next Section 4 the influence of
$\Pt^{miss}$ parameter on the selection of events with a reduced value
of the total sum of neutrino contribution into $\Pt^{Jet}$, i.e. 
$\Pt^{Jet}_{(\nu)}$.
The aim of the event selection with small $\Pt^{Jet}_{(\nu)}$
is quite obvious: we need a set of events with a reduced
$\Pt^{Jet}$ uncertainty due to a possible presence of a non-detectable
particle contribution to a jet
\footnote{In Section 8 we also underline the importance of this cut for reduction of $e^\pm$
events contribution to the background to the signal $\gamma^{dir}+jet$ events.}.

To conclude this section, let us write the basic $\Pt$-balance equation (16) of the previous section
with the notations introduced here in the form
more suitable to present the final results. For this
purpose we shall write equation (16) in the following scalar form (see also \cite{BKS_P1}, 
\cite{D0_Note}): \\[-12pt]
\begin{eqnarray}
\frac{\Pt^{\gamma}-\Pt^{Jet}}{\Pt^{\gamma}}=(1-cos\dphi) 
+ \Db/\Pt^{\gamma}, \label{eq:sc12}
\label{eq:sc_bal}
\end{eqnarray}
where
$\Db\equiv (\vec{\Pt}^{O}+\vec{\Pt}^{|\eta|>5.0)})\cdot \vec{n}^{Jet}$ ~~~~ with ~~
$\vec{n}^{Jet}=\vec{\Pt}^{Jet}/\Pt^{Jet}$.

As will be shown in Section 7, the first term on the
right-hand side of equation (\ref{eq:sc_bal}), i.e. $(1-cos\dphi)$ is negligibly
small in a case of Selection 1, as compared with the second term 
and tends to decrease fast with  growing $\Pt^{Jet}$. 
So, in this case the main contribution to the $\Pt$ disbalance in the
\gpj system is caused by the term $\Db/\Pt^{\gamma}$.

\section{ESTIMATION OF A  NON-DETECTABLE PART OF $\Pt^{Jet}$.}       

%
\it\small
\hspace*{9mm}
   The contribution to  $\Pt^{jet}$  from neutrino ($\Pt^{Jet}_{(\nu)}$)
  is estimated. It is shown that a cut imposed
  onto the value of  $\Pt^{miss}$ allows to select events with a negligibly 
  small averaged value of   $\la\Pt^{Jet}_{(\nu)}\ra_{all\; events}$.
  The values of the corresponding neutrino corrections to a  measurable    
  quantity  $\vec{\Pt}^{jet}$ 
  are given in the tables of Appendix 1. The  estimations of
  the number of events with charm and beauty quarks in different intervals
  of $\Pt^{Jet}$ as well as the averaged values of jet radius and the ratios
  of ``gluonic process'' (1a) are also included there.
\rm\normalsize
\vskip4mm

%
In Section 3.1 we have separated the transverse momentum of a jet,
i.e. $\Pt^{Jet}$, into two parts,
a detectable $\Pt^{jet}$ and a non-measurable 
($\Pt^{Jet}-\Pt^{jet}$), consisting of $\Pt^{Jet}_{(\nu)}$
and $\Pt^{Jet}_{(\mu,|\eta|>2.4)}$ (see (11).
In the same way, according to equation (15), we divided
the transverse momentum $\Pt^O$ of ``other particles'' that are ``out'' of a jet and
a direct photon system
into a detectable part $\Pt^{out}$
and a non-measurable part consisting of the sum of $\Pt^O_{(\nu)}$
 and $\Pt^O_{(\mu,|\eta|>2.4)}$.

We shall estimate here what part of $\Pt^{Jet}$ may be
carried out by non-detectable particles
\footnote{First we shall consider the case of switched-off decays of
 $\pi^\pm$ and $K^\pm$ mesons
(according to the PYTHIA default agreement $\pi^\pm$ and  $\;K^{\pm}$ mesons 
are stable).}.
For this aim we shall use the bank of the signal \gpj events, i.e. caused by subprocesses
(1a) and (1b),
generated for three $\Ptg$ intervals:
$40<\Ptg<50$, $100<\Ptg<120$ and $300<\Ptg<360~GeV/c$
and selected with restrictions (17) -- (24) (Selection 1) and the following cut values:
\begin{equation}
 \Pt_{CUT}^{isol}=5 ~GeV/c,~~\epsilon^{\gamma}_{CUT}=7\%,
~~\dphi<15^\circ,~~\Pt^{clust}_{CUT}=30~ GeV/c.
\end{equation}
Here the cut $\Pt^{clust}_{CUT}=30~ GeV/c$ has the meaning of a very weak restriction
on mini-jets or clusters activity.
No restriction was imposed on the $\Pt^{out}$ value.
The results of analysis of these events are presented in Fig.~5.

The first row of Fig.~\ref{fig20-22}
contains  $\Pt^{miss}$ spectra of the \gpj events for different
$\Ptg$ intervals which demonstrate (to a good accuracy)
their practical independence of $\Ptg$.

In the second row of Fig.~\ref{fig20-22} we present the spectra of
$\Pt^{miss}$ for those events (denoted as $\Pt^{Jet}_{(\nu)}>0$) which
contain jets having neutrinos with  a non-zero $\Pt^{Jet}_{(\nu)}$
component of  $\Pt^{Jet}$. From these figures the dependence of the $\Pt^{miss}$ spectrum
on the direct photon $\Pt^{\gamma}$ 
(approximately equal to $\Pt^{Jet}$) is clearly seen:  the spectrum tails as well as
the mean values shift to a large $\Pt^{miss}$ region with growing $\Pt^{Jet}$.
(At the same time the peak position
remains in the region of $\Pt^{miss}<5 ~GeV/c$.) Comparison of the
number of entries in the second row plots of Fig.~\ref{fig20-22} with
those in the first row allows the conclusion that the part of events
with the jet having the non-zero neutrinos contribution 
($\Pt^{Jet}_{(\nu)}>0$) has practically the same
size of about $3.3\%$ in all $\Ptg$ (or $\Pt^{Jet}$) intervals.

The same spectra of $\Pt^{miss}$ for events with $\Pt^{Jet}_{(\nu)}>0$
show how many of these events shall  remain after imposing a cut on
$\Pt^{miss}$  (defined by (7))  in every $\Ptg$ interval.( As it will be shown 
in Section 8 $\Pt^{miss}_{CUT}$ cut also reduces the contribution 
to background from the decay subprocesses ~$q\,g \to q' + W^{\pm}$~ and
~$q\bar{~q'} \to g + W^{\pm}$~  with the subsequent decays
$W^{\pm} \to e^{\pm}\nu$ that lead to a substantial $\Pt^{miss}$ value
and are the  main source of electrons $e^\pm$ that may appear  as 
direct photon candidates.)

The important thing is that such a reduction of the number of events
with $\Pt^{Jet}_{(\nu)}>0$~  leads  to reduction  of 
the mean value of the $\Pt^{Jet}_{(\nu)}$, i.e. the value averaged over all
collected events $\la\Pt^{Jet}_{(\nu)}\ra_{all\; events}$, in every $\Ptg$ interval.

This value, found from PYTHIA generation, may serve as a model correction
$\Delta_\nu$ and it has to be estimated for proper determination
of the total $\Pt^{Jet}$. 
So, due to practical coincidence of directions of vectors of these terms
we can write $\Pt^{Jet}=\Pt^{jet} + \Delta_\nu + \Delta_\mu $, ($|\eta^{\mu}|>2.4$)
where $\Delta_\nu=\la\Pt^{Jet}_{(\nu)}\ra_{all\; events}$ and
 $\Delta_\mu=\la{\Pt}^{Jet}_{(\mu,|\eta^{\mu}|>2.4)}\ra_{all\;events}$. ( As 
we plan to use in this paper only events with jets belonging to Barell part
of calorimeter,  $\Delta_\mu$ is not so important for our analysis.)

The effect of imposing a general $\Pt^{miss}_{CUT}$  in each event of our
sample is shown in the third row of Fig.~\ref{fig20-22}. The upper
cut 
$\Pt^{miss}_{CUT}=1000 ~GeV/c$, as is seen from the comparison with the second row
pictures, means the absence of any upper limit for $\Pt^{Jet}_{(\nu)}$.
The most important illustrative fact that in the absence of any restriction
on $\Pt^{miss}$ the total neutrino $\Pt$ inside the jet
averaged over all events can be as large as
$\Pt^{Jet}_{(\nu)}\approx 1~GeV/c$
at $\Ptg\geq 300~ GeV/c$ comes from the right-hand plot of the third row in Fig.~5.
From the comparison of the plots in the second row with the corresponding plots
in the third row
\footnote{This row includes the values of $\Pt^{miss}_{CUT}$ and the corresponding
number of entries remained after imposing $\Pt^{miss}_{CUT}$, as well as the mean
value of $\la \Pt^{Jet}_{(\nu)}\ra$  (i.e. averaged 
over the number of the remaining entries) denoted as ``Mean''.}
we see that the first essential cut $\Pt^{miss}_{CUT}=20 ~GeV/c$
reduces the number of entries for the first $40<\Ptg<50 ~GeV/c$ interval by less than
$0.4\%$  and the mean value of $\Pt^{Jet}_{(\nu)}$ by less than $10\%$.
A more restrictive cut $\Pt^{miss}_{CUT}=5 ~GeV/c$ reduces the value of
 $\la \Pt^{Jet}_{(\nu)}\ra$ by a factor of three and leads to an approximate twofold
drop of the number of events.

From the right-hand plots in Fig.~5 we see that for the $300<\Ptg<360 ~GeV/c$ interval the number of
events with jets containing neutrinos (second row)
is about $3.3\%$ (Entries=3001) of the total number
of the generated \gpj events (Entries=89986, see first row). A very restrictive
$\Pt^{miss}_{CUT}$=5 $GeV/c$ cut leads to the reduction
factor of about 50 for $\la\Pt^{Jet}_{(\nu)}\ra(\equiv$Mean$)$ and to
about $30\%$ reduction of the number of events. As is seen from the
plot in the bottom right-hand corner of Fig.~\ref{fig20-22}, 
the $\Pt^{Jet}_{(\nu)}$ spectrum for the remaining events (Entries=57475) 
finishes at $\Pt^{Jet}_{(\nu)}=10~ GeV/c$ and sharply peaks (log scale!) at
$\Pt^{Jet}_{(\nu)}=0$. The averaged value of $\Pt^{Jet}_{(\nu)}$ under this
peak is equal to 0.022 $GeV/c$. So, with this cut $\Pt^{miss}_{CUT}$=5 $GeV/c$
the neutrinos make a negligible contribution to $\Pt^{Jet}$.
At the same time we see that a moderate cut
$\Pt^{miss}_{CUT}=10 ~GeV/c$ in the
$300<\Pt^{Jet}<360~GeV/c$ interval strongly reduces
(by a factor of 20) the mean value of $\Pt^{Jet}_{(\nu)}$ 
(from $1~ GeV/c$ to $\la\Pt^{Jet}_{(\nu)}\ra=0.046~GeV/c$) 
at about  less than $10\%$ reduction of
the  number of events in this $\Pt^{Jet}$ (or $\Ptg$) interval.

In the $100<\Pt^{Jet}<120~GeV/c$ interval,
as we see from the third row of Fig.~\ref{fig20-22},
the same cut $\Pt^{miss}_{CUT}=10 ~GeV/c$ reduces
the mean value of $\Pt^{Jet}_{(\nu)}$ by a factor of 5 (from $0.5~ GeV/c$ 
to $\la\Pt^{Jet}_{(\nu)}\ra=0.09~GeV/c$) with  $10\%$
reduction of the total number of events.

It should be noted that in the $40<\Pt^{Jet}<50~GeV/c$ interval,
which is less dangerous from the point of view of the
neutrino $\Pt$ content in a jet, we have already a very small mean value of 
~\\[-8mm]
\begin{center}
\begin{figure}[htbp]
\vspace{-3.3cm}
\hspace{.0cm} \includegraphics[width=15.9cm]{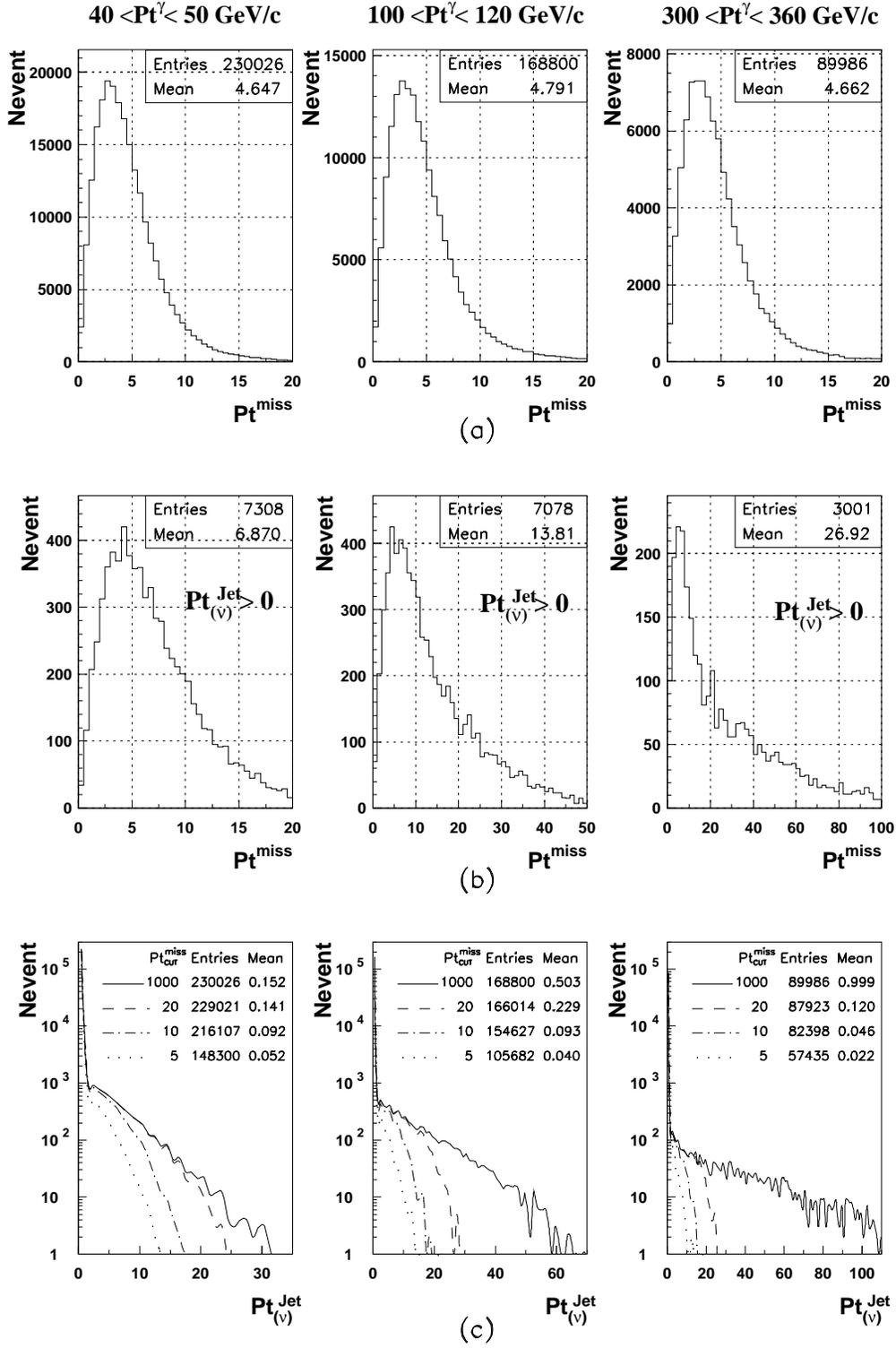}
\vspace{-0.5cm}
\caption{\hspace*{0.0cm} a) $\Pt^{miss}$ spectra in all events;
b) $\Pt^{miss}$ spectra in events having jets with non-zero $\Pt$
neutrinos, i.e. $\Pt^{Jet}_{(\nu)}>0$;~ c) $\Pt^{Jet}_{(\nu)}$ spectra
behaviour for different values of $\Pt^{miss}_{CUT}$ values in various
$\Pt^{Jet}(\approx\Pt^\gamma)$ intervals. $\Pt^{clust}_{CUT}=30 ~GeV/c$.}
\label{fig20-22}
\end{figure} 
\begin{figure}[htbp]
\vspace{-2.9cm}
\hspace{.0cm} \includegraphics[width=15.9cm]{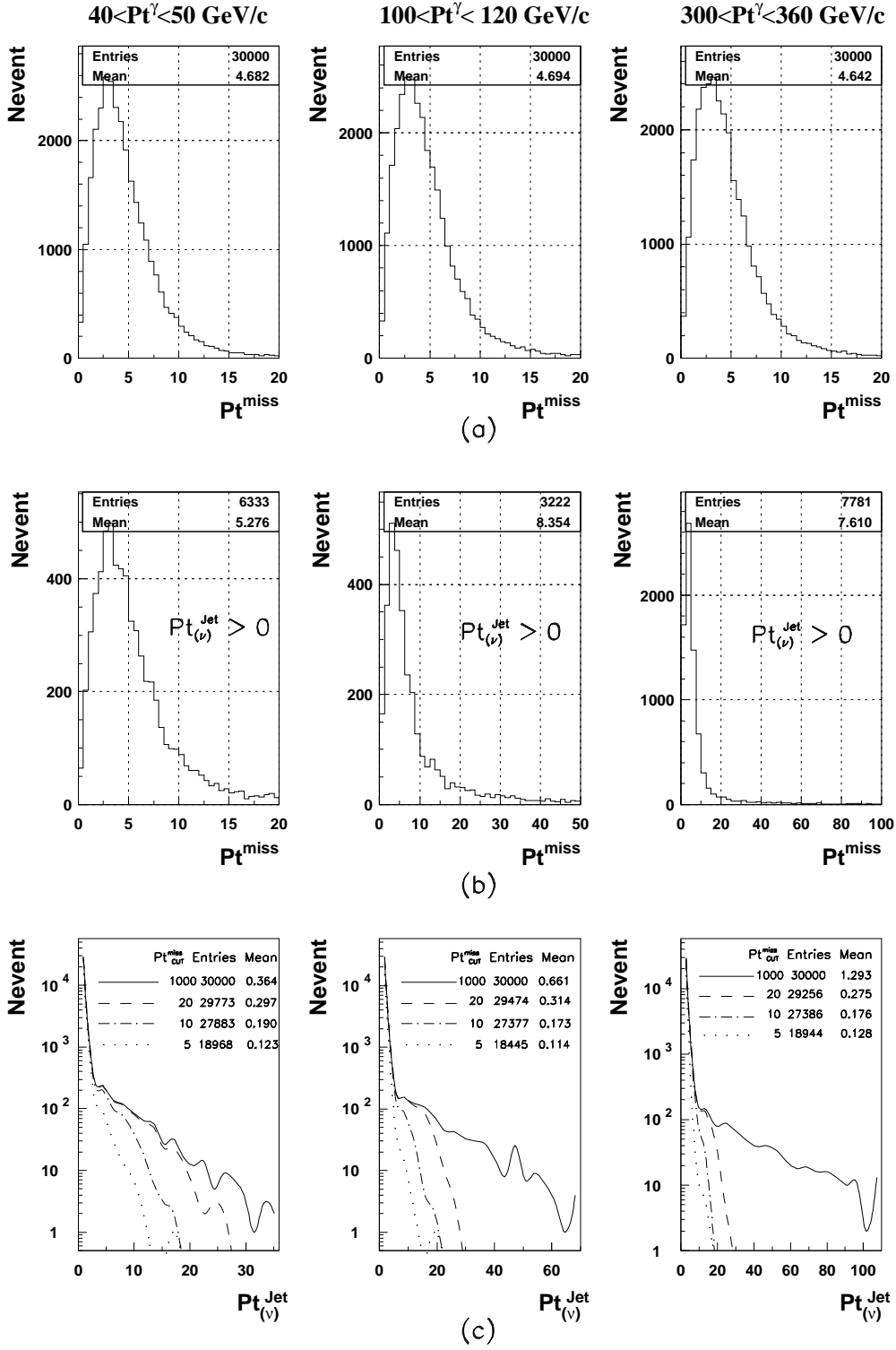}
\vspace{-0.5cm}
\caption{\hspace*{0.0cm} $K^{\pm},\pi^{\pm}-$decays are allowed inside the solenoid 
of $R=129~cm$ and $L=317~cm$.
a) $\Pt^{miss}$ spectra in all events with ;
b) $\Pt^{miss}$ spectra in events having jets with non-zero $\Pt$
neutrinos, i.e. $\Pt^{Jet}_{(\nu)}>0$;~ c) $\Pt^{Jet}_{(\nu)}$ spectra
behaviour for different values of $\Pt^{miss}_{CUT}$ values in various
$\Pt^{Jet}(\approx\Pt^\gamma)$ intervals. $\Pt^{clust}_{CUT}=30 ~GeV/c$.}
\label{fig23}
\end{figure} 
\end{center}

~\\[-20mm]
\noindent
$\la\Pt^{Jet}_{(\nu)}\ra$ equal to $0.152 ~GeV/c$ even without imposing any $\Pt^{miss}_{CUT}$.

The analogous (to neutrino) situation holds for the $\Pt^{Jet}_{(\mu)}$
contribution (as they originate mostly from the same decays).
 
The detailed information about the values of non-detectable
$\Pt^{Jet}_{(\nu)}$ averaged over all events
(no cut on $\Pt^{miss}$ was used) as well as about mean $\Pt$ values
of muons belonging to jets $\la \Pt^{Jet}_{(\mu)}\ra$ is presented
in Tables 1--8 of Appendix 1 for the sample of events with jets that are
entirely contained in the barrel region of the
HCAL ($|\eta^{jet}|<1.4$, ``HB-events'', see Section 6 and \cite{BKS_P1})
and found by UA1 and LUCELL jetfinders.                           
In these tables the ratio of the number of events with non-zero
$\Pt^{Jet}_{(\nu)}$ to the total number of events is denoted by
$R^{\nu \in Jet}_{event}$ and
the ratio of the number of events with non-zero $\Pt^{Jet}_{(\mu)}$
to the total number of events is denoted by $R^{\mu \in Jet}_{event}$.

The quantity $\Pt^{miss}$ in events with $\Pt^{Jet}_{(\nu)}>0$ is 
denoted in these tables as $\Pt^{miss}_{\nu\in Jet}$ and is
 given there for four $\Ptg$ intervals
($40<\Ptg<50$, $100<\Ptg<120$, $200<\Ptg<240$ and $300<\Ptg<360~GeV/c$)
and for other $\Pt^{clust}_{CUT}$ values ($\Pt^{clust}_{CUT}=20,15,10,5 ~GeV/c$)
complementary to $\Pt^{clust}_{CUT}=30 ~GeV/c$ used for plots
\footnote{Please, note that the values of $\Pt^{miss}$ and
$\Pt^{miss}_{\nu\in Jet}$ in the plots of Fig.~5 
are slightly different from those of Appendix 1 as the former were 
found for events in the whole $|\eta|\lt5$ region.}
of Fig.~5. From Tables 1, 2 we see that the averaged value of
$\Pt^{miss}$ calculated by using only the events with
$\Pt^{Jet}_{(\nu)}>0$, i.e. $\la\Pt^{miss}_{\nu \in Jet}\ra$,
is about 6.6 -- 7.0 $GeV/c$ for the $40<\Ptg<50 ~GeV/c$
interval. It increases to about 32 $GeV/c$ for the $300<\Ptg<360 ~GeV/c$ 
interval (see Tables 7, 8).

It should be noted that the  averaged values of the modulus of 
$\Pt^{Jet}_{(\nu)}$ (see formula (8)) presented in the third lines of 
Tables 1--8 from Appendix 1 coincide with the averaged values 
of the difference $\la\Pt^{Jet}\!-\!\Pt^{jet}\ra
\equiv \Delta_\nu$ (see Section 3.2 and second lines of Tables 1--8 ) 
to three digits, i.e. $<\!\!\Pt^{Jet}_{(\nu)}\!\!>=\Delta_\nu$.
This is because the $\vec{\Pt}^{Jet}$ and $\vec{\Pt}^{jet}$ vectors 
are practically collinear  and because we consider here the ``HB-events''
in which all jet muons are supposed to be also detected by the barrel muon system.

We underline that the $\la \Pt^{Jet}_\nu \ra$ value estimated in Tables 1--8
has a meaning of the correction $\Delta_{\nu}$ that should be added 
to $\Pt^{jet}$ in order to take into account the $\Pt$ carried away by
non-detectable particles, i.e. for a case of ``HB'' events when jet fits 
into Barrel and jet muons are measurebale, we have:  
$\la\Pt^{Jet}\ra =\la\Pt^{jet}\ra + \Delta_\nu$. It should be noted 
that latter on in Section 8 when we shall discuss the $Pt$ balance of
photon and jet we shall present in the tables of Appenicies 2-5 the 
values of  $\Pt^{Jet}$ just obtained from the values of $\Pt^{jet}$
by adding of this correction. Here in Appendix 1 we present the values 
of these corrections calculated without an application of  $\Pt^{miss}_{CUT}$,
as the aim of these tables consists in part in showing of typical
values  $\Pt^{miss}$.

Let us mention also that Tables 1--8 contain an additional information on
the numbers of \gpj events with jets produced by $c$ and $b$ quarks,
i.e. $Nevent_{(c)}$ and $Nevent_{(b)}$ (see also \cite{BKS_P1,BKS_GLU,MD2,BALD02}),
given for the integrated luminosity
$L_{int}=3~ fb^{-1}$ for different $\Pt^{Jet}$($\approx\Ptg$) intervals.
They also show the ratio (``$29sub/all$'')
of the number of events caused by gluonic subprocess (1a) to the number of events due to the sum of
subprocesses (1a) and (1b) and averaged jet radii
$<\!\!R_{jet}\!\!>$.

It was already mentioned in the Introduction that we are planning to carry
out a more  detailed analysis based on the GEANT package.
To have an idea of what changes can be expected, we shall consider now 
the case with  allowed $K^{\pm}$ and $\pi^{\pm}$ decays (as the main source of 
neutrinos and muons).
The averaged values of $\Pt^{Jet}_{(\nu)}$ for different $\Ptg$-intervals
with switched on $K^{\pm},\pi^{\pm}$ decays are given in Fig.~\ref{fig23} 
with the same meaning of all notations as in Fig.~5. Here $K^{\pm}$ and 
$\pi^{\pm}$ decays are allowed inside the solenoid volume with the barrel
radius $R_B=129~ cm$ and the distance from the interaction vertex to 
End-cap along the $z$-axis $L=317~cm$ (CMS geometry).

From the first row of Fig.~\ref{fig23} we see that in a case of allowed
$K^{\pm},\pi^{\pm}$ decays the $\Pt^{miss}$ spectrum and the position of 
the mean value of $\Pt^{miss}$ for all events practically does not change 
with growing $\Pt^{Jet}(\approx \Ptg)$  in complete analogy to the first 
rows in Fig.~5.

At the same time the tail of $\Pt^{miss}$ spectra for events that
contain neutrinos in the jet (second row of Fig.~6) changes quite noticeably.
It should be noted that the number of such
events grows to $20-25\%$ as compared with $3\%$ in
the case considered in Fig.~5, but the mean values of
$\Pt^{miss}$ do not grow so much with $\Pt^{\gamma}$, as is seen in Fig.~5.
Now we compare the third row pictures in Figs.~5 and 6.
We see that in Fig.~6 the mean value of $\Pt^{Jet}_{(\nu)}$ carried away by
neutrinos of the  jet grows from $\la\Pt^{Jet}_{(\nu)}\ra\approx0.36~GeV/c$ for
$40\!<\!\Ptg\!<\!50~ GeV/c$ to $\la\Pt^{Jet}_{(\nu)}\ra\approx1.3~GeV/c$ for
$300\!<\!\Ptg\!<\!360~ GeV/c$, i.e. its contribution into $\Pt^{Jet}$ varies as $1\%\to0.4\%$.
From the same pictures of Fig.~6 we see that the first essential cut
$\Pt^{miss}_{CUT}=20 ~GeV/c$ would reduce the contribution of neutrinos to
$\Pt^{Jet}$ to $\la\Pt^{Jet}_{(\nu)}\ra\approx0.3~GeV/c$ in all
$\Pt^{Jet}$ intervals, while the cut $\Pt^{miss}_{CUT}=10 ~GeV/c$ would lead to
 $\la\Pt^{Jet}_{(\nu)}\ra\approx0.20~GeV/c$
(which is at least twice larger than analogous values in Fig.~5 but still is
quite acceptable) with only $\approx 8\%$ reduction of the number of events.

\section{EVENT RATES FOR DIFFERENT $\Pt^{\gamma}$ AND $\eta^{Jet}$ INTERVALS.}

\it\small
\hspace*{9mm}
The number of \gpj events distribution over $\Pt^{\gamma}$ and $\eta^{\gamma}$ is studied here.
It is found that in each interval of the $\Delta\Pt^{\gamma}=10~ GeV/c$ width 
the rates decrease by a factor more than 2.
The number of events with jets which transverse momentum are completely (or with $5\%$ accuracy)
contained in HB, HE and HF regions are presented in Tables \ref{tab:sh1}--\ref{tab:sh4}
for integrated luminosity $L_{~int}=3~fb^{-1}$.
\rm\normalsize
\vskip4mm

\subsection{Dependence of distribution of the number of events
on the ``back-to-back'' angle $\phigj$ and on $\Pt^{ISR}$. }            

The definitions of the physical variables introduced in
Sections 2 and 3  allow to study
a possible way to select the events with a good $\Pt^{\gamma}$ and $\Pt^{Jet}$ balance.
Here we shall be interested to get (by help of PYTHIA generator and the theoretical models 
therein) an idea about the form of the spectrum of the variable $\Pt{56}$ 
(which is approximately proportional to $\Pt^{ISR}$ up to the value of intrinsic parton
transverse momentum $k_t$ inside a proton) at different values of $\Ptg$.
For this aim four samples of \gpj events were generated by using PYTHIA
with 2 QCD subprocesses (1a) and (1b) being included simultaneously. In what follows we shall call these events as 
``signal events''. The generations were done with the values of the PYTHIA parameter CKIN(3)($\equiv\pth$)
 equal to  20, 50, 100, 150 $GeV/c$ in order to cover
four $\Pt^{\gamma}$ intervals: 40--50, 100--120, 200--240, 300--360 $GeV/c$, respectively.
Each sample in these $\Ptg$ intervals had a size of $5\cdot 10^6$ events.
The cross sections for the two subprocesses were found to be as given in Table~\ref{tab:cross4}.\\[-2mm]
\begin{table}[h]
\begin{center}
\caption{The cross sections (in $microbarn$) of the $qg\to q+\gamma$ and $q\overline{q}\to g+\gamma$ subprocesses
for four $\Pt^{\gamma}$ intervals.}
\normalsize
\vskip.1cm
\begin{tabular}{||c||c|c|c|c|}                  \hline \hline
\label{tab:cross4}
Subprocess& \multicolumn{4}{c|}{$\Pt^{\gamma}$ interval ($GeV/c$)} \\\cline{2-5}
   type   & 40 -- 50 & 100 -- 120 & 200 -- 240  & 300 -- 360 \\\hline \hline
$qg\to q+\gamma$           & 1.19$\cdot10^{-1}$& 6.70$\cdot10^{-3}$& 6.09$\cdot10^{-4}$& 1.36$\cdot10^{-4}$ \\\hline
$q\overline{q}\to g+\gamma$& 0.10$\cdot10^{-1}$& 0.69$\cdot10^{-3}$& 0.77$\cdot10^{-4}$& 0.20$\cdot10^{-4}$ \\\hline
Total                      & 1.29$\cdot10^{-1}$& 7.39$\cdot10^{-3}$& 6.86$\cdot10^{-4}$& 1.56$\cdot10^{-4}$ \\\hline
\end{tabular}
\end{center}
\vskip-7mm
\end{table}

 For our analysis  we used ``Selection 1''  (formulae (17)--(24))
defined in Sections 3.2 and the values of cut parameters (29).

In Tables \ref{tab:pt56-1}, \ref{tab:pt56-2} and \ref{tab:pt56-4},
\ref{tab:pt56-5}  we present $\Pt56$ spectra for
two most illustrative cases of
$\Pt^{\gamma}$ intervals $40\lt\Pt^{\gamma}\lt50 ~GeV/c$ (Tables 2 and 5) and
$200<\Pt^{\gamma}<240 ~GeV/c$ (Tables 3 and 6). The distributions of the
 number of events for the integrated luminosity $L_{int}=3\,fb^{-1}$
in different $\Pt56$ intervals ($\la k_t\ra$ was taken to be
fixed at the PYTHIA default value, i.e. $\la k_t\ra=0.44\,GeV/c$) and for different ``back-to-back'' angle intervals
$\phigj=180^\circ \pm \dphi~$ ($\dphi\leq15^\circ,\,10^\circ$ and $5^\circ$ as well as without 
any restriction on $\dphi$, i.e. for the whole $\phi$ interval $\dphi\leq180^\circ$)
\footnote{The value $\Delta\phi=5^\circ$ approximately
coincides with one CMS  HCAL tower size in the $\phi$-plane.
}
are given there. The LUCELL jetfinder was used for determination of jets and clusters
\footnote{More details connected with UA1 and UA2 jetfinders application can be found in Section 7
and Appendices 2--5 for a jet contained in CC region.}.
Tables \ref{tab:pt56-1} and \ref{tab:pt56-2} correspond to $\Pt^{clust}\lt30\,GeV/c$ 
and serve as an illustration since it is rather a weak cut condition, while
Tables~\ref{tab:pt56-4} and \ref{tab:pt56-5}  correspond to a more
restrictive selection cut $\Pt^{clust}_{CUT}=5\,GeV/c$
(which leads to about twofold reduction of the number of events for $\dphi\leq15^\circ$; 
see summarizing Tables \ref{tab:pt56-3} and \ref{tab:pt56-6}).

First, from the last summary lines of Tables
\ref{tab:pt56-1}, \ref{tab:pt56-2} and \ref{tab:pt56-4}, \ref{tab:pt56-5}
we can make a general conclusion about the $\dphi$-dependence
of the event spectrum.
Thus, in the case of weak restriction  $\Pt^{clust}\lt30 ~GeV/c$
we can see from Table \ref{tab:pt56-1} that for the $40\leq \Pt^{\gamma}\leq 50 ~GeV/c$
interval about 66$\%$ of events are concentrated
in the $\Delta\phi\lt15^\circ$ range, while 32$\%$ of events are
in the $\Delta\phi\lt5^\circ$ range.
At the same time the analogous summary line of Table \ref{tab:pt56-2}
shows us that for higher $\Ptg$ interval $200\leq \Pt^{\gamma}\leq 240\, GeV/c$ the $\Pt56$ spectrum
for the same restriction $\Pt^{clust}\!\lt30 ~GeV/c$
moves (as compared with low $\Ptg$ intervals) to the small $\dphi$ region: 
more than 99$\%$ of events have
$\Delta\phi\lt15^\circ$ and 79$\%$ of them have $\Delta\phi\lt5^\circ$.
%
\def\baselinestretch{0.98}
\begin{table}[htbp]
\begin{center}
\vskip-1.2cm
\caption{Number of events dependence on $\Pt56$ and
$\Delta\phi$ for}
\vskip-3pt
{\footnotesize $40\leq Pt^{\gamma}\leq 50 \, GeV/c$ and
$\Pt^{clust}_{CUT}= 20 \, GeV/c$ for $L_{int}$=3$fb^{-1}$.}
\vskip.2cm
\begin{tabular}{||c||r|r|r|r||} \hline \hline
\label{tab:pt56-1}
 $\Pt{56}$&\multicolumn{4}{|c||}{ $\dphi_{max}$} \\\cline{2-5}
 $(GeV/c)$  &$180^\circ$&$15^\circ$ & $10^\circ$&$5^\circ$ \\\hline\hline
    0 --   5 &   1103772 &   1049690 &   1006627 &    849706 \\\hline
    5 --  10 &   1646004 &   1564393 &   1403529 &    812304 \\\hline
   10 --  15 &   1331589 &   1122473 &    771060 &    380122 \\\hline
   15 --  20 &    992374 &    568279 &    365329 &    179767 \\\hline
   20 --  25 &    725537 &    282135 &    183406 &     91113 \\\hline
   25 --  30 &    559350 &    169186 &    112308 &     58395 \\\hline
   30 --  40 &    911942 &    265961 &    178048 &     89867 \\\hline
   40 --  50 &    388950 &     94112 &     62068 &     31000 \\\hline
   50 -- 100 &     91248 &     19442 &     12973 &      6234 \\\hline
  100 -- 300 &        34 &         0 &         0 &         0 \\\hline
  300 -- 500 &         0 &         0 &         0 &         0 \\\hline \hline
    0 -- 500 &   7750799 &   5135671 &   4095348 &   2498507 \\\hline \hline
\end{tabular}
\vskip0.2cm
\caption{Number of events dependence on $\Pt56$ and
$\Delta\phi$ for}
\vskip-3pt
{\footnotesize $200\leq Pt^{\gamma}\leq 240 \, GeV/c$ and
$\Pt^{clust}_{CUT}= 20 \, GeV/c$ for $L_{int}$=3$fb^{-1}$.}
\vskip0.2cm
\begin{tabular}{||c||r|r|r|r||} \hline \hline
\label{tab:pt56-2}
 $\Pt{56}$  &\multicolumn{4}{c||}{ $\dphi_{max}$} \\\cline{2-5}
 $(GeV/c)$  &\aaa $180^\circ$\aaa&\aaa$15^\circ$\aaa&\aaa$10^\circ$\aaa&\aaa$5^\circ$\aaa \\\hline\hline
    0 --   5 &      1429 &      1429 &      1427 &      1380 \\\hline
    5 --  10 &      3266 &      3266 &      3264 &      3150 \\\hline
   10 --  15 &      3205 &      3205 &      3200 &      3069 \\\hline
   15 --  20 &      2827 &      2827 &      2819 &      2618 \\\hline
   20 --  25 &      2409 &      2408 &      2393 &      1918 \\\hline
   25 --  30 &      2006 &      2006 &      1982 &      1300 \\\hline
   30 --  40 &      2608 &      2605 &      2533 &      1411 \\\hline
   40 --  50 &      1237 &      1230 &      1067 &       586 \\\hline
   50 -- 100 &      1066 &      1018 &       842 &       536 \\\hline
  100 -- 300 &       313 &       307 &       293 &       221 \\\hline
  300 -- 500 &         0 &         0 &         0 &         0 \\\hline \hline
    0 -- 500 &     20366 &     20301 &     19820 &     16189 \\\hline \hline
\end{tabular}
\vskip0.2cm
\caption{Number of events dependence on $\dphi_{max}$ and on
$\Pt^{\gamma}$ for $L_{int}=3\,fb^{-1}$.}
\vskip-3pt
{\footnotesize $\Pt^{clust}_{CUT}=20 ~GeV/c$ (summary).}
\vskip0.2cm
\begin{tabular}{||c||r|r|r|r||} \hline \hline
\label{tab:pt56-3}
 $\Pt^{\gamma}$  &\multicolumn{4}{c||}{ $\dphi_{max}$} \\\cline{2-5}
 $(GeV/c)$  &$180^\circ$&$15^\circ$&$10^\circ$&$5^\circ$ \\\hline\hline
 40 -- 50  &7750799 &   5135671 &   4095348 &   2498507 \\\hline
100 -- 120 & 323766 &    297323 &    258691 &    176308 \\\hline
200 -- 240 &  20366 &     20301 &     19820 &     16189 \\\hline
300 -- 360 &   3638 &      3638 &      3627 &      3323 \\\hline\hline
\end{tabular}
\end{center}
\end{table}
\begin{table}[htbp]
\begin{center}
\caption{Number of events dependence on $\Pt56$ and
$\Delta\phi$ for}
\vskip-3pt
{\footnotesize $40\leq Pt^{\gamma}\leq 50 \, GeV/c$ and
$\Pt^{clust}_{CUT}= 5 \, GeV/c$ for $L_{int}$=3$fb^{-1}$.}
\vskip0.2cm
\begin{tabular}{||c||r|r|r|r||} \hline \hline
\label{tab:pt56-4}
 $\Pt{56}$  &\multicolumn{4}{c||}{ $\dphi_{max}$} \\\cline{2-5}
 $(GeV/c)$  &$180^\circ$&$15^\circ$&$10^\circ$&$5^\circ$ \\\hline\hline
    0 --   5 &    331522 &    331321 &    329876 &    295759 \\\hline
    5 --  10 &    319153 &    318581 &    299960 &    187089 \\\hline
   10 --  15 &     88603 &     82586 &     60537 &     32335 \\\hline
   15 --  20 &     21244 &     15327 &     11663 &      6924 \\\hline
   20 --  25 &      8101 &      5681 &      4639 &      2992 \\\hline
   25 --  30 &      4739 &      3395 &      2823 &      1949 \\\hline
   30 --  40 &      3495 &      2790 &      2555 &      1714 \\\hline
   40 --  50 &      1647 &      1277 &      1042 &       471 \\\hline
   50 -- 100 &       101 &        67 &        67 &        67 \\\hline
  100 -- 500 &         0 &         0 &         0 &         0 \\\hline \hline
    0 -- 500 &    778606 &    761026 &    713161 &    529299 \\\hline \hline
\end{tabular}
\vskip0.2cm
\caption{Number of events dependence on $\Pt56$ and
$\Delta\phi$ for}
\vskip-3pt
{\footnotesize $200\leq Pt^{\gamma}\leq 240 \, GeV/c$ and
$\Pt^{clust}_{CUT}= 5 \, GeV/c$ for $L_{int}$=3$fb^{-1}$.}
\vskip0.2cm
\begin{tabular}{||c||r|r|r|r||} \hline \hline
\label{tab:pt56-5}
 $\Pt{56}$  &\multicolumn{4}{c||}{ $\dphi_{max}$} \\\cline{2-5}
 $(GeV/c)$  &\aaa $180^\circ$\aaa&\aaa$15^\circ$\aaa&\aaa$10^\circ$\aaa&\aaa$5^\circ$\aaa \\\hline\hline
    0 -   5 &       369 &       369 &       369 &       369 \\\hline
    5 -  10 &       563 &       563 &       563 &       562 \\\hline
   10 -  15 &       217 &       217 &       217 &       217 \\\hline
   15 -  20 &        56 &        56 &        56 &        56 \\\hline
   20 -  25 &        20 &        20 &        20 &        18 \\\hline
   25 -  30 &         9 &         9 &         9 &         7 \\\hline
   30 -  40 &         7 &         7 &         7 &         6 \\\hline
   40 -  50 &         6 &         6 &         6 &         5 \\\hline
   50 - 100 &        10 &        10 &        10 &        10 \\\hline
  100 - 300 &         8 &         8 &         8 &         8 \\\hline
  300 - 500 &         0 &         0 &         0 &         0 \\\hline \hline
    0 - 500 &      1264 &      1264 &      1264 &      1257 \\\hline \hline
\end{tabular}
\vskip0.2cm
\caption{Number of events dependence on $\dphi_{max}$ and on
 $\Pt^{\gamma}$ for $L_{int}=3\,fb^{-1}$.}
\vskip-3pt
{\footnotesize $\Pt^{clust}_{CUT}= 5 \, GeV/c$ (summary).}
\vskip0.2cm
\begin{tabular}{||c||r|r|r|r||} \hline \hline
\label{tab:pt56-6}
 $\Pt^{\gamma}$  &\multicolumn{4}{c||}{ $\dphi_{max}$} \\\cline{2-5}
 $(GeV/c)$  &$180^\circ$&$15^\circ$&$10^\circ$&$5^\circ$ \\\hline\hline
\hline
 40 -- 50  &778606 &    761026 &    713161 &    529299 \\\hline
100 -- 120 & 22170 &     22143 &     22038 &     20786 \\\hline
200 -- 240 &  1264 &      1264 &      1264 &      1257 \\\hline
300 -- 360 &   212 &       212 &       212 &       212 \\\hline
\end{tabular}
\end{center}
\end{table}

\def\baselinestretch{1.0}

A tendency of distributions of the number of signal \gpj events to be
very rapidly concentrated in a rather narrow
back-to-back angle interval $\Delta\phi<15^\circ$ as $\Pt^{\gamma}$ grows
becomes more distinct with a more restrictive cut
$\Pt^{clust}_{CUT}= 5\,GeV/c$ (see Tables \ref{tab:pt56-4}, \ref{tab:pt56-5} and
\ref{tab:pt56-6}). From
the last summary line of Table \ref{tab:pt56-4} we see for this cut
that in the case of
$40\leq \Pt^{\gamma}\leq 50\, GeV/c~$ more than
$97\%$ of the events have $\Delta\phi<15^\circ$, while $~68\%$ of them are
in the $\Delta\phi<5^\circ$ range.
For $200\leq \Pt^{\gamma}\leq 240\, GeV/c$
(see Table \ref{tab:pt56-5}) more than  $99\%$ of the events subject to the cut
$\Pt^{clust}_{CUT}= 5 ~GeV/c$ have $\Delta\phi<5^\circ$.
It means that while suppressing
cluster or mini-jet activity by imposing $\Pt^{clust}_{CUT}= 5 ~GeV/c$
we can select the sample of events with a clean
``back-to-back'' (within 15$^\circ$) topology of $\gamma$ and jet orientation.
(Unfortunately, as it will be discussed below basing on the information from Tables 5 and 6,
it does not mean that this cut allows to suppress completely the ISR).
\footnote{See also the event spectra over $\Pt^{clust}$ in Fig.~7 of the following Section 6.}.

So, one can conclude that PYTHIA simulation predicts that
at LHC  energies most of the \gpj events (more than $66\%$)
may have the vectors $\vec{\Pt}^{\gamma}$ and $\vec{\Pt}^{jet}$ being back-to-back
within $\Delta\phi\lt15^\circ$ after imposing $\Pt^{clust}_{CUT}=20~GeV/c$.  
The cut $\Pt^{clust}_{CUT}=5~GeV/c$ significantly improves
\footnote{An increase in \ptg produces the same effect, as is seen
from comparison of Tables \ref{tab:pt56-1} and \ref{tab:pt56-2} and
will be demonstrated in more detail in Section 6 and Appendices 2--5.}
this tendency.

It is worth mentioning that this picture
reflects the predictions of one of the generators
based on the approximate  LO values for
the cross section. It may change if the
next-to-leading order or soft physics
\footnote{We thank E.~Pilon and J.~Ph.~Jouliet for the information
about new LHC data on this subject and for clarifying the importance
of NLO corrections and soft physics effects.}
effects are included.

The other lines of Tables \ref{tab:pt56-1}, \ref{tab:pt56-2} and
\ref{tab:pt56-4}, \ref{tab:pt56-5} contain the information
about the $\Pt56$ spectrum
or, up to intrinsic transverse parton momentum $\langle k_t \rangle=0.44\;GeV/c$,
about $\Pt^{ISR}$ spectrum).

From Tables 2 and 3 one can see that in the case when there are no restrictions
on $\Pt^{clust}$ the $\Pt56$ spectrum becomes a bit wider for larger values of
$\Pt^{\gamma}$. 

At the same time, one can conclude from the comparison of Table \ref{tab:pt56-1}  with  Table \ref{tab:pt56-4}
that for lower $\Ptg$ intervals the width of the most populated part of the $\Pt56$ (or $\Pt^{ISR}$) 
spectrum reduces two-fold with restricting $\Pt^{clust}_{CUT}$. So, for $\dphi_{max}=15^\circ$ we
see that it drops from $0\lt\Pt{56}\lt25\;GeV/c$ $\!$ for $\!$ $\Pt^{clust}_{CUT}=20\;GeV/c$ to a
$\!$ narrower interval $\,$ of $\!$
$0\lt\Pt{56}\lt10\,GeV/c$ $\!$ for $\!$ the $\Pt^{clust}_{CUT}=5\,GeV/c$.
At higher $\Ptg$ intervals (Tables \ref{tab:pt56-2} and \ref{tab:pt56-5})
for the same value $\dphi_{max}=15^\circ$ the reduction factor of the $\Pt56$ spectrum width 
(from the $0\lt\Pt56\lt50~ GeV/c$ interval for $\Pt^{clust}_{CUT}=20\,GeV/c$
to the $0\lt\Pt56\lt15 ~GeV/c$ interval for $\Pt^{clust}_{CUT}=5 ~GeV/c$)
is more than two. But the tails of $\Pt^{ISR}$ spectra still 
remain to be quite long in a case of high $\Ptg$ intervals.

Thus, we can summarize that the PYTHIA  generator predicts
an increase in the $\Pt^{ISR}$ spectrum with growing $\Pt^{\gamma}$ (compare Tables 2 and 3), but this increase
can be reduced by imposing a restrictive cut on 
$\Pt^{clust}$ (for more details see Sections 6 and 7).

So, the $\Pt{56}$ spectra presented in Tables
\ref{tab:pt56-1}, \ref{tab:pt56-2} and \ref{tab:pt56-3},
\ref{tab:pt56-4} show PYTHIA prediction that the ISR effect is a large one at LHC
energies. Its $\Pt$ spectrum continues at least up to $\Pt{56}= 15 ~GeV/c$ 
in the case of $\Ptg$ (or $\Pt^{jet}$) $\approx 200 ~GeV/c$
(and up to higher values as $\Ptg$ grows) even for $\Pt^{clust}_{CUT}=5 ~GeV/c$. It cannot be
completely suppressed  by $\Delta\phi$ and $\Pt^{clust}$ cuts alone.
(In Section 8 the effect of the additional $\Pt^{out}_{CUT}$ will be discussed)
Therefore we prefer to use the $\Pt$ balance equation for
the event as a whole (see equations (16) and (28) of Sections 3.1 and 3.2), i.e. an equation
that takes into account the ISR and FSR effects,
rather than balance equation (2) for fundamental processes (1a) and (1b) as discussed in Section 2.1.
(In Section 6 we shall study a behavior of each term that enter equation (28) in order to find the criteria that 
would allow to select events with a good balance of \Ptgj).

Since the last lines in Tables \ref{tab:pt56-1}, \ref{tab:pt56-2} and
\ref{tab:pt56-4}, \ref{tab:pt56-5} contain an illustrative
information on $\Delta\phi$ dependence of the total number of events, we add also the
summarizing Tables \ref{tab:pt56-3} and \ref{tab:pt56-6}. They include more $\Pt^{\gamma}$ 
intervals and contain analogous numbers of events that can be collected
in different $\Delta\phi$ intervals for two different
 $\Pt^{clust}$ cuts at $L_{int}=3\,fb^{-1}$.

\subsection{$\Pt^{\gamma}$ and $\eta^{\gamma}$ dependence of event rates.}

~\\[-12mm]

\def\baselinestretch{1.0}
\begin{flushleft}
\parbox[r]{.5\linewidth}{
Here we shall present
the number of events for different $\Pt^{\gamma}$ and $\eta^{\gamma}$ intervals as
predicted by PYTHIA simulation with weak cuts defined mostly by (29)
with only change of $\Pt^{clust}_{CUT}$ value from 30 to 10$~GeV/c$.
The lines of Table \ref{tab:pt-eta} correspond to  $\Pt^{\gamma}$ intervals
and the columns to $\eta^{\gamma}$ intervals. The last column of this table contains the total number
of events  (at $L_{int}=3\,fb^{-1}$) in the whole ECAL $\eta^{\gamma}$-region
$|\eta^{\gamma}|\lt2.61$ for a given $\Pt^{\gamma}$ interval.
We see that the number of events  decreases fast
with growing $\Pt^{\gamma}$ (by more than 50$\%$ for each subsequent interval). For the fixed
$\Pt^{\gamma}$ interval the dependence on $\eta^{\gamma}$ 
is given in lines of Table 8 and illustrated by Fig.~7.
}
\end{flushleft}
\begin{flushright}
\begin{figure}[h]
\vskip-90mm
  \hspace{85mm} \includegraphics[width=105mm,height=78mm]{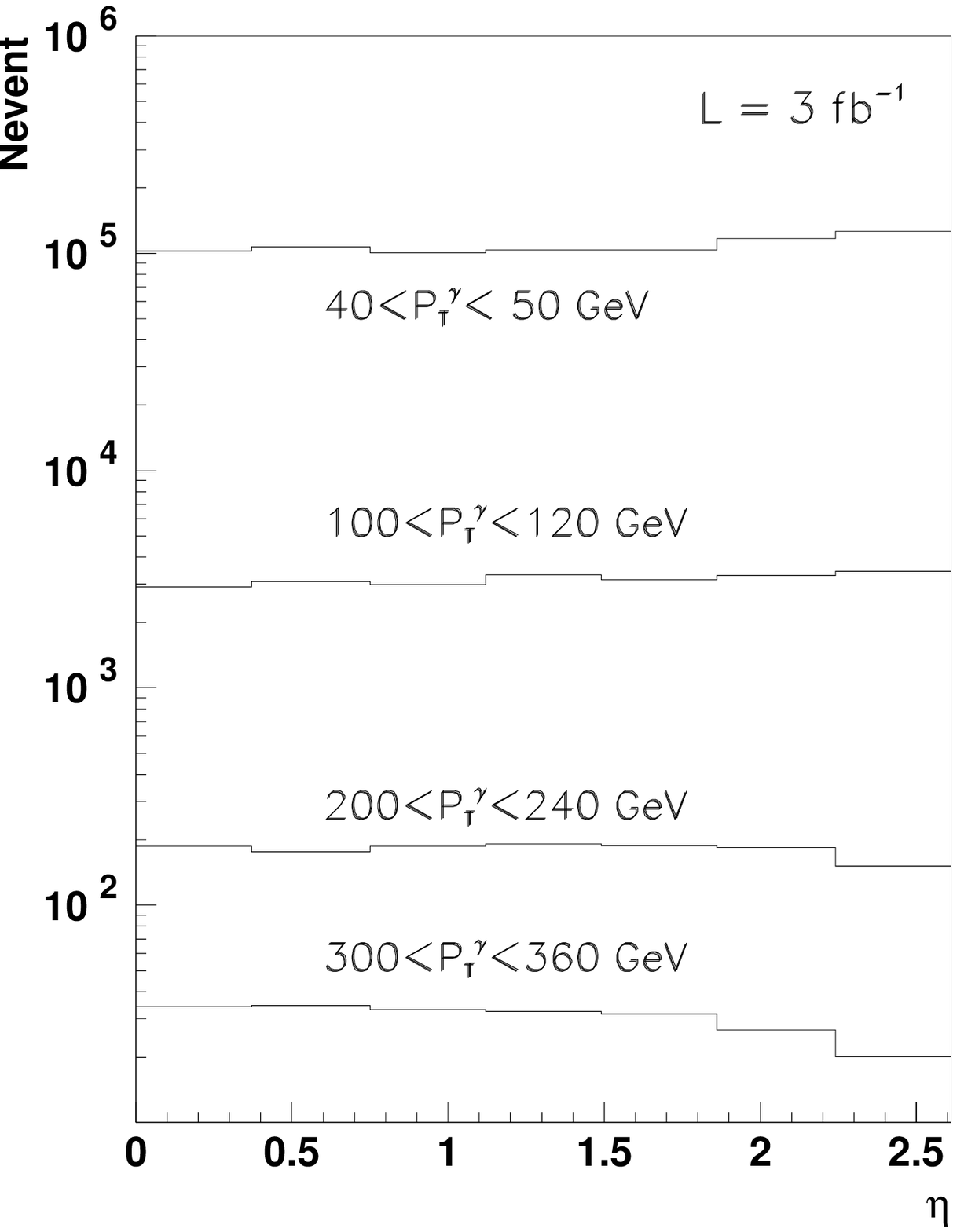}
  \label{fig:eta}
\end{figure}
\end{flushright}
\vspace{-1.7cm}
\hspace*{10cm} {\footnotesize {Fig.~7: $\eta$-dependence of rates for
different\\}}
\hspace*{11cm} {\footnotesize  {$\Pt^{\gamma}$ intervals.}}
\setcounter{figure}{5}
\def\baselinestretch{0.99}
\begin{table}[h]
\begin{center}
\caption{Rates for $L_{int}=3\,fb^{-1}$ for different $\Pt^{\gamma}$ and $\eta^{\gamma}$ intervals 
($\Pt^{clust}_{CUT}= 5 \, GeV/c$ and $\Delta\phi \leq 15^\circ$).}
\vskip0.2cm
\begin{tabular}{||c||r|r|r|r|r|r|r||r||} \hline \hline
\label{tab:pt-eta}
$\Pt^{\gamma}$ &\multicolumn{7}{c||}{$\eta^{\gamma}$~~ intervals}
&all ~~$\eta^{\gamma}$ \\\cline{2-9}
$(GeV/c)$ & 0.0-0.4 & 0.4-0.7 & 0.7-1.1 & 1.1-1.5 & 1.5-1.9 & 1.9-2.2 & 2.2-2.6 &0.0-2.6
 \\\hline
 \hline
 40 --  50& 102656& 107148& 100668& 103903& 103499& 116674& 126546& 761027\\\hline
 50 --  60&  43905&  41729&  41074&  45085&  42974&  47640&  50310& 312697\\\hline
 60 --  70& 18153 & 18326 & 19190&  20435&  20816 & 19432&  23650& 140005\\\hline
 70 --  80&   9848 & 10211&  9963 & 10166&  9951 & 11397 & 10447&  71984\\\hline
 80 --  90&   5287&  5921&  5104&  5823&  5385&  6067&  5923&  39509\\\hline
 90 -- 100&   2899&  3033&  3033&  3326&  3119&  3265&  3558&  22234\\\hline
100 -- 120&   2908&  3091&  2995&  3305&  3133&  3282&  3429&  22143\\\hline
120 -- 140&   1336  & 1359  & 1189  & 1346  & 1326  & 1499  & 1471&  9525\\\hline
140 -- 160&    624&   643&   626&   674&   706&   614&   668&  4555\\\hline
160 -- 200&    561&   469&   557&   555&   519&   555&   557&  3774\\\hline
200 -- 240&   187   & 176   & 186   & 192   & 187   & 185   & 151  & 1264\\\hline
240 -- 300&   103&    98&    98&    98   & 100&    92&    74&   665\\\hline
300 -- 360&     34&    34&    33&    32&    31&    27&    20&   212\\\hline \hline
 40 -- 360& 188517& 192274& 184734& 194957& 191761&  210742&  226819& 1389484\\\hline
\hline
\end{tabular}
\end{center}
\end{table}

~\\[-7mm]
\normalsize
\subsection{Estimation of \gpj event rates for different calorimeter regions.} 
Since a jet is a wide-spread object, the $\eta^{jet}$ dependence of rates
for different $\Pt^{\gamma}$ intervals will be presented in a different way than in Section 5.2.
Namely, Tables \ref{tab:sh1}--\ref{tab:sh4} include the rates of events 
($L_{int}=3~fb^{-1}$) for different $\eta^{jet}$ intervals, covered by the 
barrel, endcap and forward (HB, HE and HF) parts of the calorimeter and 
for different $\Ptg(\approx\Pt^{jet})$ intervals.
The selection cuts are as those of Section 3.2  specified by the following values 
of the cut parameters: \\[-7mm]
\begin{eqnarray}
\Pt^{isol}_{CUT}=5\;GeV/c; \quad
{\epsilon}^{\gamma}_{CUT}=7\%; \quad
\Delta \phi<15^{\circ}; \quad
\Pt^{clust}_{CUT}=5\;GeV/c.
\end{eqnarray}

No restrictions on other parameters are used. The first columns of these 
tables $HB$ give the number of events with the jets (found by the LUCELL jetfinding algorithm 
of PYTHIA), all particles of which are comprised (at the particle level of simulation)
entirely (100$\%$) in the HB part and there is a
$0\%$ sharing of  $\Pt^{jet}$ ($\Delta \Pt^{jet}=0$) between the HB and
the neighboring HE part of the calorimeter. The second columns of the 
tables $HB+HE$ contain the number of events in which $\Pt$ of the jet is shared
between the HB and HE regions. The same sequence of restriction conditions takes
place in the next columns. Thus, the $HE$ and $HF$ columns include the number of events 
with jets entirely contained in these regions, while the $HE+HF$ column gives 
the number of  events where the jet covers both the HE and HF regions. From these tables we can see what 
number of events can, in principle, most suitable for the precise jet energy absolute scale 
setting, carried out separately for the HB, HE and HF parts of the calorimeter in different 
$\Ptg$ intervals.

Less restrictive conditions, when up to $10\%$ of the jet $\Pt$
are allowed to be shared between the HB, HE and HF parts of the calorimeter, are given in 
Tables \ref{tab:sh2} and \ref{tab:sh4}. Tables \ref{tab:sh1}  and \ref{tab:sh2} 
correspond to  the case of Selection 1.
Tables \ref{tab:sh3} and \ref{tab:sh4} contain the  number of events collected
with Selection 2 criteria (defined in Section 3.2), i.e.
they include only the events with ``isolated jets'' satisfying the isolation criterion
$\epsilon^{jet}\lt2\%$. A reduction factor of 4 for the  number of  events can be 
found by comparing those tables with Tables 9, 10. This is the cost of passing to Selection 2.

\begin{table}[htbp]
\vskip-8mm
Table \ref{tab:sh1} corresponds to the most restrictive selection
$\Delta\Pt^{jet}=0$ and gives the number of events most suitable for 
jet energy calibration. From its last summarising line we see 
that for the entire interval $40<\Pt^{\gamma}<360\; GeV/c$~ PYTHIA predicts
around half a million events for HB and a quarter of a million events
for HE per month of continuous data taking at
low LHC luminosity, while for HF the expected value is 75 000
events per month.
\begin{center}
\caption{Selection 1. $\Delta \Pt^{jet} / \Pt^{jet} = 0.00$}
\vskip0.2cm
\begin{tabular}{||c||c|c|c|c|c||} \hline \hline
\label{tab:sh1}
$\Pt^{\gamma}$&   HB   &   HB+HE &     HE  &   HE+HF &     HF\\\hline \hline
 40 --  50& 260259&  211356&  141759&  102299& 45354 \\\hline
 50 --  60& 108827& 89126& 55975& 41553& 17216 \\\hline
 60 --  70&  49585& 40076& 25172& 18153&  7019 \\\hline
 70 --  80&  25506& 20897& 12881&  9679&  3021 \\\hline
 80 --  90&  14083& 11720&  7529&  4873&  1304 \\\hline
 90 -- 100&   7261&  7054&  4142&  2924&   853 \\\hline
100 -- 120&   7703&  6913&  4013&  2926&   588 \\\hline
120 -- 140&   3372&  2977&  1805&  1164&   207 \\\hline
140 -- 160&   1650&  1481&   865&   509&    50 \\\hline
160 -- 200&   1493&  1137&   708&   396&    40 \\\hline
200 -- 240&    503&   406&   242&   107&     6 \\\hline
240 -- 300&    287&   215&   122&    40&     1 \\\hline
300 -- 360&     96&    73&    35&     8&     0 \\\hline \hline
 40 -- 360& 480538&393378&255266&184642& 75660 \\\hline \hline
\end{tabular}
\vskip0.4cm
\caption{Selection 1. $\Delta \Pt^{jet} / \Pt^{jet} \leq 0.10$}
\vskip0.2cm
\begin{tabular}{||c||c|c|c|c|c||} \hline \hline
\label{tab:sh2}
$\Pt^{\gamma}$  &    HB   &   HB+HE &     HE  &   HE+HF &     HF\\\hline \hline
 40 --  50& 341043& 55160& 263629&26653& 74534\\\hline
 50 --  60& 144955& 20396& 108765& 9300& 29281\\\hline
 60 --  70&  65525&  8541& 49412&  3907& 12621\\\hline
 70 --  80&  34155&  4093& 25918&  1957&  5860\\\hline
 80 --  90&  19224&  1961& 14741&   804&  2778\\\hline
 90 -- 100&  10258&  1304&  8394&   536&  1742\\\hline
100 -- 120&  10859&  1043&  8357&   545&  1338\\\hline
120 -- 140&   4618&   509&  3675&   178&   546\\\hline
140 -- 160&   2325&   222&  1751&    90&   168\\\hline
160 -- 200&   1971&   147&  1458&    52&   147\\\hline
200 -- 240&    685&    61&   472&    20&    26\\\hline
240 -- 300&    383&    32&   234&     7&     9\\\hline
300 -- 360&    129&    10&    72&     1&     0\\\hline \hline
 40 -- 360& 636418& 93480&486788& 44052&129050\\\hline \hline
\end{tabular}
\end{center}
\end{table}

\begin{table}[htbp]
\hspace*{9mm} One should keep in mind that the last columns in Tables
\ref{tab:sh1}--\ref{tab:sh4} cannot be taken as the final
 result here because we have not defined the meaning of sharing
$\Pt^{jet}$ between the HF regions and the region with $|\eta|>5$, i.e.
close to a ``beam-pipe'' region. More accurate estimation can be done
here by finding events with jets in a wider region than the HF volume restricted by
$3<|\eta^{HF}|<5$ and by calculating the number of events in which jets 
are entirely contained in HF.
\hspace*{9mm} An additional information on the number of ``HB-events'' 
(i.e. events, corresponding to $HB$ column of Table 11) 
with jets produced by $c$ and $b$ quarks in gluonic subprocess (1a), i.e.
\begin{center}
\caption{Selection 2. $\Delta \Pt^{jet}/\Pt^{jet}= 0.00$, ~$\epsilon^{jet}<2\%$.}
\vskip0.2cm
\begin{tabular}{||c||c|c|c|c|c||} \hline \hline
\label{tab:sh3}
$\Pt^{\gamma}$  &    HB   &   HB+HE &     HE  &   HE+HF &     HF\\\hline \hline
 40 --  50&  46972& 32954& 26114& 16208& 10041\\\hline
 50 --  60&  23717& 18911& 13448&  8367&  5047\\\hline
 60 --  70&  14384&  9751&  7469&  4703&  2386\\\hline
 70 --  80&   8546&  6733&  4627&  2960&  1206\\\hline
 80 --  90&   5653&  4386&  3107&  1925&   573\\\hline
 90 -- 100&   3326&  3119&  1900&  1377&   390\\\hline
100 -- 120&   4157&  3435&  2271&  1467&   324\\\hline
120 -- 140&   2183&  1786&  1185&   710&   134\\\hline
140 -- 160&   1175&  1005&   635&   362&    31\\\hline
160 -- 200&   1179&   905&   565&   314&    25\\\hline
200 -- 240&    442&   353&   212&    97&     5\\\hline
240 -- 300&    273&   200&   116&    37&     1\\\hline
300 -- 360&     94&    71&    35&     7&     0\\\hline \hline
 40 -- 360& 112111& 83617& 61686& 38535& 20163\\\hline \hline
\end{tabular}
\vskip0.4cm
\caption{Selection 2. $\Delta \Pt^{jet}/\Pt^{jet}\leq 0.10$, ~$\epsilon^{jet}<2\%$.}
\vskip0.2cm
\begin{tabular}{||c||c|c|c|c|c||} \hline \hline
\label{tab:sh4}
$\Pt^{\gamma}$  &    HB   &   HB+HE &     HE  &   HE+HF  &    HF\\\hline \hline
 40 --  50&  60113&  7986& 45388&  3909& 14894\\\hline
 50 --  60&  31495&  3631& 25134&  1971&  7259\\\hline
 60 --  70&  18326&  2248& 13139&   968&  4011\\\hline
 70 --  80&  11385&  1243&  8741&   573&  2132\\\hline
 80 --  90&   7614&   633&  5957&   292&  1145\\\hline
 90 -- 100&   4544&   536&  3886&   280&   865\\\hline
100 -- 120&   5771&   481&  4434&   278&   689\\\hline
120 -- 140&   2909&   272&  2370&    94&   352\\\hline
140 -- 160&   1648&   138&  1246&    65&   111\\\hline
160 -- 200&   1560&   113&  1162&    38&   115\\\hline
200 -- 240&    600&    53&   416&    17&    23\\\hline
240 -- 300&    362&    30&   220&     6&     8\\\hline
300 -- 360&    126&    10&    71&     1&     0\\\hline \hline
 40 -- 360& 146468& 17374&112177&  8492& 31603\\\hline \hline
\end{tabular}
\end{center}
\end{table}

\normalsize
\def\baselinestretch{1.0}

\noindent
 $Nevent_{(c)}$ and $Nevent_{(b)}$ 
(given for the integrated luminosity $L_{int}=3~ fb^{-1}$)
for different $\Pt^{Jet}$($\approx\Ptg$) intervals $40-50, 100-120, 200-240$ and $300-360~GeV/c$
are contained in Tables 1--12 of Appendix 1
\footnote{Analogous estimations were done in \cite{BKS_GLU,MD1},\cite{MD_}--\cite{BALD02}.}. 
The ratio (``$29sub/all$'') of the number of events caused by gluonic subprocess (1a) $(=29sub)$, 
summed over quark flavours, to the number of events due to the sum of subprocesses (1a) and (1b) $(=all)$, 
also averaged over all quark flavours, is also shown there.

\section{FEATURES OF ~\gpj EVENTS IN THE CENTRAL CALORIMETER REGION.}

\it\small
\hspace*{9mm}
The influence of $\Pt^{clust}_{CUT}$ parameter (defining the upper limit on $\Pt$ of clusters or mini-jets 
in the event) on the variables characterizing the \ptgj balance  as well as on the 
$\Pt$ distribution in jets and out of them is studied.
\rm\normalsize
\vskip3mm

In this section we shall study the specific sample of events considered in 
the previous section that may be most suitable for the jet energy calibration in the HB region,
with jets entirely (100$\%$) contained in this region, i.e.
having 0$\%$ ~sharing of $\Pt^{jet}$ (at the PYTHIA particle level of simulation) with HE.
{\it Below we shall call them ''HB-events''}. The $\Ptg$ spectrum for this particular
set of events for $\Pt^{clust}=5 ~GeV/c$ was presented in the second column (HB) of Table \ref{tab:sh1}.
Here we shall use three different jetfinders, namely, LUCELL from PYTHIA
and UA1 and UA2 from CMSJET \cite{CMJ}. The  $\Pt^{clust}$ distributions for generated events found by 
the all three jetfinders in two $\Pt^{\gamma}$ intervals, $40\lt\Pt^{\gamma}\lt50~GeV/c$
and $\!$ $300<\Pt^{\gamma}<360\,GeV/c$, are shown in Fig.~8 for $\Pt^{clust}_{CUT}=30\;GeV/c$.
It is interesting to note an evident similarity of the $\Pt^{clust}$ spectra with $\Pt56$ spectra
(for $\dphi\leq 15^\circ$) shown in Tables 2 and 3 (see also Figs.~9, 10), 
what support our intuitive picture of ISR and cluster connection described in  Section 2.2. \\[-0.6cm]

\begin{flushleft}
\begin{figure}[htbp]
 \vskip-20mm
 \hspace{-.5cm} \includegraphics[height=74mm,width=9.7cm]{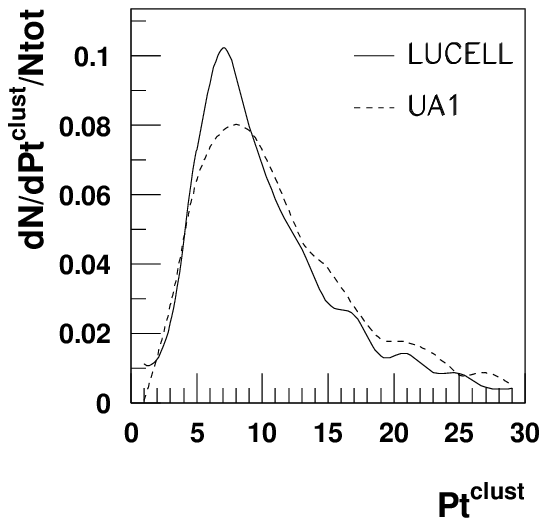}
    \label{fig9}
   \nonumber
  \end{figure}
\end{flushleft}
\begin{flushright}
\begin{figure}[htbp]
 \vskip-94mm
  \hspace{7.5cm} \includegraphics[height=74mm,width=9.7cm]{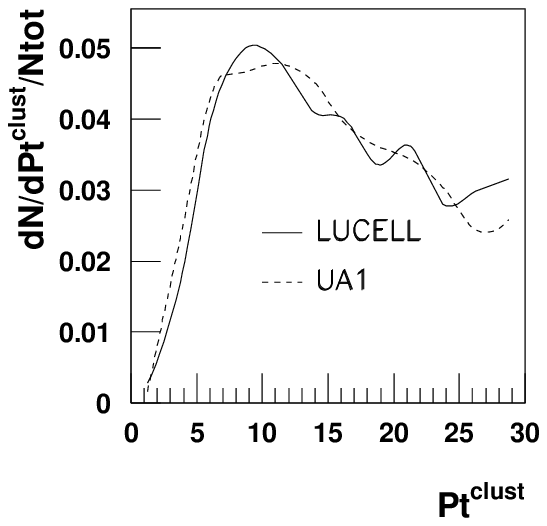}
   \nonumber
\label{fig7}
\vskip-12mm

\hspace*{4.2cm} (a) \hspace*{7.5cm} (b)\\[7pt]
\hspace*{0.3cm}{\footnotesize Fig.~8: $\Pt^{clust}$ distribution in \gpj
events from two $\Pt^{\gamma}$ intervals:
(a) $40<\Pt^{\gamma}<50\, GeV/c$ ~and \\
\hspace*{1.3cm} (b) $300<\Pt^{\gamma}<360\,GeV/c$~ with the same cut
$\Pt^{clust}_{CUT}=30\;GeV/c$ ($\dphi\leq 15^\circ$).}\\[-14mm]
\end{figure}
\end{flushright}

\setcounter{figure}{8}
\begin{figure}[htbp]
\vspace{-3.0cm}
  \hspace{-2mm} \includegraphics{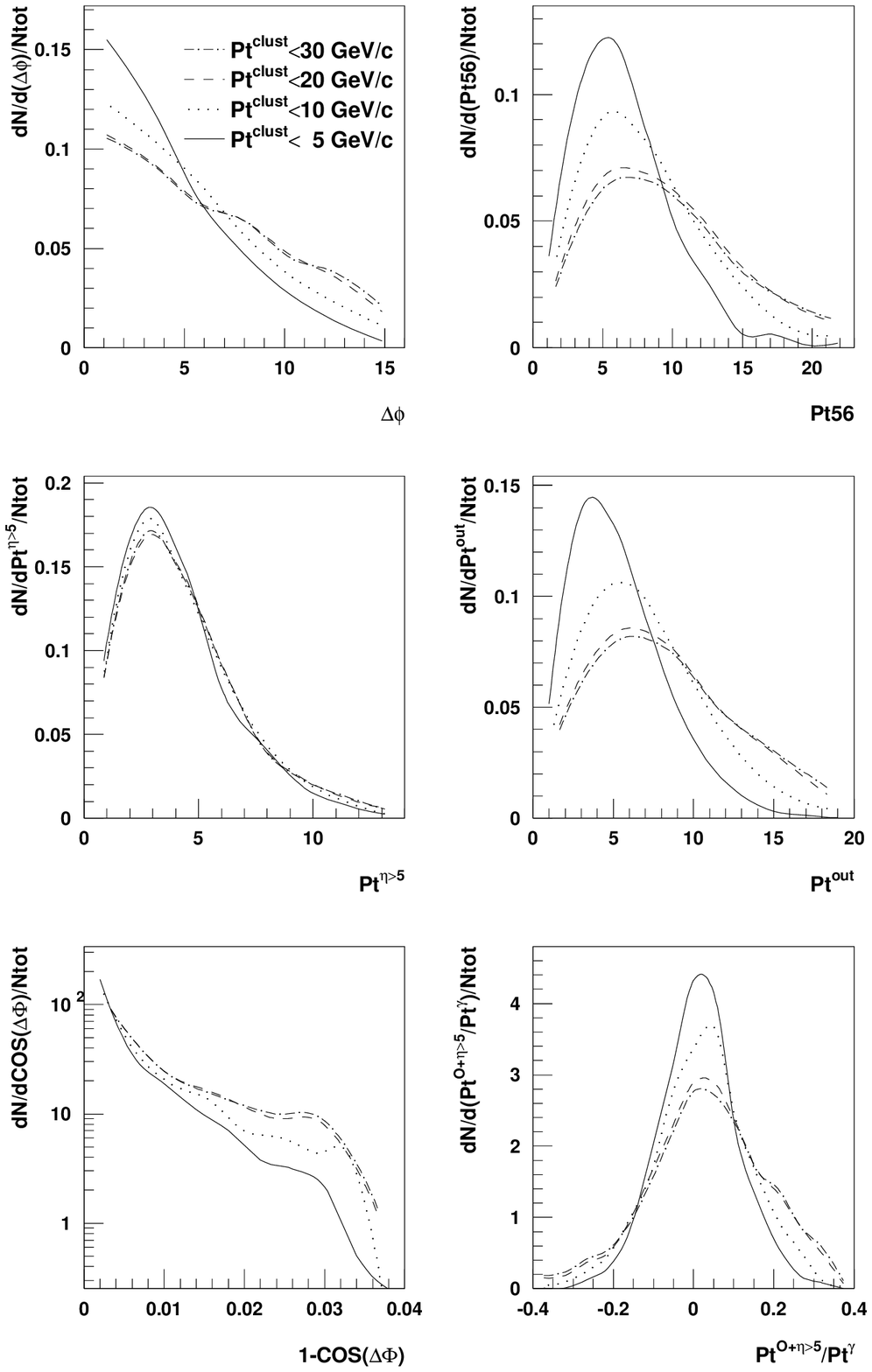} 
  \vspace{-0.5cm}
\caption{\hspace*{0.0cm} LUCELL algorithm, $\dphi<15^\circ$,
$40<\Pt^{\gamma}<50\, GeV/c$. Selection 1.}
\end{figure} 
\begin{figure}[htbp]
 \vspace{-3.0cm}
  \hspace{-2mm} \includegraphics{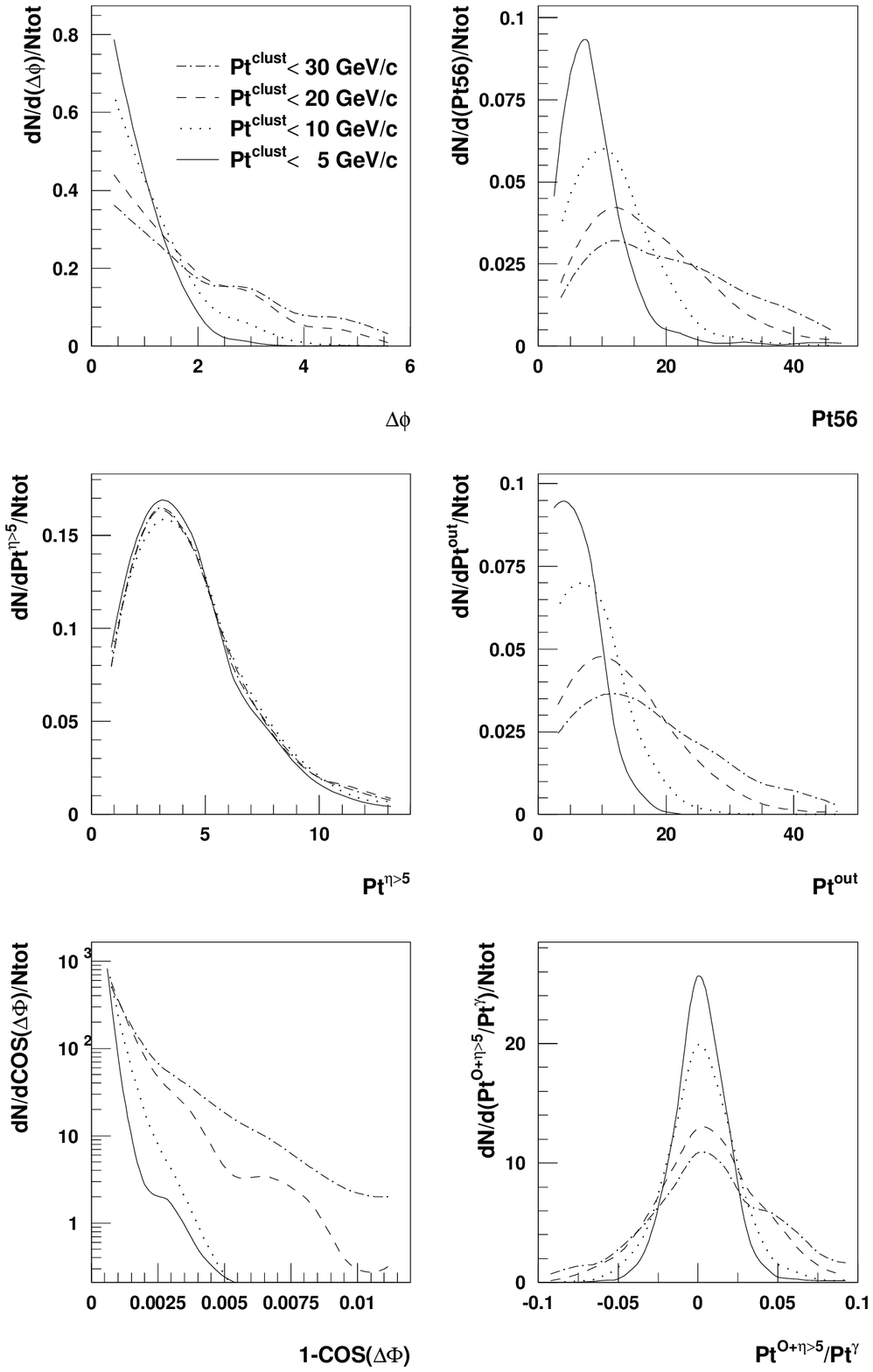}
  \vspace{-0.5cm}
\caption{\hspace*{0.0cm} LUCELL algorithm, $\dphi<15^\circ$,
$300<\Pt^{\gamma}<360\, GeV/c$. Selection 1.}
\end{figure} 

\subsection{Influence of the $\Pt^{clust}_{CUT}~$ parameter
on the photon and jet $\Pt$ balance and on the initial state radiation 
suppression.}

Here we shall study in more detail correlation of $\Pt^{clust}$ with $\Pt^{ISR}$ mentioned above.
The averaged value of intrinsic parton transverse momentum 
will be fixed at $\langle k_t \rangle= 0.44~ GeV/c$
\footnote{The influence of possible $\la k_t \ra$ variation on the
\ptgj~ balance is discussed in Section 9. See also \cite{BKS_P1}--\cite{BKS_P5}.}.

The banks of 1-jet \gpj events gained from the results of PYTHIA
generation of $5\cdot10^6$  signal \gpj events in each of four $\Pt^{\gamma}$
intervals (40 -- 50, 100 -- 120, 200 -- 240, 300 -- 360 $GeV/c$)
\footnote{they were discussed in Section 5}
will be used here. The observables defined in Sections 3.1 and 3.2  will be restricted here 
by Selection 1 cuts (17) -- (24) of Section 3.2 and the cut parameters defined by (29).

We have chosen from them two extreme $\Ptg$ intervals  to illustrate the influence of the
$\Pt^{clust}_{CUT}$ parameter on the distributions of physical variables,
 that enter the balance equation (28). These distributions are shown in Fig.~9
($40<\Pt^{\gamma}<50~ GeV/c$) and Fig.~10 ($300<\Pt^{\gamma}<360~ GeV/c$). 
 
In these figures, in addition to three variables $\Pt56$,
$\Pt^{\eta>5.0}$, $\Pt^{out}$, already explained in Sections 2.2, 3.1 and 3.2,
we present distributions of two other variables, $\Db$~ and $(1-cos\dphi)$, which
define the right-hand side of equation (28).
The distribution of the $\gamma$-jet back-to-back  angle $\dphi$ (see (22))
is also presented in Figs.~9 and 10.

The ISR describing variable $\Pt56$ (defined by formula (3))
and both components of the experimentally observable disbalance measure $\Fptgj$
(see (28)) as a sum of $(1-cos\dphi)$ and $\Db/\Ptg$, as well as two others, $\Pt^{out}$ and  $\dphi$,
show a tendency, to become smaller (the mean values and the widths)
with the restriction of the upper limit on the 
$\Pt^{clust}$ value (see Figs.~9, 10).
It means that the jet energy calibration precision may increase with decreasing
$\Pt^{clust}_{CUT}$, which justifies the intuitive choice of this new variable in Section 3.
The origin of this improvement becomes clear from the $\Pt{56}$ density plot, which demonstrates 
the decrease of $\Pt{56}$ (or $\Pt^{ISR}$) values with decrease of $\Pt^{clust}_{CUT}$.


Comparison of Fig.~9 (for $~40\!<\Pt^{\gamma}\!<50 ~GeV/c$) and Fig.~10
(for $300<\Pt^{\gamma}<360~ GeV/c$) also shows that the values of $\Delta\phi$ as a degree of
back-to-backness of the photon and jet $\Pt$ vectors in the $\phi$-plane
decreases with increasing $\Pt^{\gamma}$. At the same time $\Pt^{out}$ and $\Pt^{ISR}(=\Pt56)$ 
distributions become slightly wider. It is also seen that
the $\Pt^{\eta>5.0}$ distribution practically does not depend on
$\Pt^{\gamma}$ and $\Pt^{clust}$
\footnote{see also Appendices 2--5}.

It should be mentioned that the results presented in Figs.~9 and 10 were
 obtained with the LUCELL jetfinder of PYTHIA
\footnote{The results obtained with all jetfinders and
\ptgj ~balance will be discussed in Section 7 in more detail.}.

\subsection{Jetfinders and the $\Pt$ structure of jets in the $\eta-\phi$ space.}

In order to understand well the calibration procedure of \gpj events,
it is useful to keep control over some principal characteristics of internal jet
structure as well as over the size of $\Pt$ activity in the space
around jets. 

Let us define the coordinates of jet center of gravity in the $\eta-\phi$ space
(according to the PYTHIA's LUCELL subroutine definitions):\\[-7pt]
\begin{equation}
\eta_{gc}={\left(\sum\limits_{i=1}^{NC} \eta_i \Pt^i\right)}/\left({\sum\limits_{i=1}^{NC}\Pt^i}\right);
\quad
\phi_{gc}={\left(\sum\limits_{i=1}^{NC} \phi_i \Pt^i\right)}/\left({\sum\limits_{i=1}^{NC}\Pt^i}\right)
\label{eq:COG}
\end{equation}
The sum in formulae (\ref{eq:COG}) runs over jet cells whose total number is denoted by $NC$.

The left-hand columns of Figs.~11 and 12 present distributions
over a distance, denoted as $R^{jet}_{gc}({\eta,\phi})$,
between the center of a most remote ($mr$) cell of the jet and the jet center of gravity
($\eta_{gc}$, $\phi_{gc}$) in HB-events for the intervals $40<\Pt^{\gamma}<50\,GeV/c$
and $300<\Pt^{\gamma}<360\,GeV/c$ respectively, i.e.\\[-15pt]
\begin{equation}
R^{jet}_{gc}({\eta,\phi})=((\eta_{mr}-\eta_{gc})^2+(\phi_{mr}-\phi_{gc})^2)^{1/2},
\end{equation}
where ($\eta_{mr}$, $\phi_{mr}$) are the coordinates of the center of most remote
cell of the jet.

We choose according to Section 3.2 the jet radius counted from the initiator cell 
($ic$) to be restricted by $R^{jet}_{ic}=0.7$ for the LUCELL and UA1 jetfinders
while its value is not limitted for UA2 algorithm.
\begin{center}
\begin{figure}[htbp]
 \vspace{-3.0cm}
  \hspace{-2mm} \includegraphics{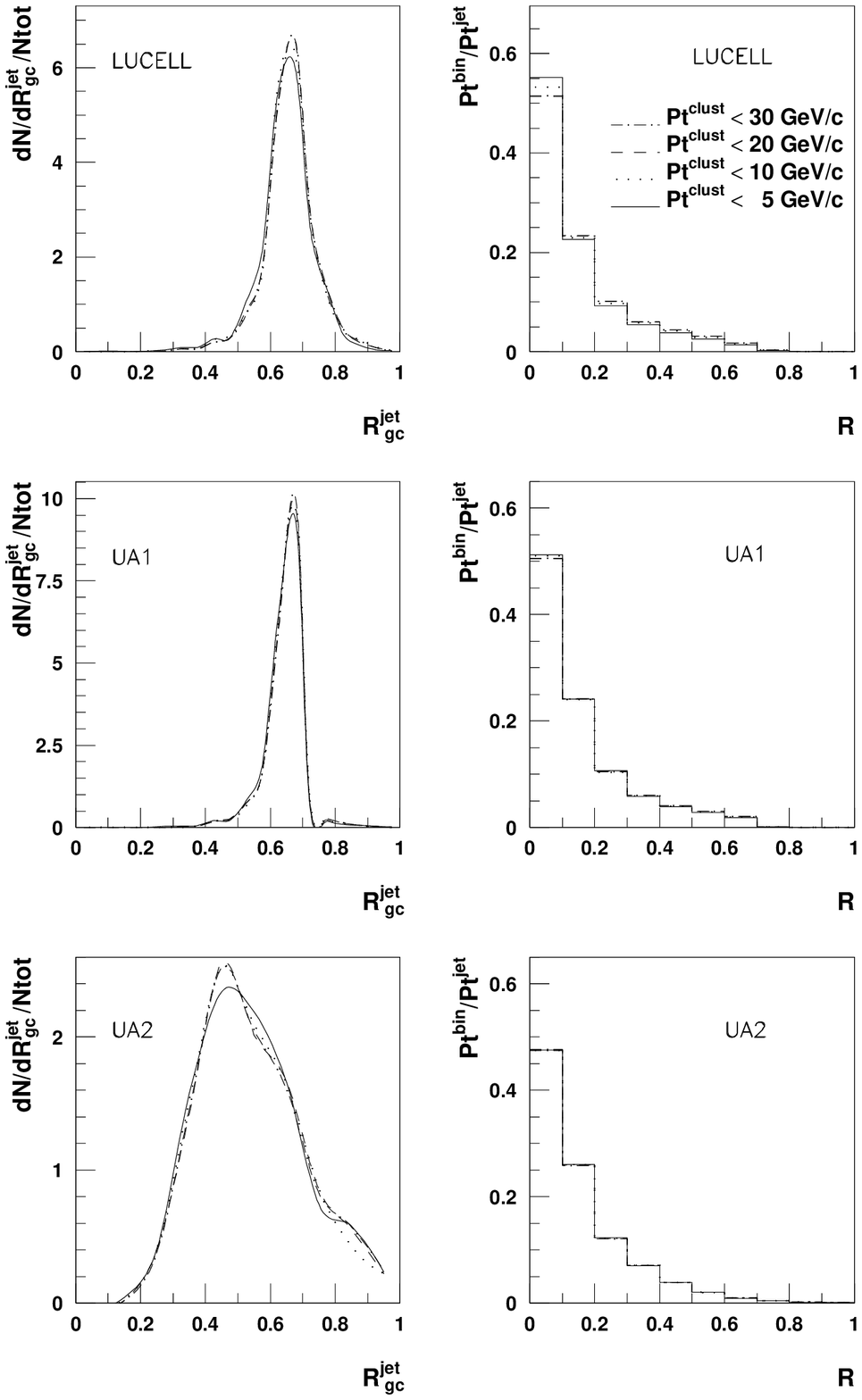} 
  \vspace{-0.5cm}
\caption{
\hspace*{0.0cm} LUCELL, UA1 and UA2 algorithms, $\dphi<15^\circ$,
$40<\Pt^{\gamma}<50\, GeV/c$. Selection 1.}
    \label{fig10} 
  \end{figure}
\begin{figure}[htbp]
 \vspace{-3.0cm}
  \hspace{-2mm} \includegraphics{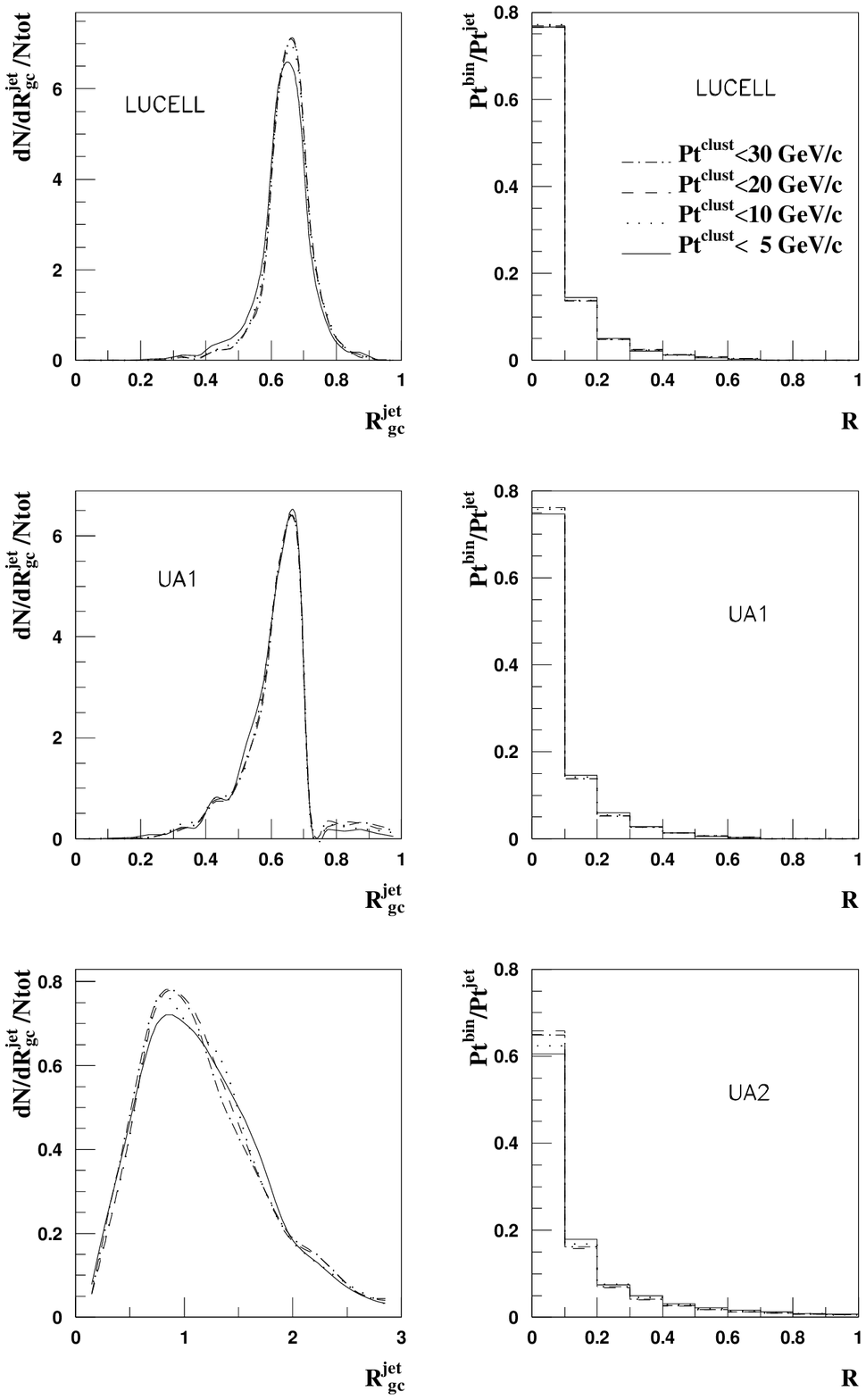}
  \vspace{-0.5cm}
    \caption{
\hspace*{0.0cm} LUCELL, UA1 and UA2 algorithms, $\dphi<15^\circ$,
$300<\Pt^{\gamma}<360\, GeV/c$. Selection 1.}
    \label{fig11}
  \end{figure}
\end{center}

From the left-hand side plots in Figs.~11 and 12 we see
that UA1 and LUCELL jetfinders give close $R^{jet}_{gc}$ distributions.

The detailed information about the averaged jet radii
for four $\Pt^{\gamma}$ intervals is presented in the tables of Appendix 1
\footnote{They show weak dependence of the jet radius on
$\Pt^{Jet}(\approx\Pt^{\gamma})$ for all algorithms.}.
\pagestyle{plain}

Now let us consider how the transverse momentum is
distributed inside a jet.
Let us divide the jet radius  $R^{jet}({\eta,\phi})\equiv R$ into a set of
$\Delta R$ bins and calculate the vector sums of cells $\Pt$ in each $\Delta R_{bin}$ ring.
Normalized to $\Pt^{Jet}$, the modulus of this vector sum, denoted
by $\Pt^{bin}$, would give the value that tells us what portion of a total
$\Pt^{Jet}$ is contained in the ring of size  $\Delta R_{bin}$.
Its variation with the distance $R$ counted from the center of
gravity of the jet is shown in the right-hand columns of Figs.~11 and 12.

From these figures we can conclude that the LUCELL, UA1 and UA2 jetfinders 
irrespective of their internal ways of jet radius calculations,
lead to more or less similar structure of $\Pt$ density 
in the central part inside a jet.


\subsection{$\Pt$ distribution inside and outside of a jet.}

\thispagestyle{plain}
Now let us see how the volume outside the jet,
(i.e. calorimeter cells outside the jet cone) may be populated by $\Pt$
in these HB \gpj events. For this
purpose we calculate a vector sum $\vec{\Pt}^{sum}$ of individual transverse
momenta of $\Delta \eta \times \Delta \phi$ cells included
by a jetfinder into a jet and of cells in a larger
volume that surrounds a jet. In the latter case this procedure can 
be viewed as straightforward enlarging of the jet radius in the $\eta -\phi$ space.
The figures that show the ratio  $\Pt^{sum}/{\Pt^{\gamma}}$
as a function of the distance $R( \eta,\phi)$ counted from the jet
gravity center towards its boundary and further into space outside the jet
are shown in the left-hand columns of
Figs.~13 and 14 for two different $\Ptg$ intervals ($40<\Ptg<50 ~GeV/c$ 
in Fig.~13 and the $300<\Ptg<360 ~GeV/c$ in Fig.~14 intervals) 
in the case when all jet particles are kept in the jet.  

From these figures we see that the space surrounding the jet is
in general far from being an empty in the case of
\gpj events considered here. We also see that an average value of
the total $\Pt^{sum}$ increases with increasing volume around the jet
and it exceeds $\Pt^{\gamma}$ at $R=0.7-0.8$ when all particles are included in the jet
(see Figs.~13 and 14). 

From the right-hand columns of Figs.~13 and 14 we see that when all
particles are included in the jet, the disbalance measure (the analog of (4))
~\\[-5pt]
\begin{equation}
\Pt^{\gamma+sum}=
\left|\vec{\Pt}^{\gamma}+\vec{\Pt}^{sum}\right|
\end{equation}
achieves its minimum at $R\approx 0.7-0.8$ for all three jetfinding algorithms
\footnote{This value is denoted as ``$\Ptg+\Pt^{sum}$'' in Figs.~13--16.}.
%

The value of $\Pt^{\gamma+sum}$ continues to grow rapidly with increasing
$R$ after the point $R=0.7-0.8$ for $40<\Pt^{\gamma}<50~ GeV/c$
(see Figs.~13, 15), while for higher
$\Pt^{\gamma}$ (see Figs.~14, 16 for the
$300<\Pt^{\gamma}<360 ~GeV/c$ interval) the ratio
$\Pt^{sum} / \Pt^{\gamma}$ and the disbalance
measure $\Pt^{\gamma+sum}$ increase more slowly with increasing
$R$ after the point $R=0.7-0.8$. This means that at higher $\Pt^{\gamma}$
(or $\Pt^{Jet}$) the topology of \gpj events becomes more
pronounced and we get a clearer picture of an "isolated" jet. This feature
clarifies the motivation of introducing by us the ``Selection 2'' criteria
in Section 3.2 (see point 9) for selection of events with ``isolated jets''.

\setcounter{figure}{12}
\begin{center}
\begin{figure}[htbp]
 \vspace{-3.0cm}
  \hspace{.0cm} \includegraphics[width=16cm]{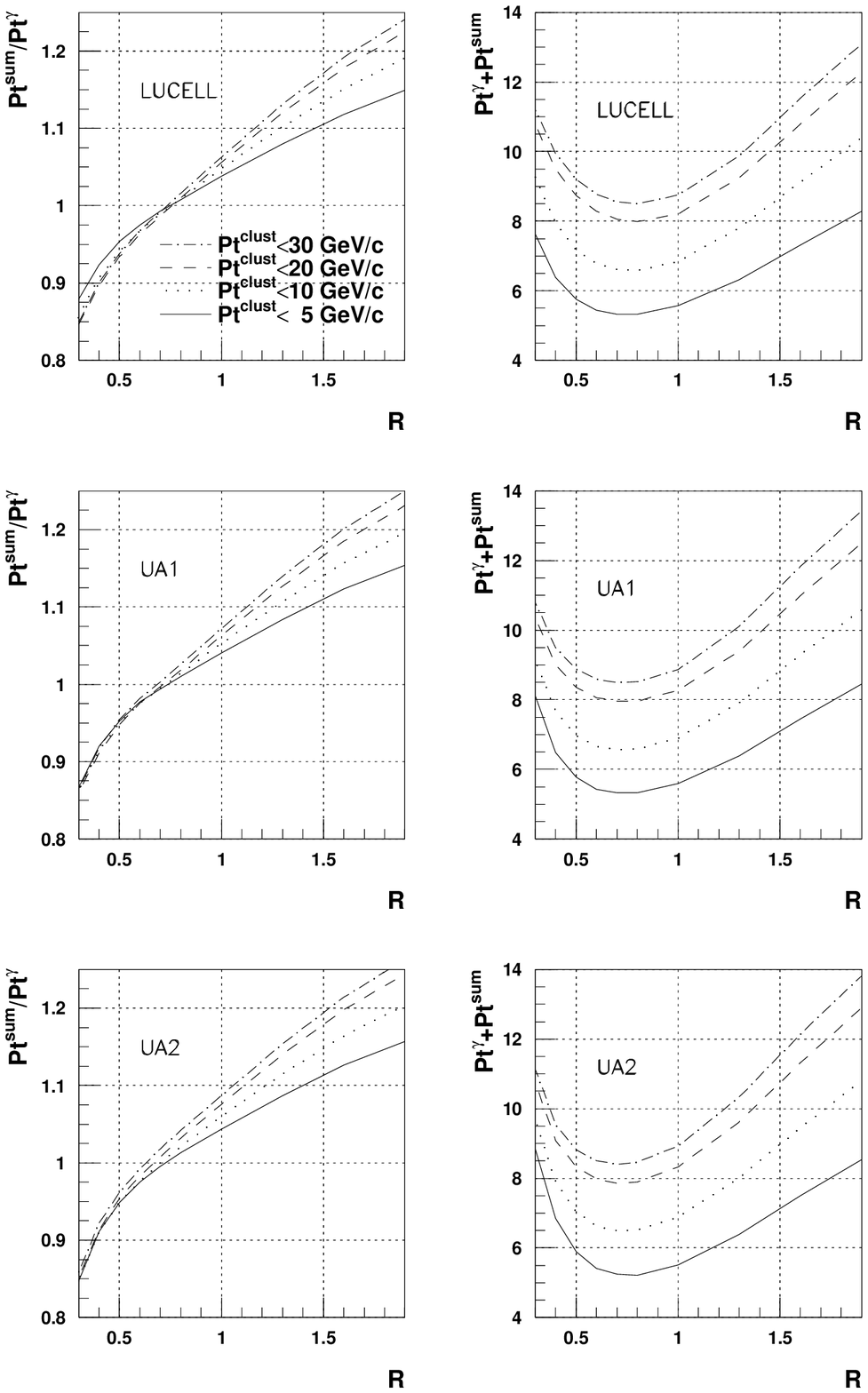}
  \vspace{-0.5cm}
    \caption{\hspace*{0.0cm} LUCELL, UA1 and UA2 algorithms, $\dphi<15^\circ$,
$40<\Pt^{\gamma}<50\, GeV/c$ (without account of magnetic field effect).}
    \label{fig18}
\end{figure}
\pagestyle{plain}

\begin{figure}[htbp]
 \vspace{-3.0cm}
  \hspace{.0cm} \includegraphics[width=24cm]{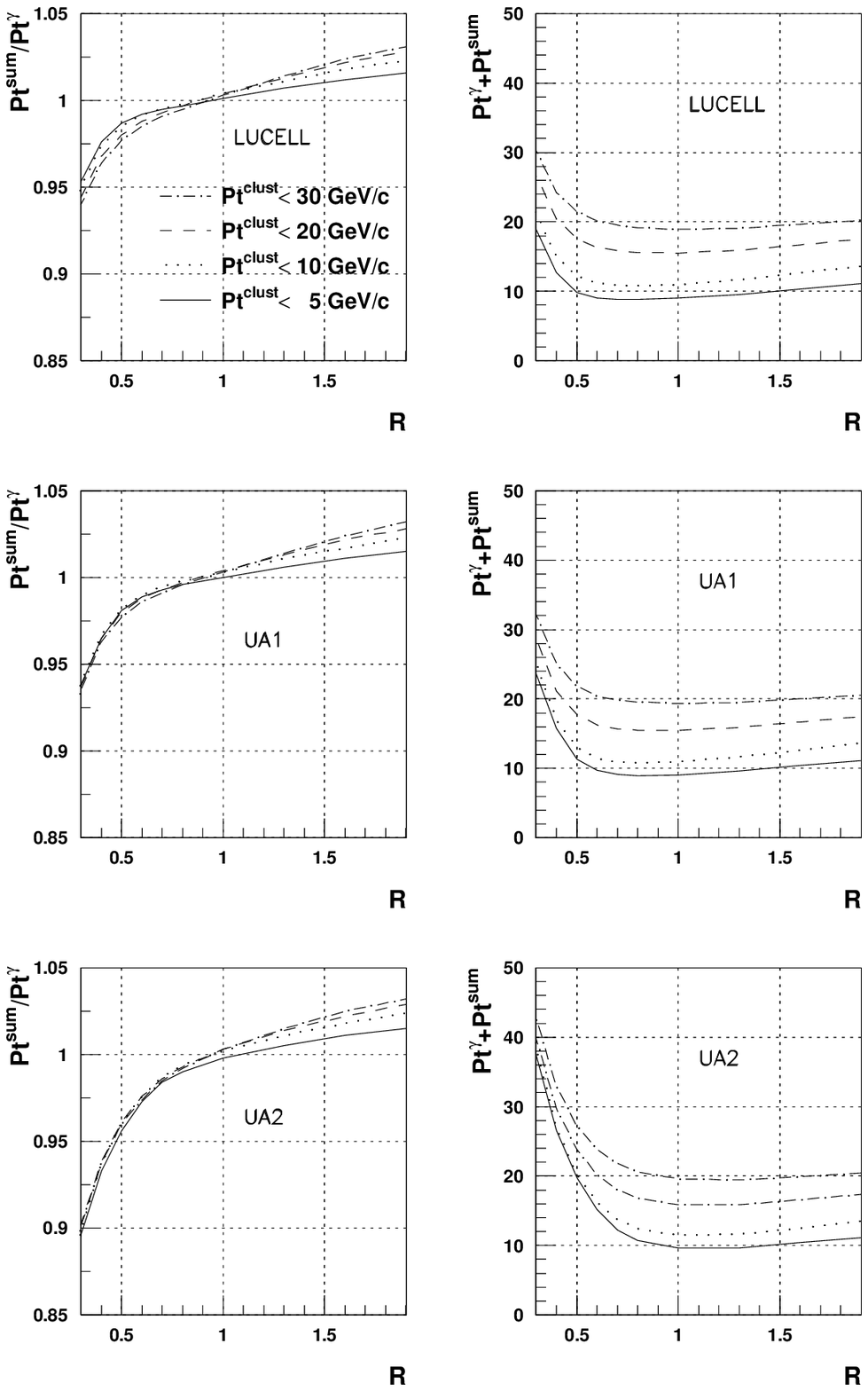}
  \vspace{-0.5cm}
    \caption{\hspace*{0.0cm} LUCELL, UA1 and UA2 algorithms, $\dphi<15^\circ$,
$300<\Pt^{\gamma}<360\, GeV/c$ (without account of magnetic field effect).}
    \label{fig19}
\end{figure}
\end{center}

\pagestyle{plain}

\section{DEPENDENCE OF THE $\Pt$-DISBALANCE IN THE \gpj SYSTEM 
ON $\Pt^{clust}_{CUT}$ and $\Pt^{out}_{CUT}$ PARAMETERS.}

\it\small
\hspace*{9mm}
It is shown that with Selection 2 
one can collect (at the particle level) a sufficient number of events with the value 
of a fractional $\Fptgj$ disbalance better than $1\%$. 
The number of events (at $L_{int}=3~fb^{-1}$)
together with other characteristics of \gpj events are presented in tables 
of Appendices 2--5 for interval $40<\Pt^{\gamma}<360~ GeV/c$. 
They show a possibility to define jet energy scale at low luminosity in few months.
\rm\normalsize
\vskip4mm

In the previous sections we have introduced physical variables
for studying \gpj events (Section 3) and discussed what cuts
for them may lead to a decrease in the disbalance of $\Pt^{\gamma}$ and $\Pt^{Jet}$  
(Sections 5, 6). One can make these cuts to be tighter if more events would be collected
during data taking.

Here we shall study in detail the dependence of the $\Pt$ disbalance
 in the \gpj system on $\Pt^{clust}_{CUT}$ and $\Pt^{out}_{CUT}$  values. 
For this aim we shall use the same samples of events as in Section 5 that were generated
by using PYTHIA with 2 QCD subprocesses (1a) and (1b) and collected 
to cover four $\Ptg$ intervals: $40-50, 100-120, 200-240, 300-360~ GeV/c$.
These events were selected with\\[-8mm]
\begin{eqnarray}
\Ptg\geq 40~ GeV/c, ~~~
\Pt^{jet}\geq 30~ GeV/c ~~~
\end{eqnarray}

~\\[-7mm]
\noindent
and with the use of the set of cut parameters defined by (29). 

The dependence of the number of events that can be gained with Selection 1 and
the above-mentioned set of cut parameters
on the value of $\Pt^{clust}_{CUT}$ is shown for the case of $\dphi\leq15^\circ$ and for four
$\Ptg$ intervals in Fig.~16 and in Fig.~18 for Selection 2 and
in Fig.~20 for Selection 3. Each of these plots is accompanied at the same page by four
additional plots (Figs.~17,19,21) that show the dependence of the fractional disbalance 
$\Fptgj$ on $\Pt^{clust}_{CUT}$ in different $\Ptg$ intervals. The dependence of this 
ratio is presented for three different jetfinders LUCEL, UA1 and UA2 used to determine
a jet in the same event. It is worth mentioning that in contrast to UA1 and LUCELL
algorithms that use a fixed value of jet radius $R^{jet}(=0.7)$, the value of
$R^{jet}$  is not restricted directly for UA2
\footnote{The only radii defining in UA2 algorithm are cone radius for preclusters search
($=0.4$) and cone radius for subsequent precluster dressing ($=0.3$) (see \cite{CMJ}).}
and, thus, it may take different values (see \cite{BKS_P2} and $R$ values in Appendices 1).
The differences in the results of these three jetfinders  application 
were discussed in Section 6.2 and in \cite{D0_Note} (see also Appendices 1--5).

\begin{figure}
\vskip-10mm
\hspace{-2mm} \includegraphics[width=16cm]{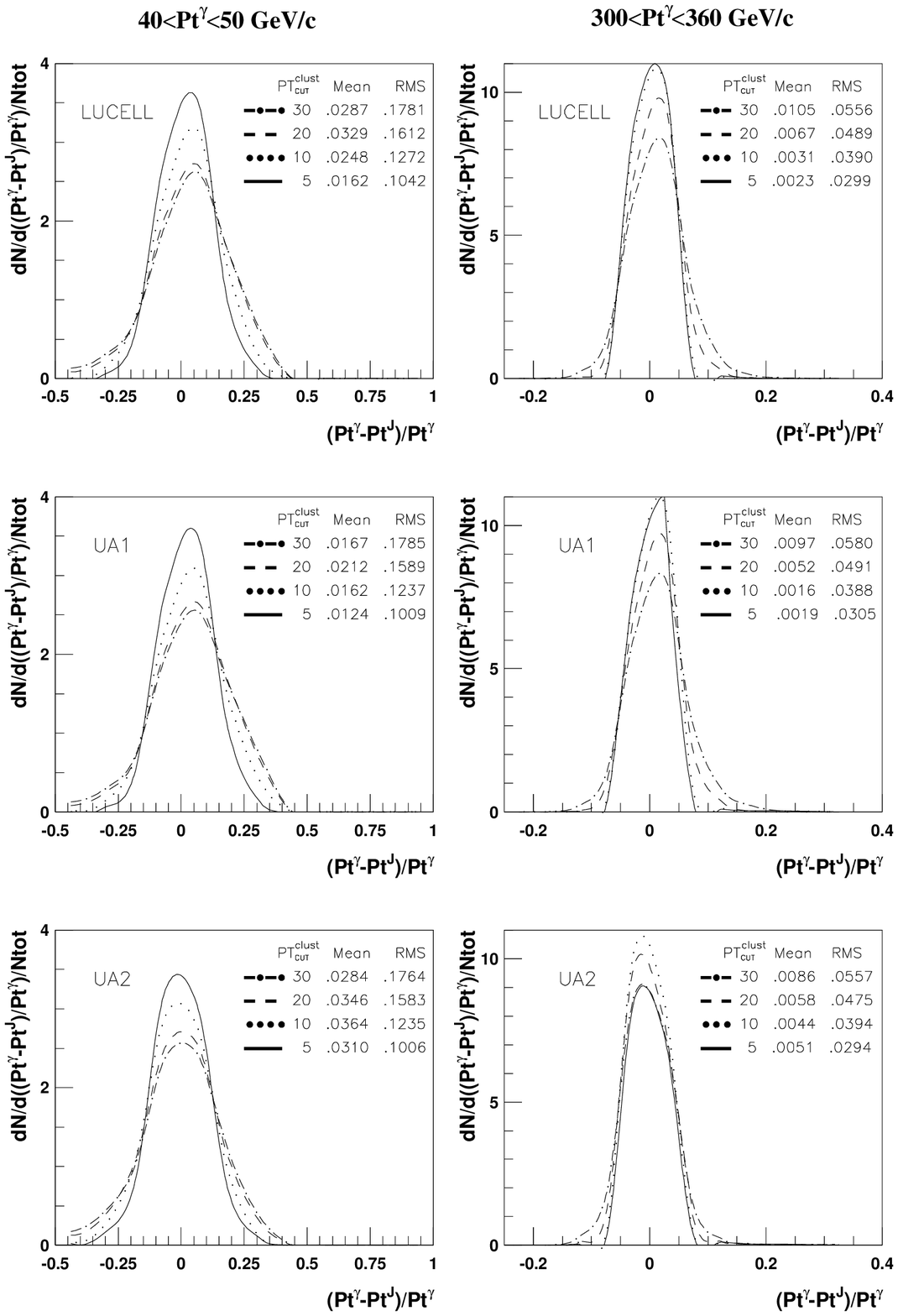} 
\caption{A dependence of $(\Pt^{\gamma}-\Pt^{J})/\Pt^{\gamma}$ on
$\Pt^{clust}_{CUT}$ for LUCELL, UA1 and UA2 jetfinding algorithms and two
intervals of \ptg. ~~The mean and RMS of the distributions are displayed on
the plots. $\dphi\lt15^\circ$. $\Pt^{out}$ is not limited. Selection 1.}
\label{fig:j-clu}
\end{figure}

The normalized event distributions over $\Fptgj$ for two most illustrative $\Ptg$ intervals
$40\lt\Ptg\lt50$ and $300\lt\Ptg\lt360 ~GeV/c$ are shown for a case of $\dphi\leq15^\circ$ 
in Fig.~\ref{fig:j-clu} in different plots for three jetfinders. These plots demonstrate
the dependence of the mean square deviations on $\Pt^{clust}_{CUT}$ value, not shown in Fig.~17.
From the comparison of Figs.~17, 19 and 21 (see also Appendices 2--5)
one can easily see that passing from Selection 1
to  Selection 2 and 3 allows to select events with a better balance of \Ptgj
(about $1\%$ and better) on the PYTHIA particle level. It is also seen that in events
with ``isolated jets'' there is no such a strong dependence on $\Pt^{clust}_{CUT}$ value
in the events with $\Ptg\gt100~GeV/c$.

More details on $\Pt^{clust}_{CUT}$ dependence of different important
features of \gpj events (as predicted by PYTHIA. i.e. without account of detector effects)
are presented in tables of Appendices 2 -- 5. They include the information about 
a topology of events and mean values of most important variables that characterize
$\Ptg-\Pt^{Jet}$ disbalance
\footnote{Please note that the information about averaged values of
jet radius as well as $\Pt^{miss}$ and non-detectable content of a jet is included in the tables of
Appendix 1 for the same $\Ptg$ intervals.}.
This information can be useful as a model guideline
while performing jet energy calibration procedure and also serve for fine tuning
of PYTHIA paramters while comparing its predictions with the collected data.

Appendix 2 contains the tables for events with $\Pt^{\gamma}$ varying from $40$ to
$50~GeV/c$. In these tables we present the values of
interest found with the UA1, UA2 and LUCELL jetfinders
\footnote{the first two are taken from CMSJET fast Monte Carlo program \cite{CMJ}}
for three different Selections
mentioned in Section 3.2. Each page corresponds to a definite value of $\Delta \phi$ 
(see (22)) as a measure of deviation from the absolute
back-to-back orientation of two $\vec{\Pt}^{\gamma}$ and $\vec{\Pt}^{Jet}$ vectors.

So, Tables 1 -- 3 on the first page of each of Appendices 2--5 correspond to 
$\dphi<180^{\circ}$, i.e. to the case when no restriction
on the back-to-back $\dphi$ angle is applied. Tables 4--6 on the second page
correspond to $\dphi<15^{\circ}$.
The third and fourth pages correspond to $\dphi<10^{\circ}$
and $\dphi<5^{\circ}$ respectively. 

The first four pages of each Appendix contain information
about variables that characterize the $\Pt^{\gamma}$ -- $\Pt^{Jet}$ balance
for Selection 1, i.e. when only cuts (17)--(24) of Section 3.2 are used.

On the fifth page of each of Appendices 2--5 we present Tables
13--15 that correspond to Selection 2 described in Section 3.2 for the cut $\dphi<15^{\circ}$.
Selection 2 differs from Selection 1 presented in Tables 1 -- 12
by addition of cut (25). It allows one to select events with the ''isolated jet'', i.e. events
with the total $\Pt$ activity in the $\Delta R = 0.3$ ring around the jet not 
exceeding $3-8\%$ of jet $\Pt$. We have limited $\epsilon^{jet}\leq 8\%$ for
$40\lt\Ptg\lt 50$ with a gradual change to $\epsilon^{jet}\leq 3\%$ for $\Ptg\geq 200 ~GeV/c$.
The best result for UA2 in the case of $40\lt\Ptg\lt 50$ is obtained with a stricter cut
$\epsilon^{jet}\leq 6\%$ (as its radius is larger)
instead of the cut $\epsilon^{jet}\leq 8\%$ chosen for UA1 and LUCELL algorithms
\footnote{In \cite{BKS_P1} -- \cite{BKS_P5} the Selection 2 criterion was considered 
with a more severe cut $\epsilon^{jet} \leq 2\%$.}.
The results obtained with Selection 3
\footnote{Selection 3 (see Section 3.2, point 10) leaves
only those events in which jets are found simultaneously by UA1, UA2 and LUCELL jetfinders
i.e. events with jets having up to a good accuracy equal coordinates of
the center of gravity, $\Pt^{jet}$ and $\phigj$.}
are given on the sixth page of Appendices 2--5. 

The columns in Tables 1 -- 18 correspond to five different
values $\Pt^{clust}_{CUT}=30,\ 20,\ 15,$ $10$ and $5 ~GeV/c$.
The upper lines of Tables 1 -- 15 in Appendices 2--5
contain the expected numbers $N_{event}$ of ``HB events''
(i.e. \gpj events in which the jet is entirely
fitted (at the particle level!) into the  barrel region of the HCAL; see Section 5))
for the integrated luminosity $L_{int}=3\;fb^{-1}$. 

In the next four lines of the tables we put the values of $\Pt56$,
$\Delta \phi$, $\Pt^{out}$ and $\Pt^{|\eta|>5.0}$
defined by formulae (3), (22), (24) and (5) respectively and
averaged over the events selected with a chosen $\Pt^{clust}_{CUT}$ value.

From the tables we see that the values of $\Pt56$, $\Delta \phi$, $\Pt^{out}$ decrease fast
with decreasing $\Pt^{clust}_{CUT}\,$, while the averaged values of
$\Pt^{|\eta|>5.0}$ show very weak dependence on it (practically constant)
\footnote{Compare also with Figs.~9 and 10.}.

The following three lines (from 6-th to 8-th) present the average values of the variables
$\gpart$,
$\Jpart$,
$\gJ$ (here J$\equiv$Jet), the first and the third of which
serve as the measures of the $\Pt$ disbalance in the \gpp and ``$\gamma+Jet$''
systems while the second one has a meaning
of the measure of the parton-to-hadrons (Jet) fragmentation effect. 

The 9-th and 10-th lines include the averaged
values of $\Db/\Pt^{\gamma}$ and $\,(1-cos(\dphi))$ quantities that appear on 
the right-hand side of equation (28), a  scalar variant of vector
equation (16) for the total transverse momentum conservation in a physical event.

The value of $\left<1-cos(\dphi)\right>$ is smaller than the value of
$\left<\Db/\Pt^{\gamma}\right>$ in the case of Selection 1 with the cut $\dphi\lt15^\circ$
and tends to decrease faster with growing energy (compare Figs.~9 and 10). 
So, we can conclude that the main contribution 
into the $\Pt$ disbalance in the \gpj system, as defined by equation (28), in the case of Selection 1
comes from the term $\Db/\Pt^{\gamma}$, while in Selections 2 and 3 the contribution of
$\left<\Db/\Pt^{\gamma}\right>$ reduces with growing $\Pt^{clust}$
to the level of that of $\left<1-cos(\dphi)\right>$ and even to smaller values.

We have estimated separately the contributuions of these two terms  
$\vec{\Pt}^{O}\cdot \vec{n}^{Jet}$ and $\vec{\Pt}^{|\eta|>5.0}\cdot \vec{n}^{Jet}$ 
(with $\vec{n}^{Jet}=\vec{\Pt}^{Jet}/\Pt^{Jet}$, see (28)) that enter $\Db$. 
Firstly from tables it is easily seen that $\Pt^{|\eta|>5.0}$ has practically the same value
in all $\Ptg$ intervals and it does not depend neither on $\dphi$ nor on $\Pt^{clust}$ values
being equal approximately to $5 ~GeV/c$.  At the same time the contribution of its projection
 $\vec{\Pt}^{|\eta|>5.0}\cdot \vec{n}^{Jet}$  shows a dependence
 on  $\Pt^{clust}, \dphi$ and $\Ptg$ (being of order of $\approx 0.1-0.2~ GeV/c$ for $\Pt^{clust}<20~GeV/c$
and $\dphi<15^\circ$). The value of the fraction $\vec{\Pt}^{|\eta|>5.0}\cdot \vec{n}^{Jet}/\Ptg$ for
$\Pt^{clust}_{CUT}=10~GeV/c$ and $\dphi<15^\circ$
is $0.006$ at $40\lt\Ptg\lt50~GeV/c$ and decreases to $0.002$ at $100\lt\Ptg\lt 120~GeV/c$. 
Among these two terms the first one, $\vec{\Pt}^{O}\cdot \vec{n}^{Jet}$, is a measurable one (its value can be
found from the numbers in lines with $\Db$). Below in this section the cuts
on the value of $\Pt^{out}$ is applied to select events with better \Ptgj balance.
Let us emphasize that it is a prediction of PYTHIA.
The second term may be reduced in the experiment by imposing a cut on $\Pt^{miss}$
(see corresponding spectra in Figs. 5, 6 of Section 4) as $\Pt^{|\eta|>5.0}$ is a part of it.

The following two lines contain the averaged values of the standard deviations
{\small $\sgmgj$} and {\small $\sgmgp$} of $\gJ (\equiv Db[\gamma,J])$ and
$\gpart(\equiv Db[\gamma,part])$ respectively.
These two variables drop 
as one goes from $\Pt^{clust}=30 ~GeV/c$ to $\Pt^{clust}=5 ~GeV/c$
for all $\Pt^{\gamma}$ intervals and for all jetfinding algorithms.

The last lines of the tables present the number of generated events (i.e. entries) 
left after cuts.

Two features are clearly seen from these tables:\\
(1) {\it parton-photon} fractional disbalance $\gpart$ in events,\\
\hspace*{5mm} being averaged over number of events selected with $\Pt^{clust}_{CUT}= 20~GeV/c$
and $\dphi<15^\circ$, \\
\hspace*{5mm} does not exceed $1\%$  and it has mainly positive sign in Selection 1;\\ 
(2) {\it parton-to-jet} hadronization/fragmentation effect $\Jpart$.
(that includes partially  \\
\hspace*{5mm} also FSR) can be. It always has a negative value. 
It means that a jet does not receive some \\
\hspace*{5mm}  part of the parent parton transverse momentum $\Pt^{part}$.  It is seen 
that in the case of Selection 1\\
\hspace*{5mm}  this effect gives a larger contribution into \Ptgj disbalance than 
the contribution from\\
\hspace*{5mm} {\it parton-photon}  disbalance even  after application of $\Pt^{clust}_{CUT}=10~GeV/c$. \\
(3) due to different signs these two effects partially compensate each other.


In a case of Selection 1 we see from Appendices 3--5 that for
$\Ptg>50 GeV/c$  the decrease in $\Pt^{clust}_{CUT}$ leads also to
a decrease in the $(\Pt^{\gamma}-\Pt^J)/\Pt^{\gamma}$
ratio, i.e. we select the events that can be lead to more precise level of
jet energy calibration accuracy. For instance, in the
case of $100<\Pt^{\gamma}<120 ~GeV/c$ the mean value of
$(\Pt^{\gamma}\!-\!\Pt^J)/\Pt^{\gamma}$ drops from $4.3\!-\!4.5\%$ to
$1.0\!-\!1.8\%$ (see Tables 4 -- 6 of Appendix 3 and Figs.~16, 17) and in the
case of $200<\Pt^{\gamma}<240 ~GeV/c$ the mean value of this variable
drops from $1.5-1.6\%$ to less than $0.5-0.8\%$ (see Tables 4 -- 6 of Appendix 4).
A worse situation is seen for the $40<\Pt^{\gamma}<50 ~GeV/c$ interval, where
the disbalance changes, i.g. for LUCELL algorithm, as $2.9\to 2.5\%$ unless
to we pass to stricter Selection 2. At the same time a reduction of $\Pt^{clust}$
leads to a reduction of $Db[\gamma,J]$. Fig.~15 serves for accumulation and illustration
of the information about the \ptgj disbalnce variation with $\Pt^{clust}_{CUT}$.

After imposing the jet isolation requirement (see Tables 13--15 of Appendices 2--5)
we observe that starting with $\Pt^{\gamma} = 100 ~GeV/c$ the mean values of
$\Pt^{\gamma}$ and $\Pt^{Jet}$ disbalance, i.e.
$(\Pt^{\gamma}\!-\!\Pt^J)/\Pt^{\gamma}$, are contained inside the $1\%$ window
(at particle level) for any $\Pt^{clust}$. In the $40<\Pt^{\gamma}<50 ~GeV/c$ interval,
where we have enough events even after passing to Selection 2, 
we see that $\Pt^{clust}_{CUT}$ works more effectively
\footnote{The same is true for Selection 3, see \cite{BKS_P1} -- \cite{BKS_P5}. 
In those papers the Selection 2 criterion was considered for a more severe 
cut $\epsilon^{jet} = 2\%$.}.
Thus, $\Pt^{clust}_{CUT}=20 ~ GeV/c$
allows $(\Pt^{\gamma}\!-\!\Pt^J)/\Pt^{\gamma}$ to be reduced to less than $1.5\%$
while a stricter cut $\Pt^{clust}_{CUT}=10 ~ GeV/c$
makes it less than $1\%$. The Selection 2 criterion (by $\Pt^{clust}_{CUT}=10 ~GeV/c$ 
for instance) leaves quite a sufficient number of events:
about 350--750 thousand for different jetfinders (the lower value correspond to UA2 algorithm)
for the $40<\Pt^{\gamma}<50 ~GeV/c$ interval and
about 25 thousand for the $100<\Pt^{\gamma}<120 ~GeV/c$ interval (see Tables 13 -- 15
of Appendices 2, 3 and Figs.~18, 19) at $L_{int}=3 ~fb^{-1}$.


{\normalsize \it Thus, to summarize the results presented in tables of  Appendices 2--5,
we want to underline that:\\
(I) for all Selections the reduction of $\Pt^{clust}_{CUT}$ leads to lower values of mean square deviations 
of the photon-parton $Db[\gamma,part]$ and of photon-jet $Db[\gamma,J]$ balances;\\
(II) after imposing the jet isolation requirement (see Tables 13--15 of Appendices 2--5)
the mean values of $\Pt^{\gamma}$ and $\Pt^{Jet}$ disbalance, i.e.
$(\Pt^{\gamma}\!-\!\Pt^J)/\Pt^{\gamma}$, for all $\Ptg$ intervals 
are contained inside the $1\%$ window for any $\Pt^{clust}\leq10~GeV/c$.
}

The Selection 2  (with $\Pt^{clust}_{CUT}=10 ~GeV/c$, for instance) 
leaves after its application the following number of events 
with jets {\it entirely contained} (see Section 5) {\it in the HB region} 
(at $L_{int}=3 ~fb^{-1}$) :\\
(1) about 350 000 -- 750 000 for  $40<\Pt^{\gamma}<50 ~GeV/c$,\\
(2) about 25 000 for  $100<\Pt^{\gamma}<120 ~GeV/c$, \\
(3) about 2000 for  $200<\Pt^{\gamma}<240 ~GeV/c$ and \\
(4) about 500  for the $300<\Pt^{\gamma}<360 ~GeV/c$. 

\thispagestyle{plain}

For intervals with $\Ptg\geq 100~GeV/c$ these numbers can be three times higher even with larger
values of $\Pt^{clust}_{CUT}$.

The analogous results for Selection 3 are presented in Tables 16--18 of Appendices 2--5. 
Let us consider first the most difficult interval $40\lt\Ptg\lt50 ~GeV/c$.
From the tables of Appendix 2 one can see that this selection leads to approximately 
$30\%$ further reduction of the number of selected events as compared with Selection 2.
A combined usage of all three jetfinders in this $\Ptg$ interval (Tables 16--18) does not improve
the balance values. A requirement of simultaneous jet finding by two of them, namely
UA1 and LUCELL algorithms (that use fixed value of $R^{jet}=0.7$), gives
 values of the $\Fptgj$ balance and other variables, 
presented in Tables 19, 20, close to the case of Selection 2 and leads to a better result
(from point of view of the $\Fptgj$ balance values as well as from point of view of
the number of selected events) as compared with the case of combined usage of all three 
jetfinders for this aim (compare also plots A and B in Fig.~21). 
This fact stresses a good compatibility of UA1 and LUCELL jetfinders.
For other  high $\Ptg$ intervals considered here the UA1, UA2 and LUCELL algorithms give more or 
less close results. 

\thispagestyle{plain}

Let us mention
that Selections 2 and 3, besides improving the \ptgj balance value,
are also important for selecting events with a clean jet topology and for rising the confidence level 
of a jet determination and events selection.

Up to now we have been studying the influence of the $\Pt^{clust}_{CUT}$
parameter on the balance. Let us see, in analogy with Fig.~15,
what effect is produced by $\Pt^{out}_{CUT}$ variation
\footnote{This variable enters into the expression $\Db/\Pt^{\gamma}$,
which makes a dominant contribution to the right-hand side of $\Pt$ balance
equation (28), as we mentioned above.
}.
If we vary this variable from 30 to 5 $GeV/c$, keeping $\Pt^{clust}$
slightly restricted by $\Pt^{clust}_{CUT}=30~ GeV/c$ (practically
unbound), then, as can be seen from Fig.~22, the mean
and RMS values of the disbalance $(\Pt^{\gamma}\!-\!Pt^J)/\Pt^{\gamma}$
measure in the case of the LUCELL algorithm for $40<\Pt^{\gamma}<50~
GeV/c$ decrease as follows: mean from $3.7\%$  to $1.0\%$ and RMS from $16.8\%$
to $9.6\%$. For $300<\Pt^{\gamma}<360~GeV/c$ practically for
all events the mean and RMS values of $(\Pt^{\gamma}\!-\!\Pt^J)/\Pt^{\gamma}$ 
turn out to be less than $0.5\%$ and $4\%$, respectively 
starting from the cut $\Pt^{out}_{CUT}=20~ GeV/c$.
From these plots we also conclude that variation of $\Pt^{out}_{CUT}$ improves the
disbalance, in fact, in the same way as the variation of
$\Pt^{clust}_{CUT}$. It is not surprising as the  cluster $\Pt$
activity is a part of the $\Pt^{out}$ activity.

The influence of the $\Pt^{out}_{CUT}$ variation (with the fixed value
$\Pt^{clust}_{CUT}=10 ~GeV/c$) on the
distribution of $(\Pt^{\gamma}\!-\!\Pt^J)/\Pt^{\gamma}$ is shown in
Fig.~23 for Selection 1. In this case the mean value of
$(\Pt^{\gamma}\!-\!\Pt^J)/\Pt^{\gamma}$ drops
from $2\%$ to $0.9\%$ for LUCELL and UA2 algorithms (and to even less value for UA1)
for the $40<\Pt^{\gamma}<50~ GeV/c$ interval. At the same time
the RMS value improves from $13\%$ to $9\%$ for all algorithms.
For interval $300<\Pt^{\gamma}<360~ GeV/c$ the mean value and RMS of 
$(\Pt^{\gamma}\!-\!\Pt^J)/\Pt^{\gamma}$ are less then $0.4\%$ and
$3.3\%$ for all three jetfinders.

{\it 
 So, we conclude basing on the analysis of PYTHIA (as a model) simulation 
 that the new cuts $\Pt^{clust}_{CUT}$ and $\Pt^{out}_{CUT}$ introduced in Section 3
as well as introduction of a new object, the ``isolated jet'', are found as those that may be
very efficient tools to improve the jet calibration accuracy }
\footnote{We plan to continue this study on the level of the full event reconstruction
after CMSIM simulation.}. 
Their combined usage for this aim and for the background suppression will be a subject of a 
further more detailed study in Section 8. 
\begin{flushright}
\parbox[r]{.37\linewidth}{
\vskip29mm
{~~~
\\ \\ \\ \\ \\ \\ \\ \\ \\ \\ 
\noindent
\vskip-10mm
}}
\end{flushright}
\begin{figure}[htbp]
 \vspace{-3.7cm}
 \hspace*{15mm} \includegraphics[height=10cm,width=13cm]{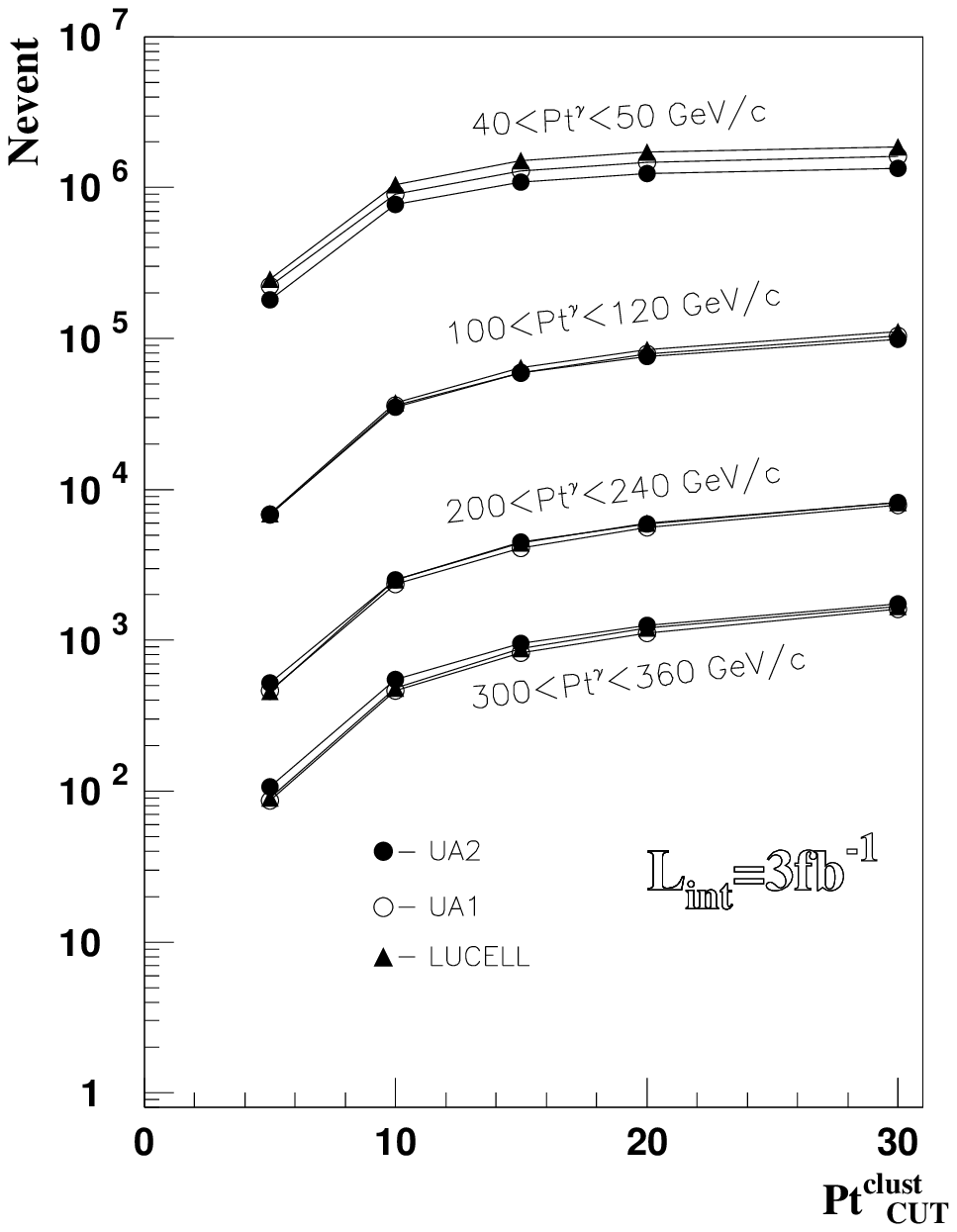}
    \label{fig01}
   \nonumber
  \end{figure}
\begin{figure}[htbp]
 \vskip-41mm
\hspace*{-3mm} \includegraphics[height=14cm,width=18cm]{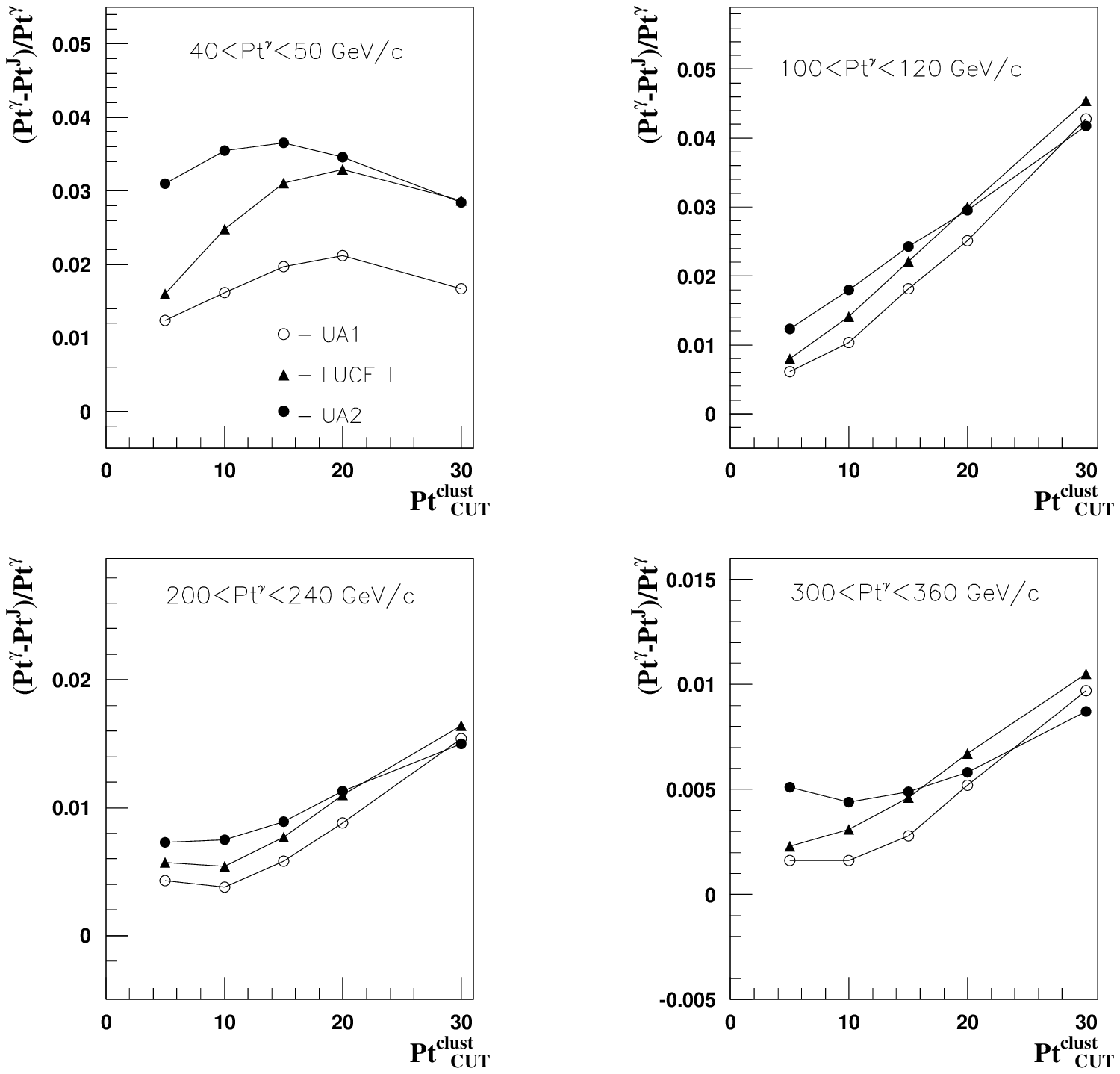}
    \label{fig02}
\nonumber
\vskip-2mm
\footnotesize{Selection 1. Dependence of the
number of events for $L_{int}=3 fb^{-1}$ (Fig.~16, top) and $(\Ptg-\Pt^{J})/\Ptg$
(Fig.~17, bottom) on $\Pt^{clust}_{CUT}$ in cases of LUCELL, UA1 and UA2 jetfinding algorithms. 
 $\Delta \phi=15^{\circ}$. $\Pt^{out}$ is not limited.}
\vskip -144mm {\hspace*{77mm}\bf {\normalsize Fig.~16}}
\vskip 144mm
\vskip -22mm {\hspace*{77mm}\bf {\normalsize Fig.~17}}
\vskip 22mm
\end{figure}

\begin{flushright}
\parbox[r]{.37\linewidth}{
\vskip29mm
{~~~
\\ \\ \\ \\ \\ \\ \\ \\ \\ \\ 
\noindent
\vskip-10mm
}}
\end{flushright}
\begin{figure}[htbp]
 \vspace{-10.5cm}
 \hspace*{15mm} \includegraphics[height=10cm,width=13cm]{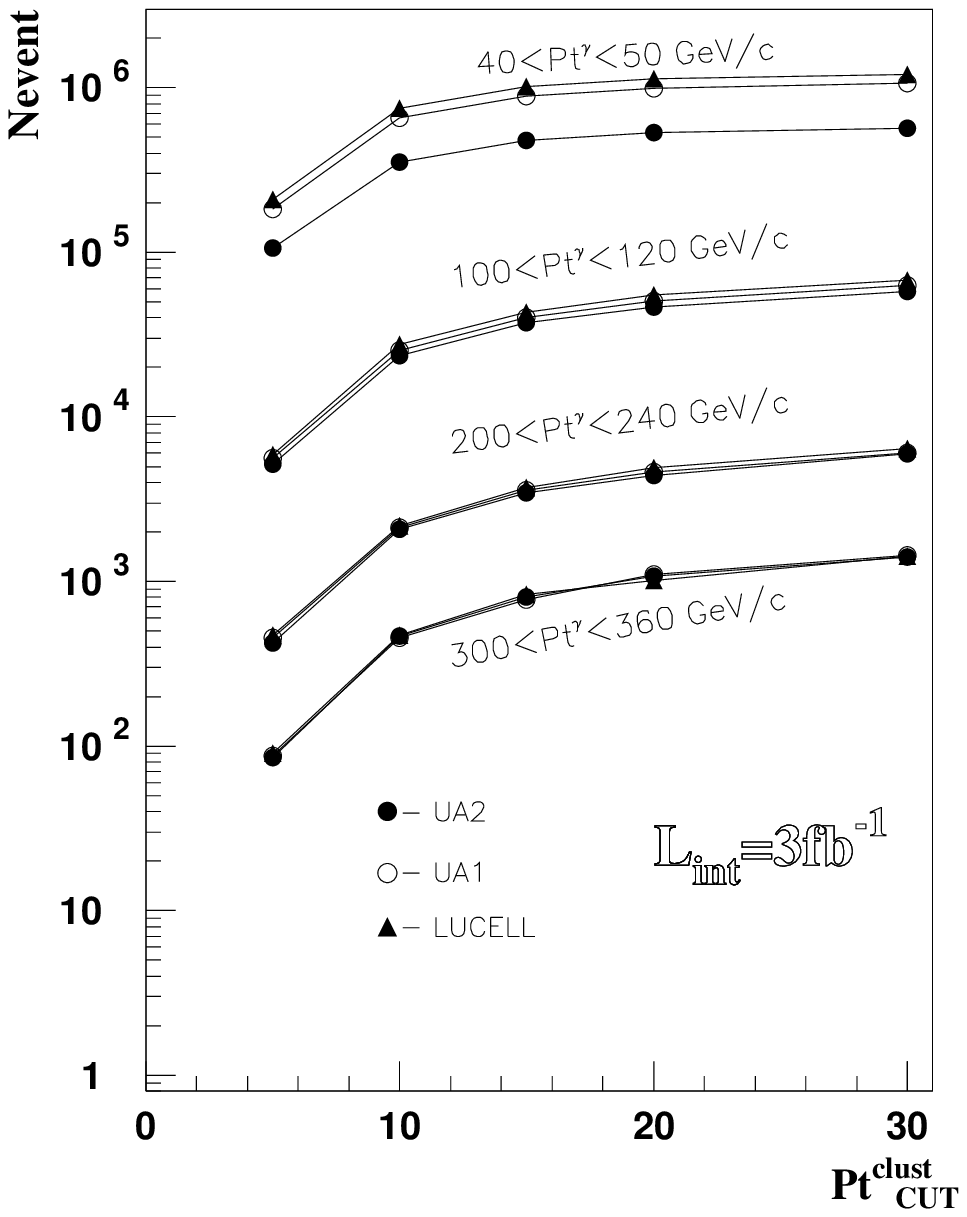}
    \label{fig01}
   \nonumber
  \end{figure}
\begin{figure}[htbp]
 \vskip-16mm
\hspace*{-3mm} \includegraphics[height=14cm,width=18cm]{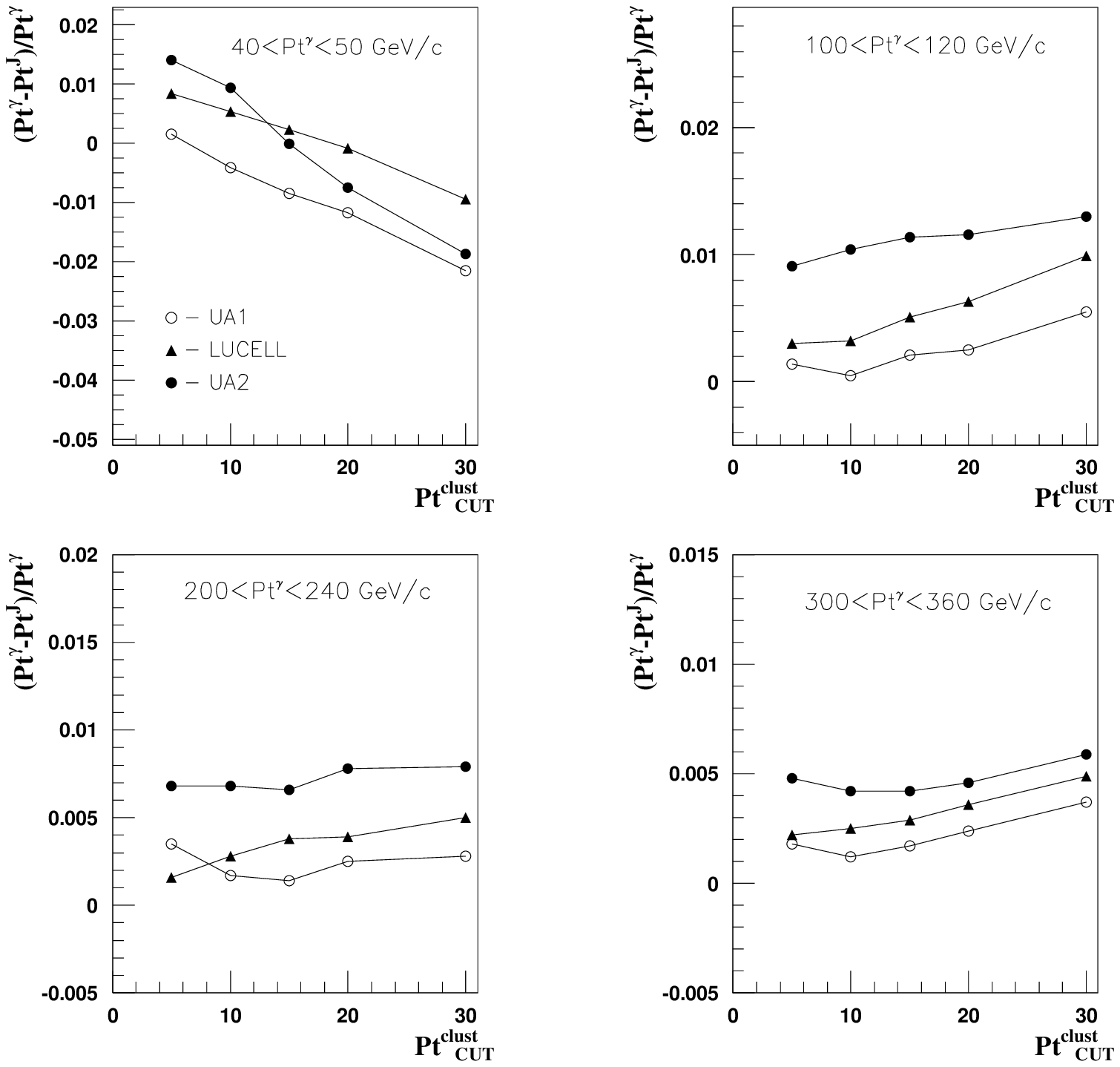}
    \label{fig02}
\nonumber
\vskip-2mm
\footnotesize{Selection 2. Dependence of the
number of events for $L_{int}=3 fb^{-1}$ (Fig.~18, top) and $(\Ptg-\Pt^{J})/\Ptg$
(Fig.~19, bottom) on $\Pt^{clust}_{CUT}$ in cases of LUCELL, UA1 and UA2 jetfinding algorithms. 
 $\Delta \phi=15^{\circ}$. $\Pt^{out}$ is not limited.\\
$\epsilon^{jet}\lt6-8\% ~(40\lt\Ptg\lt50), \lt4\% ~(100\lt\Ptg\lt120), \lt3\% ~(200\lt\Ptg\lt240),
\lt3\% ~(300\lt\Ptg\lt360)$.}
\vskip -151mm {\hspace*{77mm}\bf {\normalsize Fig.~18}}
\vskip 151mm
\vskip -25mm {\hspace*{77mm}\bf {\normalsize Fig.~19}}
\vskip 25mm
\end{figure}
\begin{figure}
\vskip-45mm
\hspace*{-30mm} \includegraphics{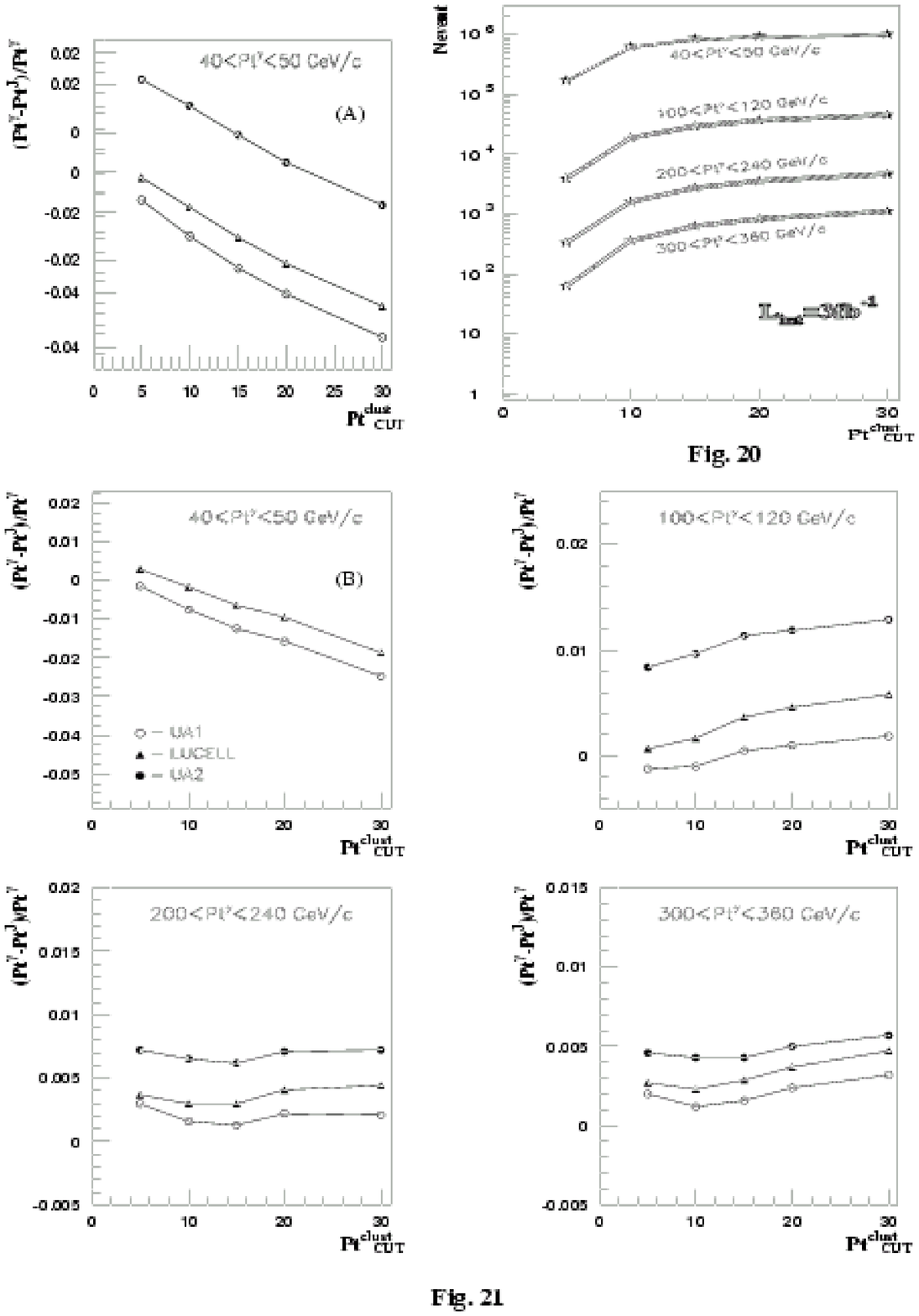} 
\vskip-1mm
\footnotesize{Selection 3. Dependence of the
number of events for $L_{int}=3 fb^{-1}$ (Fig.~20, right-hand top) and $(\Ptg-\Pt^{J})/\Ptg$
(Fig.~21, left-hand top and bottom) on $\Pt^{clust}_{CUT}$ in cases of LUCELL, UA1 and UA2 
jetfinding algorithms. $\Delta \phi=15^{\circ}$. $\Pt^{out}$ is not limited.
$\epsilon^{jet}\lt6-8\% ~(40\lt\Ptg\lt50), \lt4\% ~(100\lt\Ptg\lt120), \lt3\% ~(200\lt\Ptg\lt240),
\lt3\% ~(300\lt\Ptg\lt360)$. The \Ptgj balances for simultaneous jet finding by UA1, UA2 and LUCELL 
for $40\lt\Ptg\lt50$ are plotted in Fig.~21A and for simultaneous jet finding by only UA1 and LUCELL
for $40\lt\Ptg\lt50$ are plotted in Fig.~21B (see also text of Section 7).}
\end{figure}

\setcounter{figure}{21}
\begin{figure}
\vskip-10mm
  \hspace*{-2mm} \includegraphics[width=16cm]{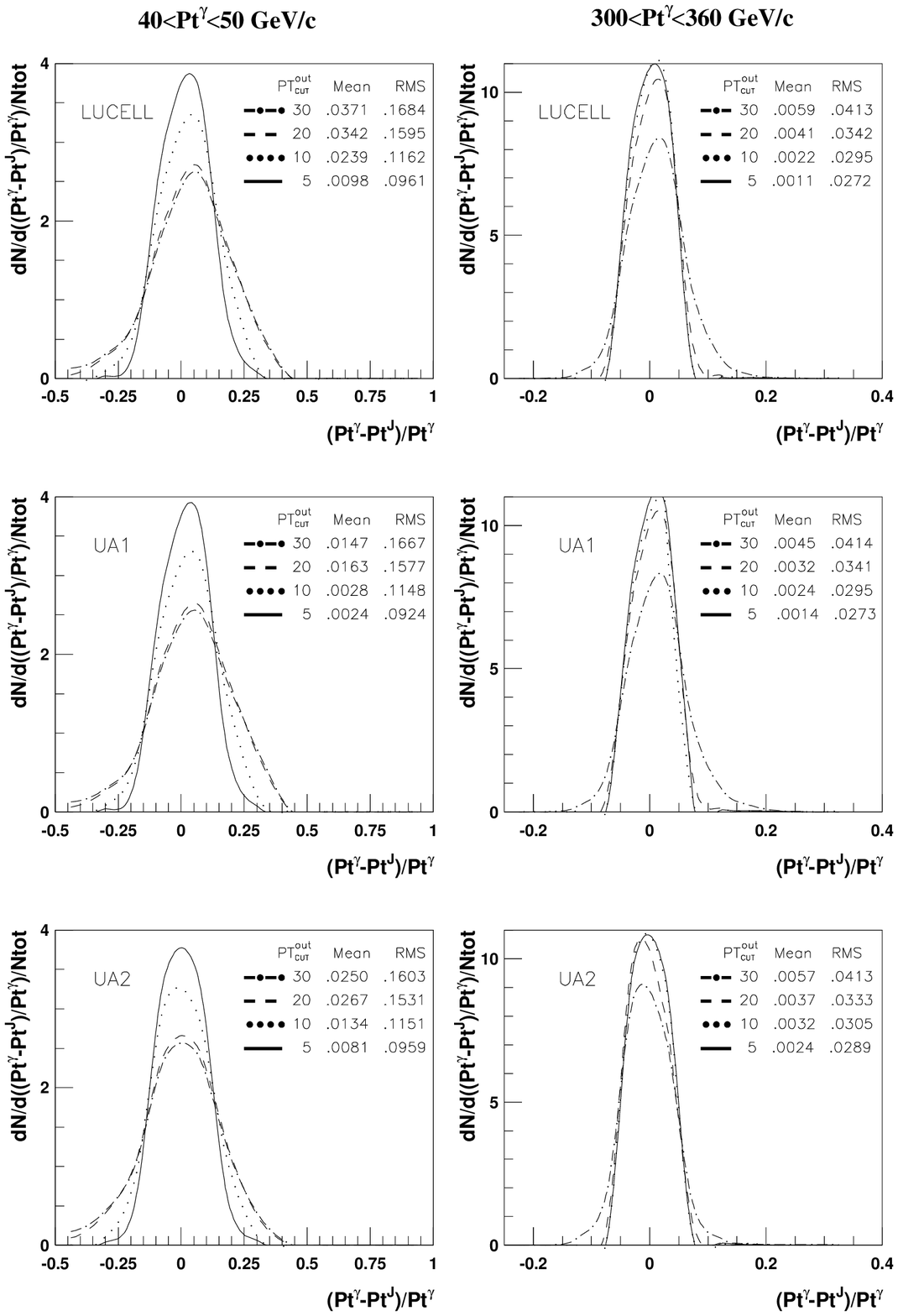} 
\caption{A dependence of $(\Pt^{\gamma}-\Pt^{J})/\Pt^{\gamma}$ on
$\Pt^{out}_{CUT}$ for LUCELL, UA1 and UA2 jetfinding algorithms and two
intervals of \ptg.  ~~The mean and RMS of the distributions are displayed on
the plots. $\dphi\leq15^\circ$, $\Pt^{clust}_{CUT}=30 ~GeV/c$. Selection 1.}
\label{fig:j-out}
\vskip25mm 
\end{figure}
\begin{figure}
\vskip-10mm
  \hspace*{-2mm} \includegraphics[width=16cm]{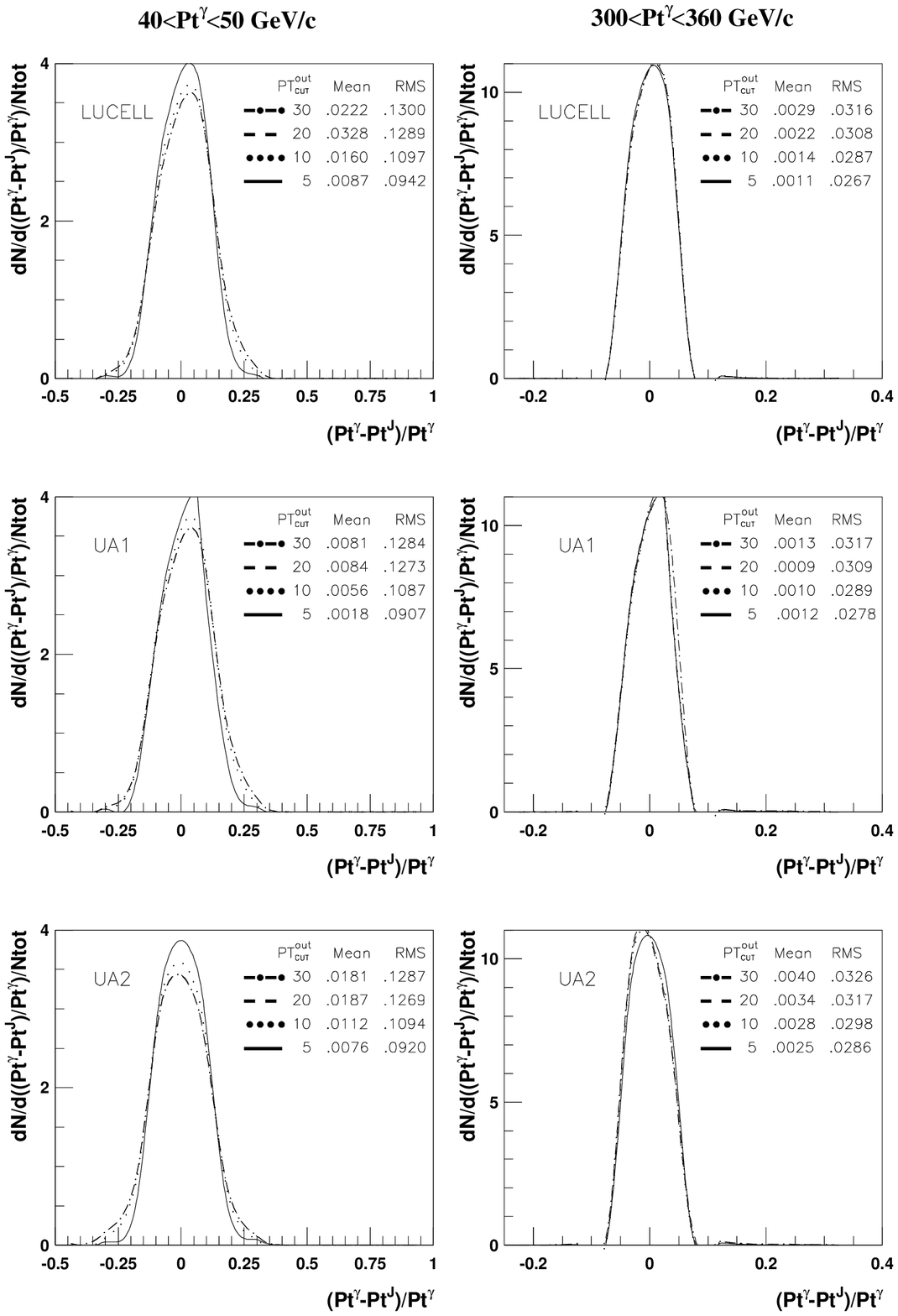} 
\caption{A dependence of $(\Pt^{\gamma}-\Pt^{J})/\Pt^{\gamma}$ on
$\Pt^{out}_{CUT}$ for LUCELL, UA1 and UA2  jetfinding algorithms and two
intervals of \ptg.  ~~The mean and RMS of the distributions are displayed on
the plots. $\dphi\leq15^\circ$, $\Pt^{clust}_{CUT}=10 ~GeV/c$. Selection 1.}
\label{fig:j-out-}
\vskip25mm 
\end{figure}


\newpage


\section{ESTIMATION OF BACKGROUND SUPPRESSION CUTS EFFICIENCY.}       

\pagestyle{plain}

\it\small
\hspace*{9mm}
The relative efficiency of ``hadronic'' cuts that are added to ``photonic'' ones, used to suppress the background
in the case of inclusive photon measurement, is estimated at the particle level. 

It is also shown that the simultaneous use of $\Pt^{out}_{CUT}$, $\Pt^{clust}_{CUT}$ together 
with imposing jet isolation criterion would lead to a substantial improvement of signal-to-background
ratio  and \ptgj balance (see Tables 14--21 and Appendix 6).

The potentially dangerous role of a new source of background 
to the signal ``$\gamma^{\,dir}+jet$'' events caused by hard
bremsstrahlung photons (``$\gamma-brem$'') is demonstrated. It is shown that at LHC energy this new
``$\gamma-brem$'' irreducible background may be compatible at low $\Ptg$ intervals with the $\pi^0$ contribution
and it may grow faster with $\Ptg$ increasing than the latter one.

\rm\normalsize
\vskip5mm

\noindent
\hspace*{8mm}
To$\!$ estimate$\!$ the$\!$ efficiency$\!$ of$\!$ the$\!$ cuts$\!$ proposed in Section 3.2 $\!$ we$\!$ 
carried$\!$ out$\!$ the$\!$ simulation
\footnote{ PYTHIA~5.7 version with default CTEQ2L parameterization
of structure functions is used here.}
with a mixture of all QCD and SM subprocesses with large cross sections existing in PYTHIA
(namely, in notations of PYTHIA, with
ISUB=1, 2, 11--20, 28--31, 53, 68). The events caused by this set of the subprocesses may give a large background 
to the ``$\gamma^{dir}+jet$'' signal events defined by  the subprocesses (1a) and (1b)
\footnote{A contribution of another possible NLO channel $gg\rrr g\gamma$
(ISUB=115 in PYTHIA) was found to be still negligible even at LHC energies.}
(ISUB=29 and 14) that were also included in this simulation.

Three generations  with the above-mentioned set of subprocesses
were done. Each of them was performed with a different value of
PYTHIA parameter $CKIN(3)\equiv\pth$  that defines the minimal value 
of $\Pt$ appearing in the final state of a hard $2\to 2$ parton level fundamental subprocess
in the case of ISR absence. These values were $\pth=40, 100$ and $200~GeV/c$. 
By 80, 50 and 80 million events were
generated for three $\pth$ values respectively.
The cross sections of the above-mentioned subprocesses define the rates of the corresponding physical
events and, thus, serve in simulation as weight factors.

We selected ``$\gamma^{dir}$-candidate +1 Jet'' events  containing 
one $\gamma^{dir}$-candidate (denoted in what follows as ${\tilde{\gamma}}$) and one jet (found by LUCELL)
with $\Pt^{Jet}> 30~ GeV/c$.
Here and below, as we work at the PYTHIA particle level of simulation,
speaking about the $\gamma^{dir}$-candidate, 
we actually mean, apart from  the $\gamma^{dir}$, a set of particles
like electrons, bremsstrahlung photons and also photons from neutral meson decays that may be registered in the
$5\times 5$ ECAL crystal cell window having the cell with the highest $\Pt$ 
($\gamma/e$) in its center.

Below we consider a set of 17 cuts that are separated into 2 subsets: a set of the
``photonic'' cuts and a set of the ``hadronic'' ones. 
The first set consists of five cuts used to select an isolated photon candidate
in some $\Pt^{\tilde{\gamma}}$ interval. The second one includes twelve cuts  applied after 
the ``photonic'' cuts. These ``hadronic'' cuts deal mostly with jets and clusters and are used 
to select events having one ``isolated jet'' and limited $\Pt$ activity out of ``${\tilde{\gamma}}+jet$'' system.

The used cuts are listed in Table 13. To give an idea
about their physical meaning and importance we have done an estimation of their possible 
influence on the signal-to-background ratios $S/B$. The letter were calculated after application of each cut.
Their values are presented in Table 14 for a case of the most illustrative intermediate interval
of event generation with $\pth=100~GeV/c$. In Table 14 the number in each line corresponds
to the number of the cut in Table 13 (the important  lines of Table \ref{tab:sb4} are darkened 
because they will be often referenced to while discussing the following Tables \ref{tab:sb1}--\ref{tab:sb3}).
\\[-6mm]
\begin{table}[h]
\caption{List of the applied cuts (will be used also in Tables \ref{tab:sb4} -- \ref{tab:sb3}).}
\begin{tabular}{lc} \hline
\label{tab:sb0}
\hspace*{-.2cm} {\bf 0}. No cuts; \\
{\bf 1}. $a)~ \Pt^{\tilde{\gamma}}\geq 40 ~GeV/c, ~~b)~|\eta^{\tilde{\gamma}}|\leq 2.61,
~~ c)~ \Pt^{jet}\geq 30 ~GeV/c,~~d) \Pt^{hadr}\!<5 ~GeV/c^{\;\ast}$;\\
{\bf 2}. $\epsilon^{\tilde{\gamma}} \leq 15\%$;
\hspace*{1.87cm} {\bf 11}. $\Pt^{clust}<20 ~GeV/c$; \\
{\bf 3}. $\Pt^{\tilde{\gamma}}\geq\pth$;
\hspace*{1.56cm} {\bf 12}. $\Pt^{clust}<15 ~GeV/c$; \\
{\bf 4}. $\epsilon^{\tilde{\gamma}} \leq 5\%$;
\hspace*{2.09cm} {\bf 13}. $\Pt^{clust}<10 ~GeV/c$; \\
{\bf 5}. $\Pt^{isol}\!\leq 2~ GeV/c$;
\hspace*{0.67cm} {\bf 14}. $\Pt^{out}<20 ~GeV/c$; \\
{\bf 6}. $Njet\leq3$;
\hspace*{1.87cm}  {\bf 15}. $\Pt^{out}<15 ~GeV/c$; \\
{\bf 7}. $Njet\leq2$;
\hspace*{1.89cm}  {\bf 16}. $\Pt^{out}<10 ~GeV/c$;\\
{\bf 8}. $Njet=1$;
\hspace*{1.90cm} {\bf 17}. $\epsilon^{jet} \leq 5\%$.\\
{\bf 9}. $\dphi<15^\circ$; \\
{\bf 10}. $\Pt^{miss}\!\leq10~GeV/c$;\\\hline
\footnotesize{${\;\ast}~\Pt$ of a hadron in the 5x5 ECAL cell window containing 
the $\gamma^{dir}$-candidate in the center.}\\
\end{tabular}
\end{table}

Line number 1 of  Table 13 contains four primary preselection criteria. It includes and specifies
our first general cut (17) of Section 3.2 as well as the cut connected with ECAL geometry and 
the cut (21) that excludes $\gamma^{dir}$-candidates accompanied by hadrons.

Line number 2 of  Table 13 fixes the values of $\epsilon^{\gamma}_{CUT}$
that, according to (19), define the isolation parameters of ${\tilde{\gamma}}$.

The third cut selects the events with $\gamma^{dir}$-candidates having $\Pt$ higher than
$CKIN(3)\equiv\pth$ threshold. We impose the third cut to select the samples of events with
$\Pt^{\tilde{\gamma}}\geq40, 100$ and $200~GeV/c$ as ISR may smear the sharp kinematical cutoff defined by
$CKIN(3)$ \cite{PYT}. This cut reflects an experimental viewpoint when one is interested in
how many events with $\gamma^{dir}$-candidates are contained in some definite interval of $\Pt^{\tilde{\gamma}}$.

The restriction $\epsilon^{\gamma}_{CUT}<5\%$, realized in the fourth line,
acts already on the events having a rather clean surrounding space near $\gamma^{dir}$-candidate
and makes the fractional isolation cut in line 2 to be tighter.

The fifth cut makes stronger the isolation criterion of $\gamma^{dir}$-candidate 
(within $R=0.7$) than it was required by the second line of Table 13. 
It should be noted that this cut includes the restriction of ``infrared'' cut (20)
of Section 3.2 which was not included to this reason into Table \ref{tab:sb0}.
\def\baselinestretch{0.97}
\begin{table}[htbp]
\small
\begin{center}
\vskip-0.2cm
\caption{Values of significance and  efficiencies for $\hat{p}_{\perp}^{\;min}$=100 $GeV/c$}
\vskip0.5mm
\begin{tabular}{||c||c|c|c|c|c|c||}                  \hline \hline
\label{tab:sb4}
Cut& $S$ & $B^\ast$ & $Eff_S(\%)$ & $Eff_{B^\ast}(\%)$  & $S/B^\ast$& $e^\pm$ \\\hline \hline
\rowcolor[gray]{\coltab}%
 0 & 19420 & 5356.E+6 &             &                 &0.00 &3.9E+6  \\\hline 
\rowcolor[gray]{\coltab}%
 1 & 19359 &  1151425 & 100.00 $\pm$ 0.00& 100.000 $\pm$ 0.000  &0.02 &47061\\\hline 
 2 & 18236 &   65839  & 94.20 $\pm$ 0.97 &   5.718 $\pm$ 0.023  &0.28 &8809 \\\hline 
 3 & 15197 &    22437 &  78.50 $\pm$ 0.85&   1.949 $\pm$ 0.013&  0.71 &2507 \\\hline 
 4 & 14140 &     9433 &  73.04 $\pm$ 0.81&   0.819 $\pm$ 0.008&  1.50 &2210 \\\hline 
 5 &  8892 &     4618 &  45.93 $\pm$ 0.59&   0.401 $\pm$ 0.006&  1.93 &1331 \\\hline 
 6 & 8572  &    3748  &  44.28 $\pm$ 0.57&   0.326$\pm$  0.005&  2.29 &1174  \\\hline 
 7 & 7663  &    2488  &  39.58 $\pm$ 0.53&   0.216$\pm$  0.004&  3.08 & 921  \\\hline 
 8 & 4844  &     813  &  25.02 $\pm$ 0.40&   0.071$\pm$  0.002&  5.96 & 505  \\\hline 
 9 & 4634  &     709  &  23.94 $\pm$ 0.39&   0.062 $\pm$ 0.002&  6.54 & 406 \\\hline 
10 &  4244 &     650  &  21.92 $\pm$ 0.37&   0.056 $\pm$ 0.002&  6.53 &  87 \\\hline 
11 &  3261 &      345 &  16.84 $\pm$ 0.32&   0.030 $\pm$ 0.002&  9.45 &53 \\\hline 
12 &  2558 &      194 &  13.21 $\pm$ 0.28&   0.017 $\pm$ 0.001& 13.19 &41 \\\hline 
13 &  1605 &       91 &   8.29 $\pm$ 0.22&   0.008 $\pm$ 0.001& 17.64 &26 \\\hline 
14 &  1568 &       86 &   8.10 $\pm$ 0.21&   0.007 $\pm$ 0.001& 18.23 &26 \\\hline 
15 &  1477 &       77 &   7.63 $\pm$ 0.21&   0.007 $\pm$ 0.001& 19.18 &25 \\\hline 
\rowcolor[gray]{\coltab}%
16 &  1179 &       52 &   6.09 $\pm$ 0.18&   0.005 $\pm$ 0.001& 22.67 &22 \\\hline 
\rowcolor[gray]{\coltab}%
17 &  1125 &       46 &   5.81 $\pm$ 0.18&   0.004 $\pm$ 0.001& 24.46 &21 \\\hline 
\hline \hline
\end{tabular}
\end{center}
\vskip-3mm
\noindent
\hspace*{5mm} \footnotesize{${\;\ast}$ The background $B^\ast$ is considered here with no
account of contribution from the ``$e^\pm$ events'' in which $e^\pm$`s appear as 
$\gamma^{dir}$-candidates, separated into the column ``$e^\pm$''.}
\vskip-3mm
\end{table}
\normalsize

The cuts considered up to now, apart from general preselection cut $\Pt^{jet}\geq30~GeV/c$
used in the first line of Table \ref{tab:sb0},
were connected with photon selection (``photonic'' cuts). Before we go further, some words of caution 
must be said here. Firstly, we want to emphasize that the starting numbers of the signal ($S$) and background ($B$)
events (first line of Table 14) may be specific only for PYTHIA generator and for the way of
preparing primary samples of the signal and background events described above. So, we want to underline here
that the starting values of $S$ and $B$ in the first columns of Table 14 are model dependent

Nevertheless, for our aim of investigation of efficiencies of new hadronic cuts 6--17 
(see \cite{9}--\cite{BKS_P5})  the important thing here is that we can use these starting model numbers of
$S$- and $B$-events  for studying the further relative  influence of these cuts on $S/B$ ratio,
choosing the conventional normalization to $100\%$ of the cut efficiencies
\footnote{In Table \ref{tab:sb4} the efficiencies $Eff_{S(B)}$ (with their errors) are defined as a ratio
of the number of signal (background) events that passed under a cut
(1--17) to the number of the preselected events (1st cut of this table).}
for $S$- and $B$-events in line 1.  

In spite of self-explaining notations of the cuts 6--9 let us mention, before passing to cuts 10--17, 
that the cuts 6--9 are connected with the selection of events having only one jet 
and the definition of jet-photon spatial orientation. Usage of these four cuts leads to the almost three-fold 
relative improvement of model $S/B$ ratio (compare lines 5 and 9 of Table 14).

\begin{figure}[htbp]
\vspace{-1.5cm}
\hspace{.0cm} \includegraphics[width=16cm,height=7.5cm]{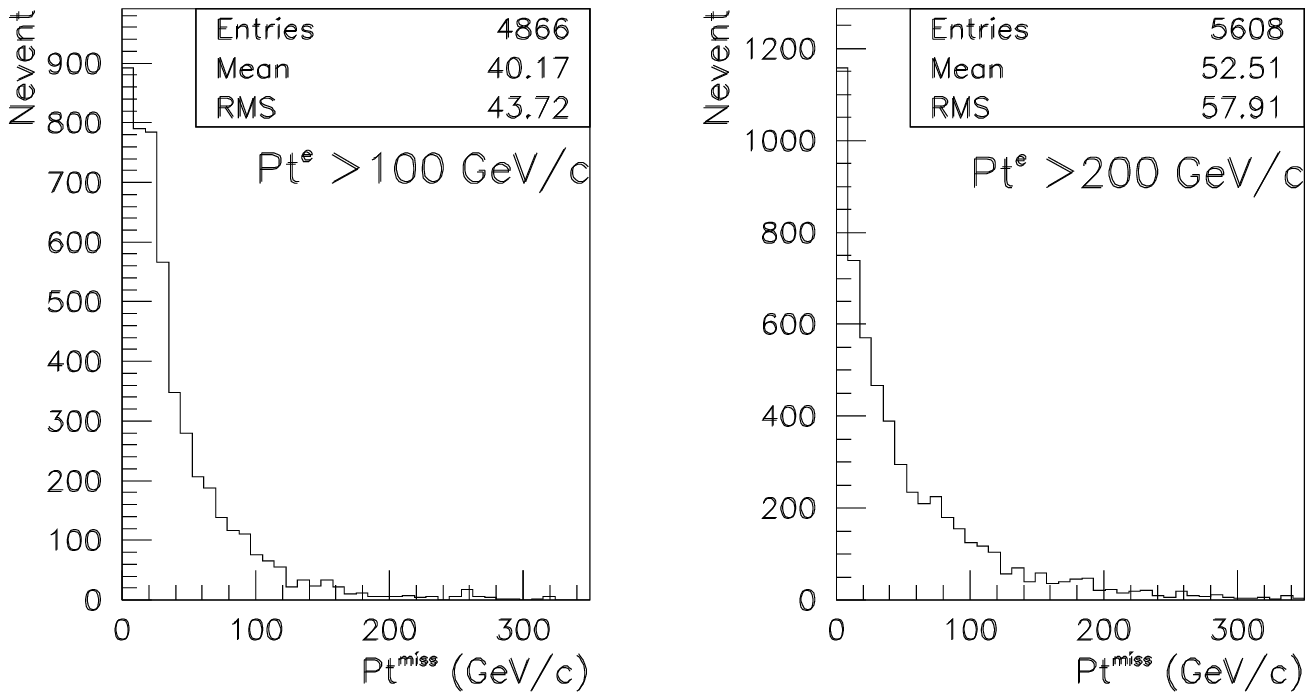}
\vspace{-1cm}
\caption{\hspace*{0.0cm} Distribution of events over $\Pt^{miss}$
in  events with energetic $e^\pm$`s appearing as direct photon candidates for 
the cases ~$\Pt^e\geq100~ GeV/c~$
and $~\Pt^e\geq200~ GeV/c$~ (here  events satisfying cuts 1--3 of
Table \ref{tab:sb0} are used).}
\label{fig:ptmiss} 
\end{figure}

In line 10 we used the $\Pt^{miss}_{CUT}$ cut, applied in Section 4,
to reduce an uncertainty of $\Pt^{Jet}$ due to a possible neutrino
contribution to a jet. Here it also reduces the contribution to background
from the decay subprocesses ~$q\,g \to q' + W^{\pm}$~ and ~$q\bar{~q'} \to g + W^{\pm}$~ 
with the subsequent decay $W^{\pm} \to e^{\pm}\nu$ that leads to a substantial $\Pt^{miss}$ value.
It is clear from the distributions over $\Pt^{miss}$ for two
$\Pt^e$ intervals presented in Fig.~\ref{fig:ptmiss}. From the last column ($e^\pm$) of 
Table \ref{tab:sb4} one can see that $\Pt^{miss}_{CUT}$ cut
(see line 10) reduces strongly (5 times) the number of events containing $e^\pm$ as
direct photon candidates. So, $\Pt^{miss}_{CUT}$ would make a noticeable improvement
of the total $S/B$ ratio. 
\\[-2mm]

Moving further we see from Table \ref{tab:sb4} that the
cuts 10--16 of Table \ref{tab:sb0} reduce the values of $\Pt^{clust}$ and $\Pt^{out}$ down to the values 
less than $10 ~GeV/c$. The 17-th cut of Table \ref{tab:sb0} imposes the jet isolation requirement. 
It leaves only the events
with jets having the sum of $\Pt$ in a ring surrounding a jet to be less than $5\%$ of $\Pt^{Jet}$.
From comparison of the numbers in 9-th and 17-th lines we make the important conclusion that all these
new cuts (10--17), despite of model dependent nature of starting $S/B$ value in line 10, may, in principle,
lead to the following about four-fold improvement of $S/B$ ratio. 
This improvement is reached by reducing the $\Pt$ activity out of ``$\tilde{\gamma}+1~jet$'' system. 

It is also rather interesting to mention that the total effect of ``hadronic cuts'' 6--17 at $\Ptg>100 ~GeV/c$
consist of decrease of background contribution by 2 orders (!)
at the cost of eight-fold loss of signal events. So, in this sense, we may conclude that 
from the viewpoint of $S/B$ ratio the study of \gpj events may be more preferable 
as compared with a case of inclusive photon production.

Below we shall demonstrate in some plots how new selection criteria 10--17 work to choose the events
with further almost four-fold improvement of $S/B$ ratio. 
 For this reason we have built the distributions that correspond to the three above-mentioned 
values of $\pth$ and for the ``$\tilde{\gamma}+1~jet$''
events that have passed the set of cuts 1--8 defined in Table 13.
Thus, no special cuts were imposed on $\Delta\phi$, $\Pt^{out}$ and $\Pt^{clust}$
(the values of $\Pt^{clust}$ are automatically bounded from above since
we select ``$\tilde{\gamma}+1~jet$'' events with $\Pt^{jet}>30~ GeV/c$).

\begin{figure}[htbp]
 \vspace*{-0.0cm}
 \hspace*{-0.7cm}
  \includegraphics[width=16.cm,height=19cm]{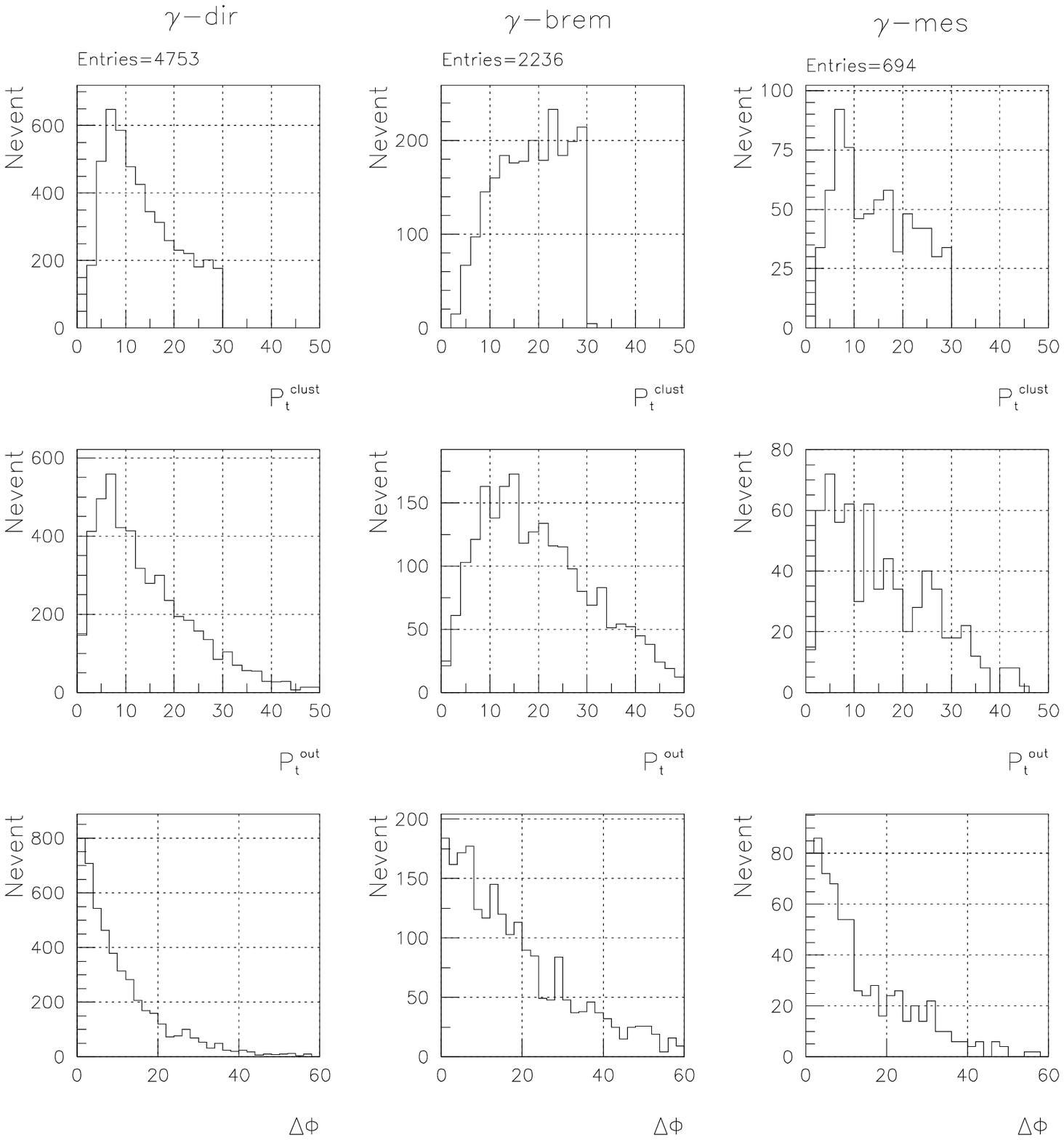}
  \vspace{-0.8cm}
    \caption{\hspace*{0.0cm}  Signal/Background: Number of events distribution
over $\Pt^{clust}$, $\Pt^{out}$, $\dphi$ ($\Pt^{\tilde{\gamma}}\geq40~GeV/c$).}
    \label{fig:1b40}
  \end{figure}

\begin{figure}[htbp]
 \vspace{-0.0cm}
 \hspace*{-0.7cm}
  \includegraphics[width=16.cm,height=19cm]{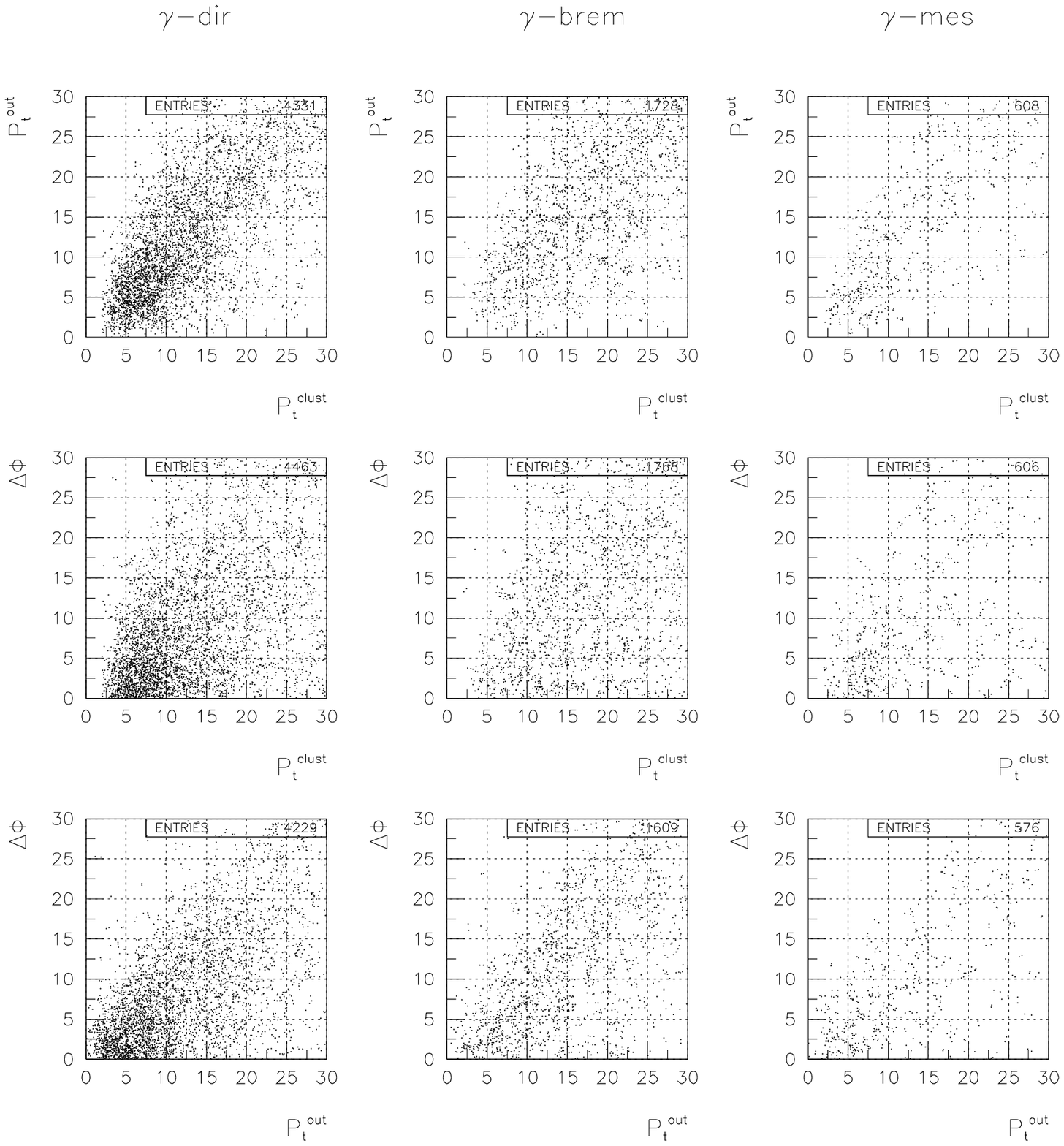}
  \vspace{-0.8cm}
    \caption{\hspace*{0.0cm} Signal/Background: $\Pt^{clust}$ vs. $\Pt^{out}$,
$\Pt^{clust}$ vs.$\dphi$, $\Pt^{out}$ vs. $\dphi$ ($\Pt^{\tilde{\gamma}}\geq 40~GeV/c$).}
    \label{fig:2b40}
  \end{figure}

\begin{figure}[htbp]
 \vspace{0.0cm}
 \hspace*{-0.7cm}
 \includegraphics[width=16cm,height=19cm]{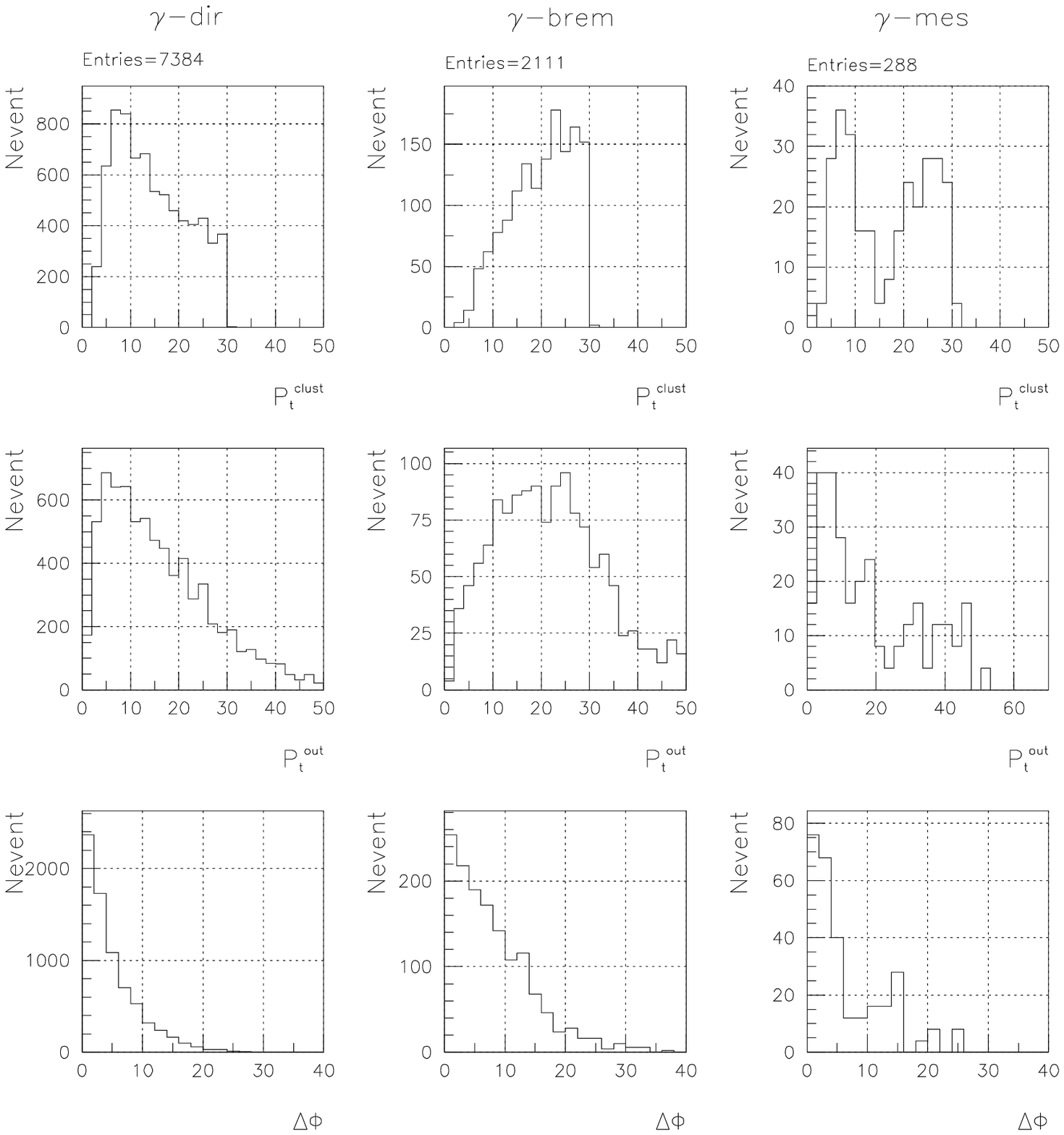}
  \vspace{-0.8cm}
    \caption{\hspace*{0.0cm}  Signal/Background: Number of events distribution
over $\Pt^{clust}$, $\Pt^{out}$, $\dphi$ ($\Pt^{\tilde{\gamma}}\geq 100~GeV/c$).}
    \label{fig:1b100}
  \end{figure}

\begin{figure}[htbp]
 \vspace{0.0cm}
 \hspace*{-0.7cm}
  \includegraphics[width=16cm,height=19cm]{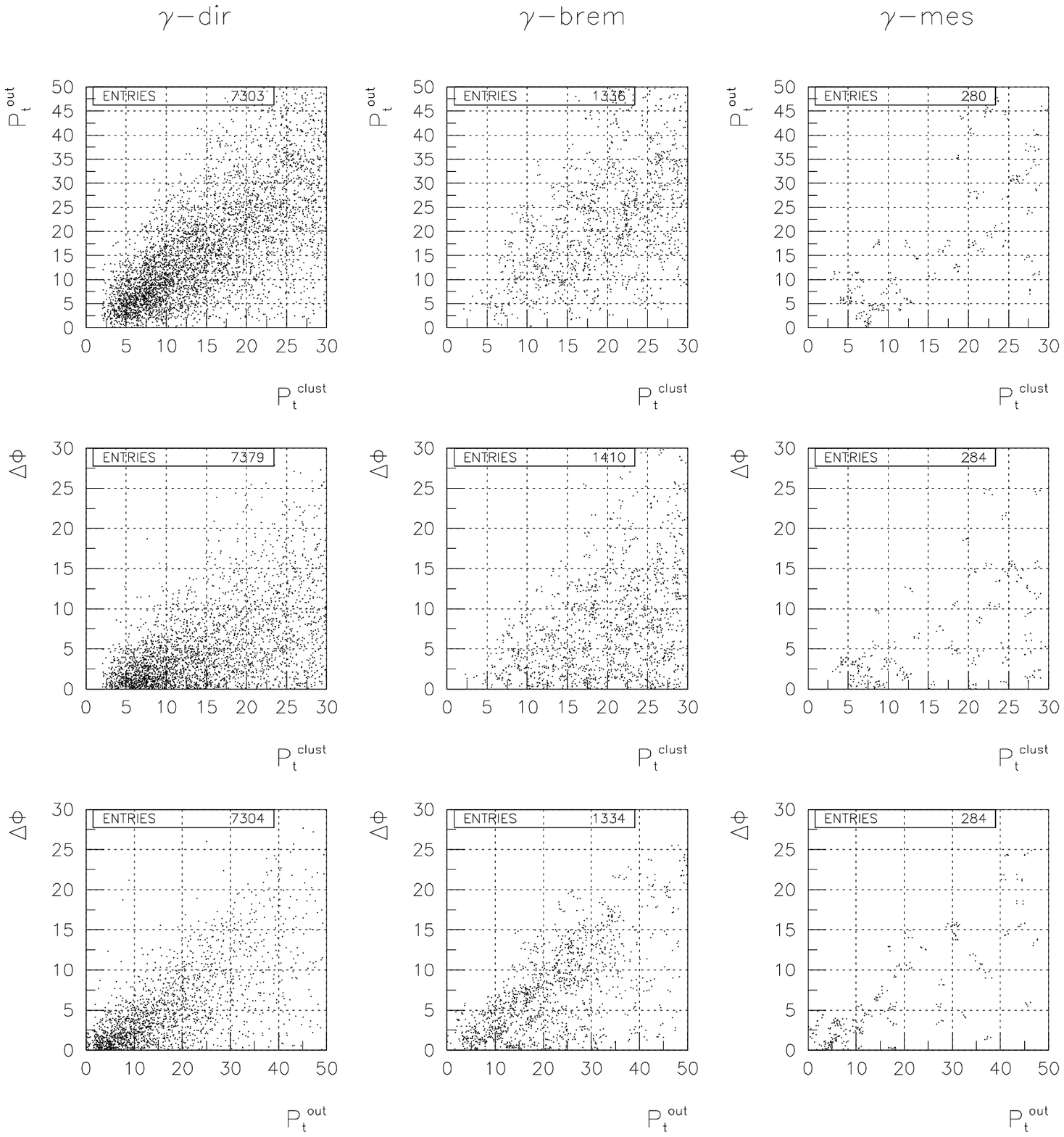}
  \vspace{-0.8cm}
    \caption{\hspace*{0.0cm}  Signal/Background: $\Pt^{clust}$ vs. $\Pt^{out}$,
$\Pt^{clust}$ vs.$\dphi$, $\Pt^{out}$ vs. $\dphi$ ($\Pt^{\tilde{\gamma}}\geq 100~GeV/c$).}
    \label{fig:2b100}
  \end{figure}

\begin{figure}[htbp]
 \vspace{0.0cm}
 \hspace*{-0.7cm}
 \includegraphics[width=16cm,height=19cm]{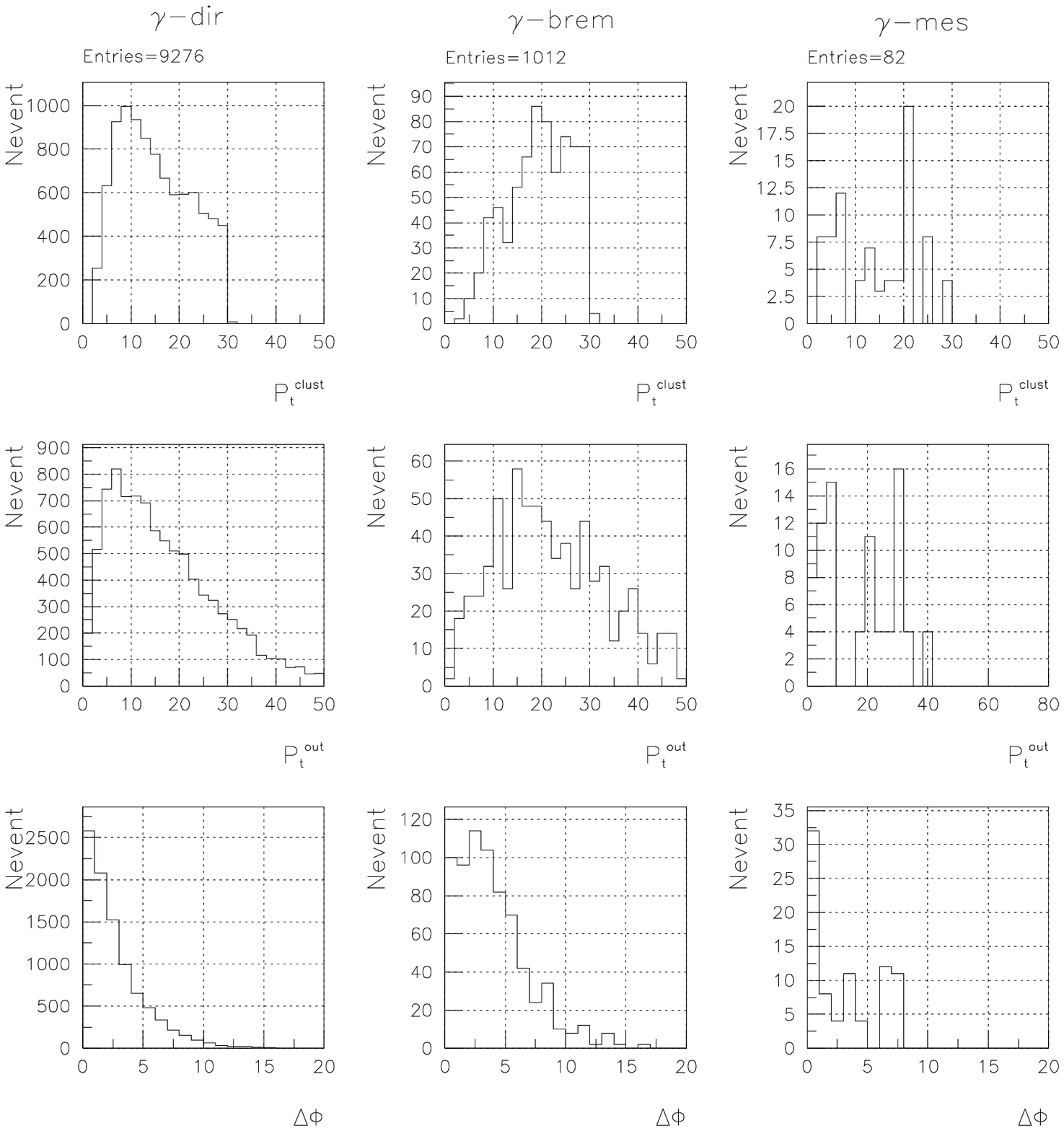}
  \vspace{-0.8cm}
    \caption{\hspace*{0.0cm} Signal/Background: Number of events distribution
over $\Pt^{clust}$, $\Pt^{out}$, $\dphi$ ($\Pt^{\tilde{\gamma}}\geq 200~GeV/c$).}
    \label{fig:1b200}
  \end{figure}

\begin{figure}[htbp]
 \vspace{0.0cm}
 \hspace*{-0.7cm}
  \includegraphics[width=16cm,height=19cm]{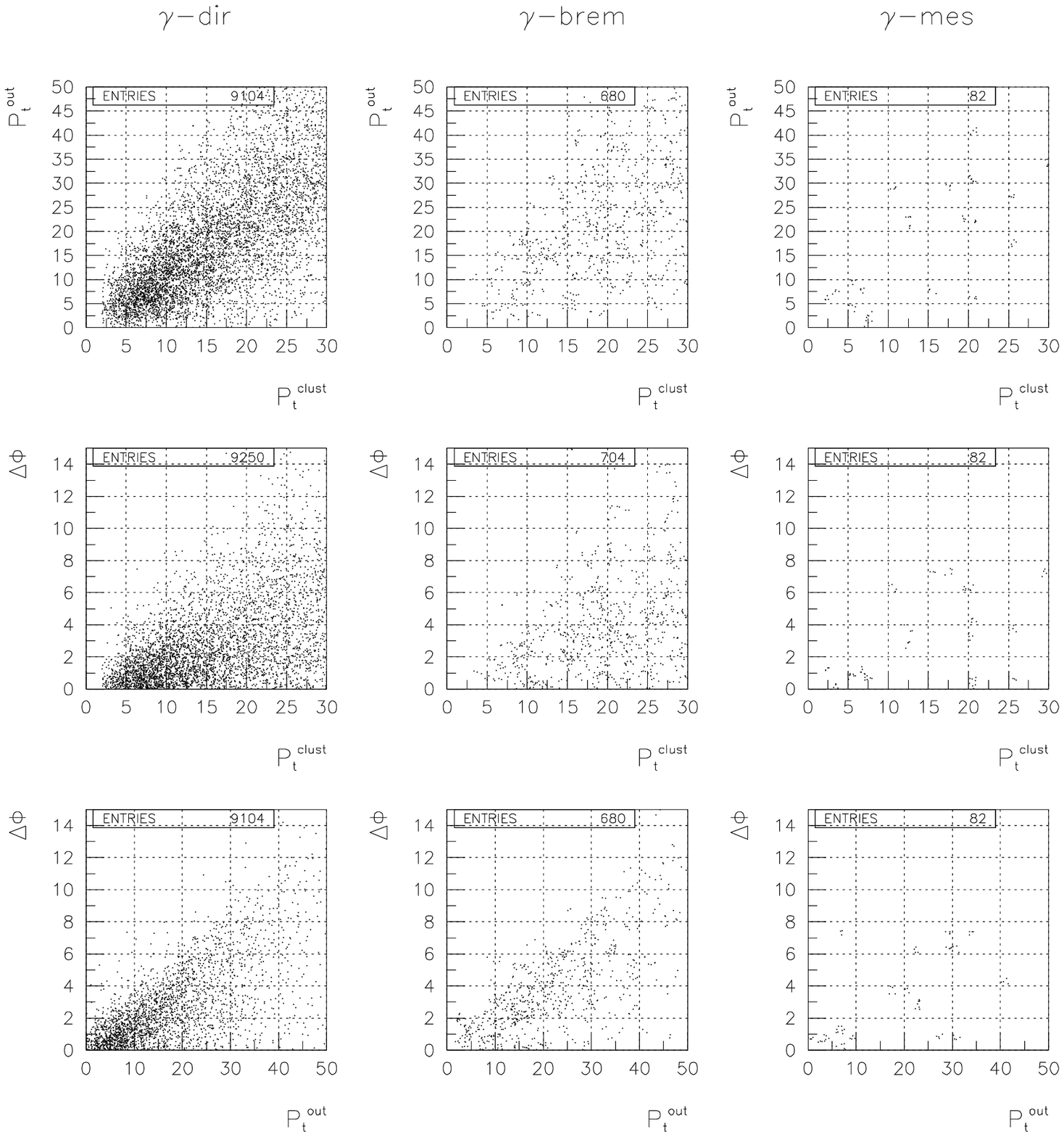}
  \vspace{-0.8cm}
  \caption{\hspace*{0.0cm} Signal/Background: $\Pt^{clust}$ vs. $\Pt^{out}$,
$\Pt^{clust}$ vs.$\dphi$, $\Pt^{out}$ vs. $\dphi$ ($\Pt^{\tilde{\gamma}}\geq 200~GeV/c$).}
  \label{fig:2b200}
\end{figure}

These distributions are given here to show the dependence of the number of events on
the physical observables $\dphi, \Pt^{out}$ and $\Pt^{clust}$
introduced in Sections 3.1 and 3.2. We present them
separately for the signal ``$\gamma$-dir'' and background events contained in each of three generated
samples. The distributions are given for three different $\Pt^{\tilde{\gamma}}$ intervals
in Figs.~\ref{fig:1b40}, \ref{fig:1b100}, \ref{fig:1b200} and are accompanied by 
scatter plots \ref{fig:2b40}, \ref{fig:2b100}, \ref{fig:2b200}. So, each pair of a figure
and a scatter plot does correspond to one $\Pt^{\tilde{\gamma}}$ interval. 
Thus, Fig.~\ref{fig:1b40} and scatter plot \ref{fig:2b40} correspond to $\Pt^{\tilde{\gamma}}\geq40~GeV/c$ and so on.

The first columns in these figures, denoted by ``$\gamma$ - dir'', show the distributions 
in the signal events, i.e. in the events corresponding to processes (1a) and (1b).
The second columns, denoted as
``$\gamma$ - brem'', correspond to the events in which the photons
were emitted from quarks (i.e. bremsstrahlung photons).
 The distributions in the third columns were built on the basis of the events containing 
``$\gamma$-mes'' photons, i.e. those photons which originate from multiphoton decays of mesons
($\pi^0$, $\eta$, $\omega$ and $K^0_S$). 




First, we see that in the case of $\Pt^{\tilde{\gamma}} \geq 100 ~GeV/c$
(see Figs.~\ref{fig:1b100}, \ref{fig:1b200}) practically all ``signal events'' are within
$\Delta \phi<15^{\circ}$. 
It is seen from Fig.~\ref{fig:1b40} that for $\Pt^{\tilde{\gamma}} \geq 40 ~GeV/c$ there is still
a large number  of signal events (about $70\%$) belonging to the
$\Delta \phi<15^{\circ}$ interval. From here and from the comparison of plots in the
``$\gamma$-dir'' and ``$\gamma$-brem'' columns  (showing the $\Delta\phi$ dependence) in the same figures
\ref{fig:1b40}--\ref{fig:2b200} we conclude that the upper cut $\Delta\phi<15^{\circ}$, used in previous sections,
is reasonable, and moreover, it does discard a lot of ``$\gamma$-brem'' background events 
in the intervals with $\Pt^{\tilde{\gamma}}\lt100 ~GeV/c$.

From the second ``$\gamma$-brem'' columns of Figs.~\ref{fig:1b40},
~\ref{fig:1b100} and ~\ref{fig:1b200} one can also see that $\Pt^{clust}$ spectra 
of the events with bremsstrahlung photons look  different from the analogous $\Pt^{clust}$ 
distributions of the signal ``$\gamma$-dir'' photons. The latter distributions have the most of the events
in the region of small $\Pt^{clust}$ values 

Since the bremsstrahlung (``$\gamma$-brem'') photons give the most sizeable background 
\footnote{The numbers in Table 15 below supports this remark. But it is also necessary to keep in mind the results 
obtained in \cite{Kirk} that the PYTHIA/JETSET fragmentation may underestimate the $\pi^0, \eta$ 
contribution to the isolated photon background.}, 
(compare the numbers of entries in the second ``$\gamma$-brem'' and the third ``$\gamma$-mes'' columns
of Figs.~24--29) the observed difference of the spectra prompts an idea of using
an upper cut for the value of $\Pt^{clust}$ to reduce
the ``$\gamma$-brem'' background which dominates at large $\Pt^{clust}$ values
(that was not a primary guideline for introduction of $\Pt^{clust}$ in Sections 2 and 3 
as a physical variable and a cut on it).

The analogous difference of $\Pt^{out}$ spectra of signal
``$\gamma$-dir'' events (which are concentrated at low  $\Pt^{out}$ values)
from those of background ''$\gamma$-brem'' events 
having longer tails at high $\Pt^{out}$ enables us to impose an upper cut on the $\Pt^{out}$ value.

Now from the scatter plots in Figs.~\ref{fig:2b40},
\ref{fig:2b100} and \ref{fig:2b200} as well as from Figs.~\ref{fig:1b40},
\ref{fig:1b100} and \ref{fig:1b200} we can conclude that the use of cuts
\footnote{rather soft here, but the results of their further restriction were already shown
in tables of Appendices 2--5 and Figs.~12--20 and will be discussed below}:
\noindent
$\Delta\phi\lt15^{\circ}$, $\Pt^{clust}_{CUT}=10\;GeV/c$, $\Pt^{out}_{CUT}=10\;GeV/c$
\noindent
would allow to keep a big number of the signal ``$\gamma$-dir'' events and
to reduce noticeably the contribution from the background ``$\gamma$-brem'' and
``$\gamma$-mes'' events in all intervals of $\Pt^{\tilde{\gamma}}$. At the same time
the Figs.~\ref{fig:1b40}--\ref{fig:2b200} give the information about what parts of different 
spectra are lost with the imposed cuts.

So, Figs.~\ref{fig:1b40}--\ref{fig:2b200} illustrate well that the new physical variables $\Pt^{clust}$ 
and $\Pt^{out}$ \cite{9}--\cite{BKS_P5}, described in Sections 3.1 and 3.2
may be useful for separation of the ``$\gamma^{dir}+jet$'' events from the background ones 
(the latter, in principle, are not supposed to have the well-balanced $\Pt^{\tilde{\gamma}}$ and $\Pt^{Jet}$).


Table \ref{tab:sb1} includes the numbers of signal and background events left in three generated event samples
after application of cuts 1--16 and 1--17. They are given for all three intervals of $\Pt^{\tilde{\gamma}}$.
Tables \ref{tab:sb1} and \ref{tab:sb4} are complementary to each other.
The summary of Table \ref{tab:sb4} is presented in the middle section ($\pth=100 ~GeV/c$)
of Table \ref{tab:sb1} where the line ``Preselected'' corresponds to the cut 1 of Table \ref{tab:sb0} 
and, respectively, to the line number 1 of  Table  \ref{tab:sb4} presented above.
The line ``After cuts'' corresponds to the line 16 of  Table \ref{tab:sb4} and 
line ``+jet isolation'' corresponds to the line 17 of  Table \ref{tab:sb4}. 

Table \ref{tab:sb1} is done to show in more detail the origin of $\gamma^{dir}$-candidates.
The numbers in the  ``$\gamma-direct$'' column correspond to the respective  numbers  of 
signal events left in each of $\Pt^{\tilde{\gamma}}$ intervals after application of the cuts defined
in lines 1, 16 and 17 of Table \ref{tab:sb0} (and in column ``$S$'' of Table \ref{tab:sb4}). Analogously
the numbers in the ``$\gamma-brem$'' column of Table  \ref{tab:sb1} correspond to the numbers
of events with the photons radiated from quarks
participating in the hard interactions. Their $\Pt^{clust}$ and $\Pt^{out}$ distributions
were presented in the central columns of Figs.~\ref{fig:1b40} -- \ref{fig:2b200}.
Columns 5 -- 8 of Table \ref{tab:sb1} illustrate the numbers of the ``$\gamma-mes$''  events with photons
originating from $\pi^0,~\eta,~\omega$ and $K^0_S$ meson decays.
Their distributions were shown in the right-hand columns of 
Figs.~\ref{fig:1b40} -- \ref{fig:2b200}. In a case of $\Pt^{\tilde{\gamma}}\gt100~GeV/c$ 
the total numbers of background events,
i.e. a sum over the numbers presented in columns 4 -- 8 of Table \ref{tab:sb1}, 
are shown in the lines 1, 16 and 17 of column ``$B^\ast$'' of Table \ref{tab:sb4}.
The other lines of Table \ref{tab:sb1} for $\pth\!=40$ and
$~200 ~GeV/c~$ have the meaning analogous to that described above for $\pth=100 ~GeV/c$.

~\\[-11mm]
\begin{table}[h]
\small
\begin{center}
\caption{Number of signal and background events remained after cuts ~~(\bf I)}
\vskip.5mm
\begin{tabular}{||c|c||c|c|c|c|c|c|c||}                  \hline \hline
\label{tab:sb1}
\hmm$\pth$\hmm& &$\gamma$ & $\gamma$ &\multicolumn{4}{c|}{  photons from the mesons}  &
\\\cline{5-8}
\Gvc& Cuts&\hmm direct\hmm &\hmm brem\hmm & $\;\;$ $\pi^0$ $\;\;$ &$\quad$ $\eta$ $\quad$ &
$\omega$ &  $K_S^0$ &\hmm $e^{\pm}$\hmm \\\hline \hline
    &Preselected&\hmm12394&\hmm 20952& 166821& 66533& 17464& 23942&\hmm 6684\hmm  \\\cline{2-9}
 40 &After cuts &\hmm 1718&\hmm 220&     146&     56&     2& 15&\hmm   10\hmm\\\cline{2-9}
    &+ jet isol.  &\hmm 1003&\hmm 102&    59&     26&     2&  7&   8\\\hline  \hline
    &Preselected&\hmm19359  &\hmm90022 &658981 &247644 &69210  & 85568 &\hmm47061\hmm\\\cline{2-9}
100 &After cuts&\hmm 1179 &\hmm34 &13 &4 &1  & 0 &\hmm 22\hmm \\\cline{2-9}   
    &+ jet isol. &\hmm 1125 &\hmm32 &9 &4 &1  & 0 &\hmm 21\hmm \\\hline \hline
    &Preselected&\hmm55839 &\hmm354602 &1334124 &393880 &141053 & 167605 &\hmm153410\hmm\\\cline{2-9}
200 &After cuts&\hmm 1837 &\hmm 30& 4 &6 &0  & 0 &\hmm 17\hmm\\\cline{2-9}
    &+ jet isol. &\hmm1828 &\hmm 30& 4 &6 &0  & 0 &\hmm 17\hmm\\\hline \hline
\end{tabular}
\vskip0.2cm
\caption{Efficiencies and significance values in events without jet isolation cut ~~(\bf I)}
\vskip0.1cm
\begin{tabular}{||c||c|c|c|c|>{\columncolor[gray]{\coltab}}c|c||} \hline \hline
\label{tab:sb2}
$\pth$ \Gvc& $S$ & $B^\ast$ & $Eff_S(\%)$  & $Eff_B^\ast(\%)$  & $S/B^\ast$& $S/\sqrt{B^\ast}$
\\\hline \hline
40  & 1718& 439 & 13.86 $\pm$ 0.36 & 0.149 $\pm$ 0.007&  3.9 & 82.0
 \\\hline
100 & 1179& 52 & 6.09 $\pm$ 0.18 & 0.005 $\pm$ 0.001& 22.7 & 163.5
 \\\hline  
200 & 1837& 40 & 3.29 $\pm$ 0.08 & 0.002 $\pm$ 0.000& 45.9 & 290.5 
\\\hline \hline
\end{tabular}
\vskip0.2cm
\caption{Efficiencies and significance values in events with jet isolation cut ~~(\bf I)}
\vskip0.1cm
\begin{tabular}{||c||c|c|c|c|>{\columncolor[gray]{\coltab}}c|c||}  \hline \hline
\label{tab:sb3}
$\pth$ \Gvc& ~~$S$~~ & ~~$B^\ast$~~ & $Eff_S(\%)$ & $Eff_B^\ast(\%)$  & $S/B^\ast$& $S/\sqrt{B^\ast}$
 \\\hline \hline
40  & 1003& 196 & 8.09 $\pm$ 0.27 & 0.066 $\pm$ 0.005&  5.1 & 71.6 \\\hline 
100 & 1125&46 & 5.81 $\pm$ 0.18 & 0.004 $\pm$ 0.001& 24.5 & 165.9 \\\hline  
200 & 1828& 40 & 3.28 $\pm$ 0.08 & 0.002 $\pm$ 0.000& 45.7 & 289.0 
\\\hline \hline
\end{tabular}
\vskip-5mm
\end{center}
\end{table}

The last column of Table \ref{tab:sb1} shows the number of preselected events with
$e^\pm$ (see our notes above while discussing the tenth cut of Table \ref{tab:sb0}).

The numbers in Tables \ref{tab:sb2} (without jet isolation cut) and \ref{tab:sb3}
(with jet isolation cut) accumulate in a compact form the final information of 
Tables \ref{tab:sb0} -- \ref{tab:sb1}. 
Thus, for example, the columns $S$ and $B$ of the  line that corresponds to $\pth=100 ~GeV/c$ 
contain the total numbers of the selected signal and background events taken at the level of 16-th  (for Table
\ref{tab:sb2}) and 17-th (for Table \ref{tab:sb3}) cuts from Table \ref{tab:sb4}. 

It is seen from Table \ref{tab:sb2}  that in the case of Selection 1 the ratio $S/B$ grows   
from 3.9 to 48.4 while $\Pt^{\tilde{\gamma}}$ increases from
$\Pt^{\tilde{\gamma}}\geq 40 ~GeV/c$ to $\Pt^{\tilde{\gamma}}\geq 200 ~GeV/c$ interval.

The jet isolation requirement (cut 17 from Table \ref{tab:sb0})
noticeably improves the situation at low $\Pt^{\tilde{\gamma}}$ (see Table \ref{tab:sb3}).
After application of this criterion the value of $S/B$ increases from 3.9 
to 5.1 at $\Pt^{\tilde{\gamma}}\geq 40 ~GeV/c$ 
and from 22.7 to 24.5 at $\Pt^{\tilde{\gamma}}\geq 100 ~GeV/c$.
Remember on this occasion the conclusion  that the sample of events
selected with our criteria has a tendency to contain more events with an isolated jet
as $\Pt^{\tilde{\gamma}}$ increases (see Sections 5--7 and Appendices 2--5).
Thus, from Appendices 4 and 5 it can be seen that the main part of jets with $\Pt^{jet}\geq 100 ~GeV/c$
appears to be  isolated (compare also the last two lines in each $\pth$ section of Table \ref{tab:sb1}).
%
%

Let us underline here that, in contrast to other types of background, ``$\gamma-brem$'' background
has an irreducible nature. So, the number of ``$\gamma-brem$'' events
should be carefully estimated for each $\Pt^{\tilde{\gamma}}$ interval using the particle level
of simulation in the framework of event generator like PYTHIA.
They are also have to be taken into account in experimental analysis
of the prompt photon production data at high energies.

Table \ref{tab:bg_or_gr} 
shows  the relative contributions of fundamental QCD subprocesses  (having the largest cross sections)
with ISUB=11, 12, 28, 53 and 68 (see \cite{PYT})
which define the main production of ``$\gamma\!-\!brem$'' background in event samples
selected with  criteria 1--13 of Table 13 in three $\Pt^{\tilde{\gamma}}$ intervals.

We found from the PYTHIA event listing analysis that
in the main part of selected ``$\gamma\!-\!brem$'' events
these photons are produced in the final state of the fundamental $2\to2$ subprocess
\footnote{i.e. from lines 7, 8 in Fig.~3}.
Namely, they are mostly radiated from the outgoing quarks 
in the case of the first three sets of subprocesses (ISUB=28, 11, 12 and 53).
They may also appear as a result of string breaking in a final state of 
$gg\to gg$ scattering (ISUB=68). But this subprocess,
naturally, gives a small contribution into ``$\tilde{\gamma}+jet$'' events production.

~\\[-12mm]
\begin{table}[h]
\begin{center}
\vskip-3mm
\caption{Relative contribution (in per cents) of different QCD subprocesses into
the ``$\gamma\!-\!brem$'' events production.}
\normalsize
\vskip.1cm
\begin{tabular}{|c||c|c|c|c|}                  \hline \hline
\label{tab:bg_or_gr}
$\Ptg$& \multicolumn{4}{c|}{fundamental QCD subprocess} \\\cline{2-5}
 \Gvc & { ISUB=28} & ISUB=11,12 & ISUB=53 & ISUB=68  
\\\hline \hline
 40--71   &  70.6$\pm$ 8.7 & 21.7$\pm$ 3.8 &  5.1$\pm$ 1.6 &  2.6$\pm$ 1.0 \\\hline 
 71--141  &  67.5$\pm$ 7.3 & 25.7$\pm$ 3.5 &  4.2$\pm$ 1.2 &  2.6$\pm$ 0.9  \\\hline 
141--283  &  58.7$\pm$ 9.0 & 39.5$\pm$ 5.7 &  1.8$\pm$ 1.0 &   ---   \\\hline\hline  
\end{tabular}
\end{center}
\vskip-7mm
\end{table}

It may be noted also from the first two columns of Table \ref{tab:bg_or_gr}
that the most of ``$\gamma\!-\!brem$'' background events  originate from  the
ISUB=28 ($fg\to fg$) and ISUB=11, 12 
($f_if_j\to f_if_j$, $f_i\bar{f_i}\to f_j\bar{f_j}$) subprocesses ($90\%$ at least).
Table  \ref{tab:bg_or_gr} shows also a tendency of 
increasing the contribution from the sum of two subprocess ``11+12'' 
(given in the second column of Table \ref{tab:bg_or_gr}) with growing $\Pt^{\tilde{\gamma}}$.

Now let us discuss how the values in Tables \ref{tab:sb4}--\ref{tab:sb3}
may change if one takes into account the real behaviour of processes in the detectors.

As for the photons from~ $\pi^0$~ decays,~ the 
the rejection efficiencies
were estimated for the Endcap \cite{CMS_EC}, \cite{EC_END}  and Barrel \cite{CMS_EC}, \cite{EC_BAR}
ECAL regions. They are of the order of 0.20 -- 0.70  for the Barrel
and 0.51 -- 0.75 for the Endcap, depending on $\Pt^\gamma$ and slightly
on $\eta^{\gamma}$, for the single photon selection efficiency
90$\%$. As for the $e^\pm$ background, we take the
electron track finding  efficiency to be $85\%$ (following \cite{CMS_Tra}
and after averaging over its $\eta$ dependence)
for $\Pt^e\geq 40 ~GeV/c$.

To study the $\eta$, $\omega$, $K_S^0$ mesons
contribution we carried out the CMSIM GEANT simulation of samples
consisting of 4000 decay events for each source meson from
Table \ref{tab:sb1}. We looked for the difference between the
profiles of showers produced by direct photons in the ECAL and the
profiles of the showers produced by photons originating from meson
decays. This search was performed in the
$20<\Pt^{\tilde\gamma}<100~ GeV/c$ interval. The results of studying
neutral and charged decay channels are presented in \cite{GMS}. 
It was found that the suppression factor of $\eta$, $\omega$,
$K_S^0$ mesons of the order of 0.3 -- 0.8 can be achieved for
$40<\Pt^{\tilde\gamma}<100~ GeV/c$ with a selection efficiency
of single photons taken to be 90$\%$. As for charged decay channels
of $\eta, \omega, K^0_s$ mesons the results of  \cite{GMS}
show that by chosing absolute isolation cut $\Et^{isol}\leq2~GeV/c$
in the isolation cone with $R^{\gamma}_{isol}=0.7$ and upper cut on the transverse energy
deposited in HCAL $\Et^{HCAL}_{dep}\leq\Et^{HCAL}_{thr}$ (where $\Et^{HCAL}_{thr}=2-5~GeV$
depends on $\Pt^{\tilde\gamma}$) one can suppress these decays with a very good efficiency
(at least $98\%$).

The correction of Tables  \ref{tab:sb1}--\ref{tab:sb3} with account of
the above rejection efficiencies is presented in
new Tables \ref{tab:sb1e}--\ref{tab:sb3e}.
Here the background ($B$) 
differs from the one in Tables \ref{tab:sb1}--\ref{tab:sb3} by including events
with electron candidates with the discussed above efficiency.
Comparing Tables \ref{tab:sb1e}--\ref{tab:sb3e} with Tables
\ref{tab:sb1}--\ref{tab:sb3} we observe the $50-55\%$ growth of the $S/B$ ratio for
$\Pt^{\tilde{\gamma}}\geq 40~GeV/c$ and, in practice, very small changes of the $S/B$ values at
$\Pt^{\tilde{\gamma}}\geq 100~GeV/c$.

We have not discussed here  the background that may appear due to  possible
$\gamma$/jet misidentification, because as was shown in \cite{ATL}, $\gamma$ and
jet can be discriminated with a high precision. Really, as was
mentioned  
at the beginning of this section (see also Section 3.2),
we defined the photon (or the candidate to be registered as the direct photon) as the signal
in the $5\times 5$ ECAL crystal cell window satisfying cut conditions (17) -- (22) of
Section 3.2. These conditions effectively discriminate the photons from jets (see  \cite{GMS}).

From  Tables \ref{tab:sb1} -- \ref{tab:sb3} we have seen that the cuts
listed in Table \ref{tab:sb0} (having rather moderate values of
$\Pt^{clust}_{CUT}$ and  $\Pt^{out}_{CUT}$) allow to suppress
the major part of the background events.
~\\[-20pt]
\begin{table}[h]
\small
\begin{center}
\caption{Signal vs. background ~~(\bf II)}
\vskip0.5mm
\begin{tabular}{||c|c||c|c|c|c|c|c|c||}                  \hline \hline
\label{tab:sb1e}
$\pth$& & $\gamma$ & $\gamma$ &\multicolumn{4}{c|}{  photons from the mesons}  &
\\\cline{5-8}
\Gvc& Cuts&\hmm direct\hmm &\hmm brem\hmm & $\;\;$ $\pi^0$ $\;\;$ &$\quad$ $\eta$ $\quad$ &
$\omega$ &  $K_S^0$ &\hmm $e^{\pm}$\hmm \\\hline \hline
    &Preselected&\hmm12394&\hmm 20952& 166821& 66533& 17464& 23942&\hmm 6684\hmm  \\\cline{2-9}
 40 &After cuts &\hmm 1546&\hmm 198&     54&    16&     1&     2&\hmm   2\\\cline{2-9}
    &+ jet isol.  &\hmm 903&\hmm 92&     23&     8&     1&     2&\hmm   1\\\hline  \hline
    &Preselected&\hmm 19359  &\hmm 90022 &658981 &247644 &69210  &85568 &\hmm 47061\\\cline{2-9}
100 &After cuts&\hmm 1061 &\hmm 31 & 9 &3 &1  &0 &\hmm 3 \\\cline{2-9}   
    &+ jet isol. &\hmm 1013 &\hmm29 &6 &3 &1  &0 &\hmm 3 \\\hline \hline
    &Preselected&\hmm55839 &\hmm354602 &1334124 &393880 &141053 & 167605 &\hmm153410\hmm\\\cline{2-9}
200 &After cuts&\hmm  1653 &\hmm 27& 3 &5 &0  &0 &\hmm 3\\\cline{2-9}
    &+ jet isol. &\hmm 1645 &\hmm 27& 3 &5 &0  &0 &\hmm 2\\\hline \hline
\end{tabular}
\vskip0.2cm
\caption{Values of efficiencies and significance ~~(\bf II)}
\vskip0.1cm
\begin{tabular}{||c||c|c|c|c|>{\columncolor[gray]{\coltab}}c|c||}   \hline \hline
\label{tab:sb2e}
$\pth$\Gvc& $S$ & $B$ & $Eff_S(\%)$  & $Eff_B(\%)$  & $S/B$& $S/\sqrt{B}$ \\\hline \hline
40  & 1546& 276 & 12.47$\pm$ 0.34 & 0.091 $\pm$ 0.006&  5.6 & 93.1 \\\hline
100 & 1061&  47 & 5.48 $\pm$ 0.17 & 0.004 $\pm$ 0.001& 22.6 & 154.8 \\\hline  
200 & 1653&  38 & 2.96 $\pm$ 0.07 & 0.001 $\pm$ 0.000& 43.5 & 268.2 
\\\hline \hline
\end{tabular}
\vskip0.2cm
\caption{Values of efficiencies and significance with jet isolation cut ~~(\bf II)}
\vskip0.1cm
\begin{tabular}{||c||c|c|c|c|>{\columncolor[gray]{\coltab}}c|c||}  \hline \hline
\label{tab:sb3e}
$\pth$\Gvc& ~~~$S$~~~ & $B$ & $Eff_S(\%)$ & $Eff_B(\%)$  & $S/B$& $S/\sqrt{B}$
 \\\hline \hline
40  & 903& 127 & 7.29 $\pm$ 0.25 & 0.042 $\pm$ 0.004&  7.1 & 80.1 \\\hline 
100 & 1013& 42 & 5.23 $\pm$ 0.17 & 0.004 $\pm$ 0.001& 24.1 & 156.3 \\\hline  
200 & 1645& 38 & 2.95 $\pm$ 0.07 & 0.001 $\pm$ 0.000& 43.3 & 266.9 
\\\hline \hline
\end{tabular}
\end{center}
\end{table}

\normalsize

The considered here samples of generated events with all QCD subprocesses
were used to 
study the effect of simultaneous application of 
$\Pt^{clust}_{CUT}$ and  $\Pt^{out}_{CUT}$ on: \\[1pt]
\hspace*{10mm} (a) the number of selected events (for $L_{int}=3~fb^{-1}$);\\
\hspace*{10mm} (b) the signal-to-background ratio $S/B$;\\
\hspace*{10mm} (c) the mean value of 
$(\Pt^{\tilde{\gamma}}\!-\!\Pt^{Jet})/\Pt^{\tilde{\gamma}}\equiv F$ and
its  standard deviation value $\sigma (F)$.\\[1pt]
The results are presented in Tables 1 -- 12 of Appendix 6 for 
Selection 1 and Tables 13--24 for Selection 2.

Let us emphasize that the tables of Appendix 6 include, in contrast to
Appendices 2--5, the results obtained after analyzing three generated 
samples (described in the beginning of this section)
of {\it signal and background} events. 
These events were selected with the cuts of Table \ref{tab:sb0}.

Namely, the cuts (1) -- (10) of Table \ref{tab:sb0} were applied for
 preselection of 
``$\tilde{\gamma}+1~jet$'' events. 
The jets in these events as well as clusters were found by use of only one 
jetfinder LUCELL (for the whole $\eta$ region $|\eta^{jet}|<5.0$).

Tables 1 -- 4 of Appendix 6 correspond to the
simulation with $\pth=40 ~GeV/c$. Analogously, the values of
$\pth=100 ~GeV/c$ and $\pth=200 ~GeV/c$ were used for
Tables 5 -- 8  and Tables 9 -- 12 respectively. The events used for analysis
in Tables  1 --  12 have passed the cuts defined
by Selection 1. The  rows and  columns of Tables 1 -- 12 illustrate, respectively, the influence of
$\Pt^{clust}_{CUT}$ and $\Pt^{out}_{CUT}$ on the quantities
mentioned above (in the points (a), (b), (c)).

First of all, we see from Tables 2, 6 and 10 that a noticeable reduction
of the background take place while moving along the table diagonal from the 
right-hand bottom corner to the
left-hand upper one, i.e. with reinforcing $\Pt^{clust}_{CUT}$ and $\Pt^{out}_{CUT}$. 
So, we see that for $\pth=40 ~GeV/c$  the value of 
$S/B$ ratio changes in the table cells along the diagonal
from $S/B=2.9$ (in the case of no limits on these two variables), to $S/B=5.6$ for the
cell with $\Pt^{clust}_{CUT}=10~ GeV/c$ and $\Pt^{out}_{CUT}=10 ~GeV/c$.
Analogously, for $\pth=200 ~GeV/c\,$ the value of $S/B$ changes in the same table cells
from 13.6 to 43.5 (see Tables 2, 10 of Appendix 6).

The second observation from Appendix 6. The restriction of $\Pt^{clust}_{CUT}$ and
$\Pt^{out}_{CUT}$ improves the calibration accuracy. Table 3 shows that in the interval $\Pt^{\tilde{\gamma}}\gt40~GeV/c$
the mean value of the fraction $F(\equiv (\Pt^{\tilde{\gamma}}\!-\!\Pt^{Jet})/\Pt^{\tilde{\gamma}})$
decreases from 0.031 (the bottom right-hand corner) to 0.009
for the table cell with $\Pt^{clust}_{CUT}=10~ GeV/c$ and $\Pt^{out}_{CUT}=10 ~GeV/c$.
At the same time, the both cuts lead to a noticeable decrease of
the gaussian width $\sigma (F)$ (see Table 4 and also Tables 8 and 12).  
For instance, for $\pth=40 ~GeV/c$ ~$\sigma (F)$ drops by about a factor of two: from 0.163 to 0.085.
It should be also noted that
Tables 4, 8 and 12 demonstrate that for any fixed
value of $\Pt^{clust}_{CUT}$  
further improvement in $\sigma (F)$ can be achieved  by limiting $\Pt^{out}$ 
(e.g. in line with $\Pt^{clust}_{CUT}=15~GeV/c$
$\sigma (F)$ drops by a factor of 2 with variation of $\Pt^{out}$ from $1000$ to $5~GeV/c$).

The explanation is simple. The balance~ equation (28) contains 2 terms on the right-hand
side ($1-cos\dphi$) and $\Db/\Pt^{\tilde{\gamma}}$.
The first one is negligibly small in a case of Selection 1 and tends to decrease with growing 
$\Pt^{\tilde{\gamma}}$ (see tables in Appendices 2--5). So, we see that in this case
the main source of the disbalance in  equation (28) is the term $\Db/\Pt^{\tilde{\gamma}}$.
This term can be diminished by decreasing $\Pt$ activity beyond the jet,
i.e. by decreasing $\Pt^{out}$.

\begin{figure}[htbp]
\vspace{-2.6cm}
\hspace{-.7cm} \includegraphics[width=17cm,height=20cm]{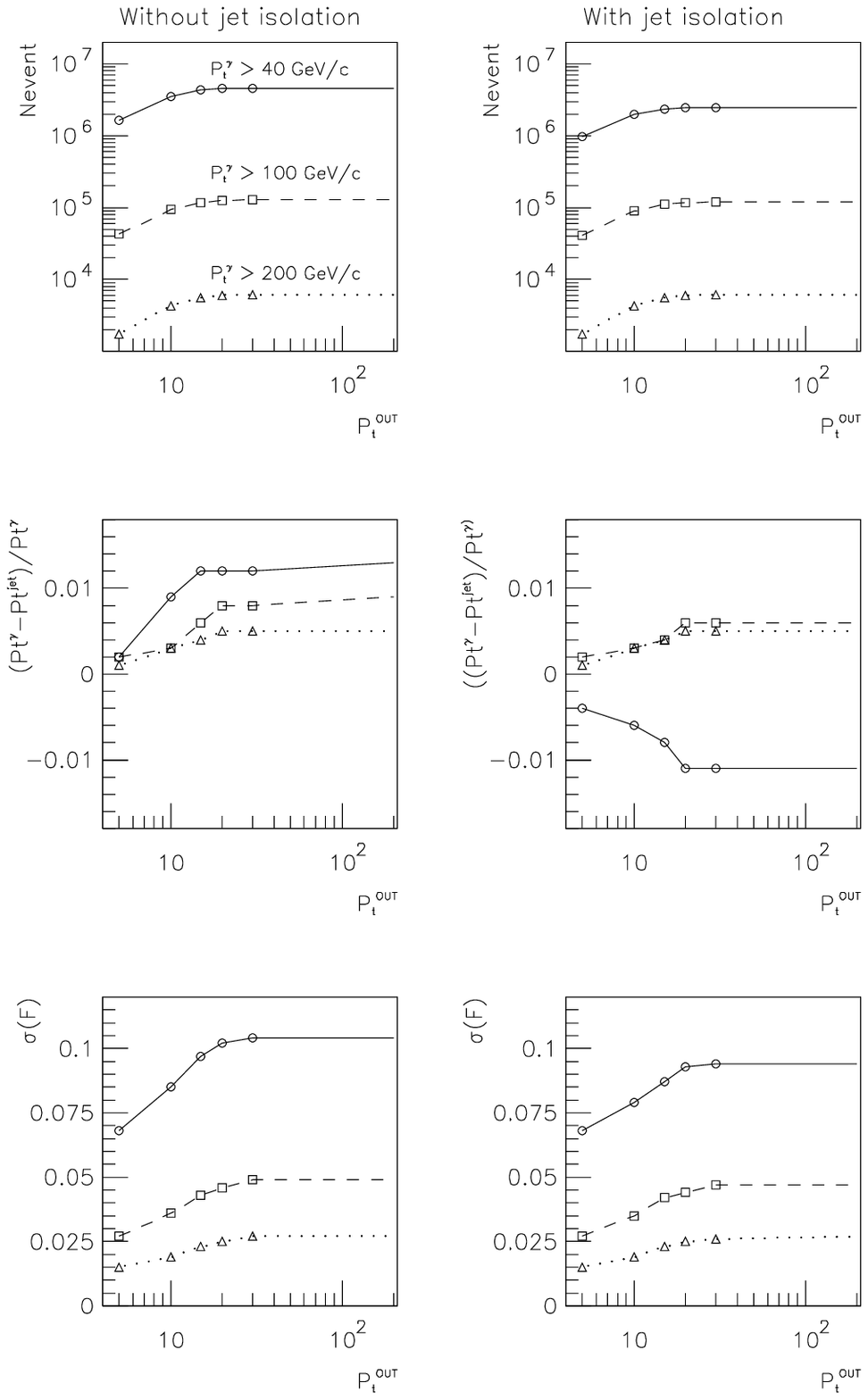}
\vspace{-0.7cm}
\caption{\hspace*{0.0cm} Number of events (for $L_{int}=3~fb^{-1}$),
 mean value of $(\Pt^{\tilde{\gamma}}-\Pt^{Jet})/\Pt^{\tilde{\gamma}}$
($\equiv F$) and its standard deviation $\sigma(F)$ distributions over $\Pt^{out}$
for the cases of nonisolated (left-hand column) and isolated (right-hand column) jet and for
three  intervals: $\Pt^{\tilde{\gamma}} > 40, 100$ and $200 ~GeV/c$. $\Pt^{clust}_{CUT}=10 ~GeV/c.$
}
\label{fig:mu-sig}
\end{figure}

The behavior of the number of selected events (for $L_{int}=3~fb^{-1}$),
 the mean values of $F=(\Pt^{\tilde{\gamma}}\!-\!\Pt^{Jet})/\Pt^{\tilde{\gamma}}$ and
its standard deviation $\sigma (F)$ as a function of $\Pt^{out}_{CUT}$ 
(with fixed $\Pt^{clust}_{CUT}=10~GeV/c$)
are also displayed in Fig.~\ref{fig:mu-sig} for events with non-isolated 
(left-hand column) and isolated jets (right-hand column, see also Tables 13--24 of Appendix 6).

{\it Thus, we can conclude that application of two criteria introduced
in Section 3.2, i.e. $\Pt^{clust}_{CUT}$ and $\Pt^{out}_{CUT}$,
results in two important consequences: significant background reduction
and essential improvement of the calibration accuracy.
}

The numbers of events for different $\Pt^{clust}_{CUT}$ and $\Pt^{out}_{CUT}$
are given in the cells of Tables 1, 5 and 9 of Appendix 6. One can see that even with such strict
$\Pt^{clust}_{CUT}$ and $\Pt^{out}_{CUT}$ values as $10 ~GeV/c$ for both, for example,
we would have a sufficient number of events
(3 million, about 80 000 and 4 000 for $\Pt^{\tilde{\gamma}}\geq40 ~GeV/c$,
$\Pt^{\tilde{\gamma}}\geq100 ~GeV/c$ and  $\Pt^{\tilde{\gamma}}\geq200 ~GeV/c$, respectively)
with low background contamination ($S/B= 5.6,~22.6,~43.5$)
and a good accuracy of the absolute jet energy scale setting
during one month of continuous LHC running (i.e. $L_{int}= 3~fb^{-1}$).

In addition, we also present in Appendix 6 Tables 13--24 obtained with Selection 2.
They contain the information analogous to that in Tables 1 -- 12
but for the case of isolated jets with $\epsilon^{jet}<5\%$.
From these tables we see that with the same cuts
 $\Pt^{clust}_{CUT}=\Pt^{out}_{CUT}=10 ~GeV/c$ one can expect about
1~700~000, 80~000 and 4~000 events for $\Pt^{\tilde{\gamma}}\geq40 ~GeV/c$,
$\Pt^{\tilde{\gamma}}\geq100 ~GeV/c$ and  $\Pt^{\tilde{\gamma}}\geq200 ~GeV/c$, 
respectively, with a much more better fractional $\Pt^{\tilde{\gamma}}-\Pt^{Jet}$ balance,
less than $F=0.5\%$ for all.

Let us mention that all these PYTHIA results give us an indication of a tendency
and may serve as a guideline for further full GEANT simulation that would allow to come to 
a final conclusion. 

To conclude this section we would like to stress, firstly, that, as is seen from
Tables~\ref{tab:sb1}, the  ``$\gamma-brem$'' background
defines a dominant part of the total background. 
Its contribution
is about 1.5 -- 3  times larger (see Tables~\ref{tab:sb1}, \ref{tab:sb1e}) than 
the combined background from neutral meson decays.
Thus, one can see from Table 17 that $\pi^0$ contribution being about a half of
``$\gamma-brem$'' background at $\pth>40~GeV/c$ becomes one order less than
``$\gamma-brem$'' background at $\pth>200~GeV/c$.
We would like to emphasize here
that this is a strong prediction of the PYTHIA generator that has to be compared with
predictions of another generator like HERWIG, for example.
%
%

Secondly, we would like to underline also that as it is seen from Table 14, 17 the photon
isolation and selection cuts 1--5, usually used in the study of inclusive photon production
(see, for instance, \cite{CDF1}, \cite{D0_1}, \cite{D0_2}), increase the $S/B$ ratio (for $\pth>100~GeV/c$)
up to 1.93 only while the other cuts 6--17, that select events with a clear \gpj topology and
limited $\Pt$ activity beyong a chosen single jet, lead to a significant improvement of
$S/B$ by about one order of magnitude to $24.46$.

The numbers in the tables of Appendix 6 were obtained with inclusion of the contribution
from the background events. The tables show that their account does not spoil the \ptgj
balance in the event samples preselected with the cuts 1--10 of Table \ref{tab:sb0}. 
The estimation of the number of these background events would be important
for the gluon distribution determination (see Section 10).

\normalsize
\def\baselinestretch{1.0}

\section{STUDY OF DEPENDENCE OF THE \Ptgj BALANCE ON PARTON $k_t$.}

\it\small
\hspace*{9mm}
It is shown that in the case of ISR presence the value of fractional disbalance $\Fptgj$
depends weakly on the variation of the average value of intrinsic parton transverse momentum 
$\la k_{~t}\ra$.
\rm\normalsize
\vskip3mm

This section is dedicated to the study (within PYTHIA simulation) of a possible influence of the intrinsic parton
 transverse momentum $k_{~t}$ on the $\Pt$ balance of the \gpj system. For this aim we consider
two samples of signal events gained by simulation  with subprocesses
(1a) and (1b) in two different ranges of $\pth$: $\pth \geq 40~ GeV/c$ and  $\pth \geq 200~ GeV/c$.
For these two $\pth$  intervals Tables \ref{ap1:tab1} and \ref{ap1:tab2}
demonstrate the average values of $\Pt56$ and $\la\Pt^{5+6}\ra$ (defined by (3)) for two different cases of generation: 
without initial state radiation 
(``ISR is OFF'') and with it (``ISR is ON''). Five different generations were done for each
$\pth$ interval. They correspond 
~\\[-8mm]
\begin{table}[h]
\small
\begin{center}
\caption{Effect of $k_t$ on the $\Pt^{\gamma}$ - $\Pt^{Jet}$ balance with
$\pth\! \!=\!\! 40 ~GeV/c. ~~ F=\Fptgj $}
\vskip0.1cm
\begin{tabular}{||c||c|c|c|c||c|c|c|c||}   \hline  \hline
\label{ap1:tab1}
$\la k_T\ra$& \multicolumn{4}{|c||}{ ISR is OFF}&\multicolumn{4}{c||}{ ISR is
ON} \\\cline{2-9}
$(GeV/c)$ &$\la \Pt56\ra$&$\la \Pt^{5+6}\ra$&$\la F\ra$&$\sigma(F)$
&$\la \Pt56\ra$&$\la \Pt^{5+6}\ra$&$\la F\ra$&$\sigma(F)$   \\\hline  \hline
 0.0 &0.0&0.0 &-0.002 &0.029&8.8 &6.9 &0.007 &0.065\\\hline
 1.0&1.8&1.3 &-0.001 &0.036&9.1 &7.0 &0.009 &0.069  \\\hline
 2.5 &4.5&3.2 &0.001 &0.054 &9.6 &7.4 &0.010 &0.074  \\\hline
 5.0 &8.7&6.1 &0.014 &0.089 &10.4&7.2 &0.015 &0.088  \\\hline
 7.0 &11.2&7.7&0.020 &0.107&11.0 &8.2 &0.022 &0.101  \\\hline  \hline
\end{tabular}
\end{center}
\end{table}
\begin{table}[h]
\small
~\\[-22mm]
\begin{center}
\caption{Effect of $k_t$ on $\Pt^{\gamma}$ -$\Pt^{Jet}$ balance with
$\pth \!\!=\!\! 200~ GeV/c.~ F=\Fptgj $}
\vskip0.1cm
\begin{tabular}[h]{||c||c|c|c|c||c|c|c|c||}                  \hline  \hline
\label{ap1:tab2}
$\la k_T\ra$& \multicolumn{4}{|c||}{ ISR is OFF}&\multicolumn{4}{c||}{ ISR is ON}
\\\cline{2-9}
$(GeV/c)$ &$\la \Pt56\ra$&$\la \Pt^{5+6}\ra$&$\la F\ra$&$\sigma(F)$
&$\la \Pt56\ra$&$\la \Pt^{5+6}\ra$&$\la F\ra$&$\sigma(F)$   \\\hline  \hline
 0.0 &0.0&0.0 &0.000 &0.010 &11.1&8.4 &-0.001 &0.027\\\hline
 1.0 &1.8&1.3 &0.000 &0.013 &11.2&8.6 &0.000  &0.028 \\\hline
 2.5 &4.5&3.1 &0.000 &0.019 &11.8&8.8 &0.001  &0.028 \\\hline
 5.0 &8.7&6.1 &0.001 &0.022 &12.7&9.3 &0.001  &0.031 \\\hline
 7.0 &11.2&7.8 &0.001 &0.029 &13.9&10.4 &0.002 &0.034  \\\hline  \hline
\end{tabular}
\end{center}
~\\[-9pt]
\hspace*{12mm}$\ast$ \footnotesize{All numbers in the tables above are given in
 $GeV/c$}.
\vskip-3mm
\end{table}
\normalsize

\noindent
to five values of parton $\la k_t\ra$
\footnote{$\equiv$ PARP(91) parameter in PYTHIA}:
$\la k_t\ra\!=\!0.0, 1.0, 2.5, 5.0$ and $7.0~GeV/c$
(the values $\la k_t\ra > 1~ GeV/c$ are given
  here only for illustration of a tendency). 

Let us consider firstly the case with ISR switched off during the simulation.
The numbers in Tables \ref{ap1:tab1} and \ref{ap1:tab2} (obtained from the set of events selected by 
the cuts $\dphi<15^\circ$, $\Pt^{out}_{CUT}=5~GeV/c$ and $\Pt^{clust}_{CUT}=10~ GeV/c$) show 
that in the case when ``ISR is OFF'' the values of $\la\Pt56\ra$ and $\la\Pt^{5+6}\ra$
grows rapidly with increasing $\la k_t\ra$ and does not depend on \ptg (or $\pth$).
In fact, the values  of $\la\Pt{56}\ra$  are proportional to the values of $\la k_t\ra$ in this case. 
\begin{figure}[t]
\begin{center}
\vspace{-3mm}
  \includegraphics[width=16.7cm,height=12cm]{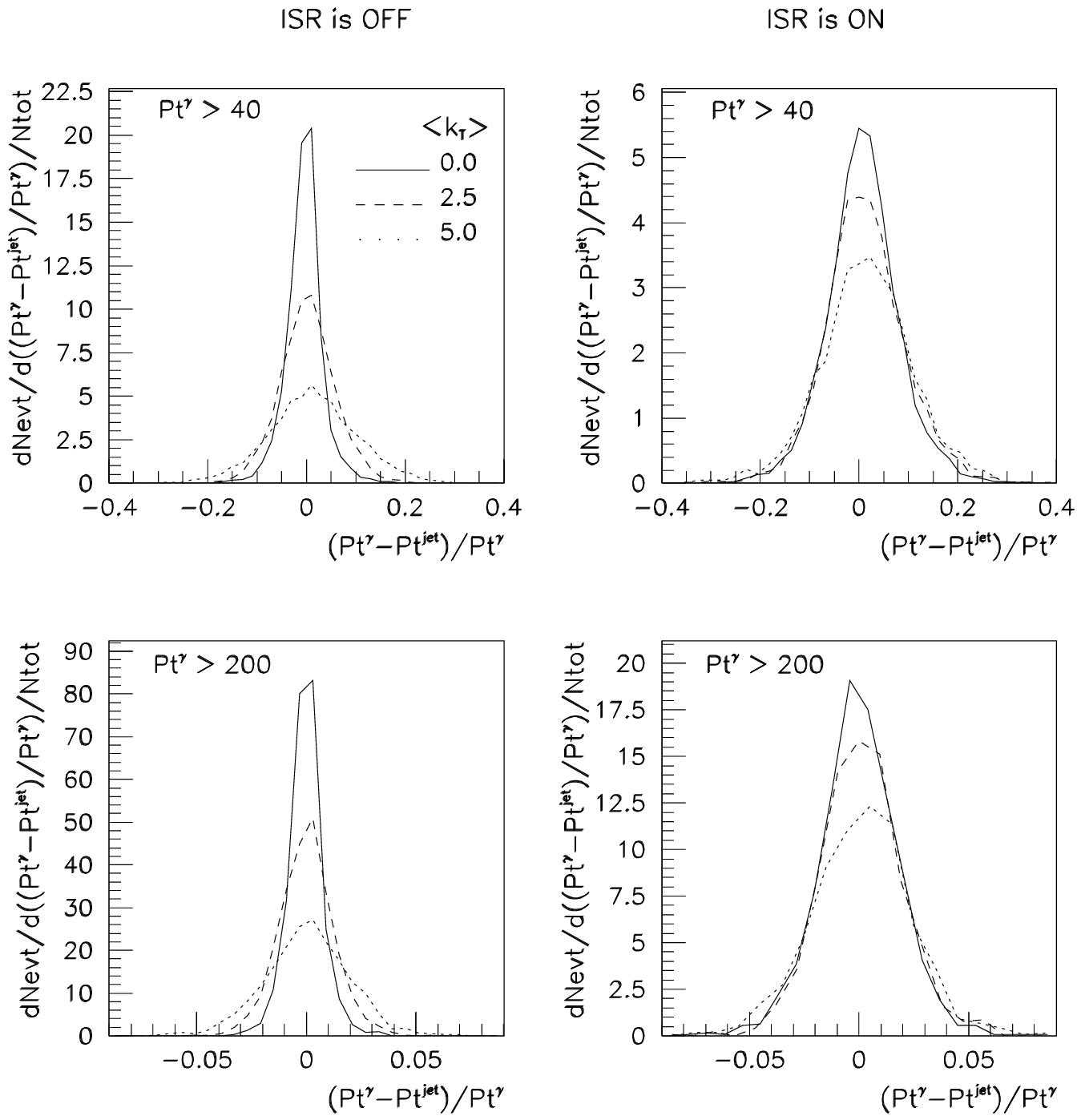}
  \vspace{-12mm}
    \caption{\hspace*{0.0cm} $\Fptgj\equiv F(\gamma,jet)$ as a function of primordial $k_t$ value
for the cases of switched off (left column) and switched on (right column) initial radiation for
$\pth = 40$ and $\pth = 200~~GeV/c$.}
\label{fig:kt}
\end{center}
\vskip-3mm
\end{figure}

The picture changes when ISR is taken into accound. In this case the variables $\la\Pt56\ra$
and $\la\Pt^{5+6}\ra$ initially get large 
value at $\la k_t\ra=0$, e.g. $\la\Pt{56}\ra=8.8~ GeV/c$  and $11.1 ~GeV/c$ 
for $\pth=40 ~GeV/c$ and $200 ~GeV/c$, respectively.
But at the same time, in contrast to the case ``ISR is OFF'',
the values of $\la\Pt56\ra$ grow more slowly with $\la k_t\ra$ when  ``ISR is ON''.
Indeed, they grow up 
from $8.8~(11.1)$ at $\la k_t\ra=0.0$ to $11.0~(13.9)$ at $\la k_t\ra=7~GeV/c$ for 
$\pth\!=\!40 ~GeV/c$~$(200 ~GeV/c)$.

The most remarkable thing, as it follows from Tables \ref{ap1:tab1} and \ref{ap1:tab2},
that $\la\Pt56\ra$ depends weakly on $\la k_t\ra$ in the range of its reasonable values 
$\la k_t\ra\leq 1~GeV/c$.

The variations of the fractional disbalance $F\!\equiv\!\Fptgj$  and its standard deviation $\sigma(F)$
with $\la k_t\ra$ are also shown in Tables \ref{ap1:tab1} and \ref{ap1:tab2} and in plots of
Fig.~\ref{fig:kt}. One can see that for reasonable values $\la k_t\ra\leq 1~GeV/c$ and for the case 
``ISR is ON'' the changes in the fractional disbalance $F$ with $k_t$ variation are very small. 
They are of order of $0.2\%$ for $\pth=40 ~GeV/c$ and of order of $0.1\%$ for $\pth=200 ~GeV/c$
\footnote{Recall that the numbers in Tables \ref{ap1:tab1} and \ref{ap1:tab2} may be 
compared with those in the tables of Appendix 6, where the same $\pth$ cuts are used,
rather than with the results of the tables of Appendices 2 -- 5, where  $\pth$ cuts 
were taken to be two times smaller (see for explanation the beginning of Section 7).}.
%

%
\section{\gpj EVENT RATE ESTIMATION FOR GLUON DISTRIBUTION $f_g(x,Q^2)$ DETERMINATION AT THE LHC.}

\it\small
\hspace*{9mm}
The number of \gpj events suitable for measurement of gluon distribution 
$f_g(x,Q^2)$ in different $x$ and $Q^{\;2}$ intervals is estimated. It is
shown that with $L_{int}=20~fb^{-1}$ it would be possible to collect about
ten million of these events. This number would allow to cover a new kinematical 
area not studied in any previous experiment
($2\cdot 10^{-4}\lt x\lt 1.0$ with $1.6\cdot 10^{3}\leq Q^2\leq8\cdot10^{4} ~(GeV/c)^2$).
This area in the region of small $x\geq10^{-4}$ has $Q^2$ by about two orders        
of magnitude higher than reached at HERA now.
\rm\normalsize
\vskip4mm

Many  of theoretical predictions on the production of new particles
(Higgs, SUSY) at the LHC are based on model estimations of the gluon density
behaviour at low $x$ and high values of square of transfered momentum 
$Q^2$. Therefore a measurement of the proton gluon density for this kinematic 
region directly in LHC experiments would be obviously useful. One of the 
promising channels for this measurement, as was shown in ~\cite{Au1},
may be a high $\Pt$ direct photon production 
$pp(\bar{p})\rightarrow \gamma^{dir} + X$.
The region of high $\Pt$, reached by UA1 \cite{UA1}, UA2 \cite{UA2},
CDF \cite{CDF1} and D0 \cite{D0_1} extends up to $\Pt \approx 80~ GeV/c$ and 
recently moved  up to $\Pt=105~GeV/c$ \cite{D0_2}. 
These data together with the later ones (see references in \cite{Fer}--\cite{Fr1})
and recent E706 \cite{E706} and UA6 \cite{UA6} results give an opportunity for 
studying deeply the gluon distribution in proton (for data analysis see 
\cite{Au2}, \cite{Vo1}, \cite{Mar}). The rates and 
estimated cross sections of inclusive direct photon production at 
the LHC were given in \cite{Au1} (see also \cite{AFF}).

Here for the same aim we shall consider 
the process $p p\rightarrow \gamma^{dir}\, +\, 1\,jet\, + \,X$
defined in the leading order by two QCD subprocesses (1a) and (1b)
(for experimental results see \cite{ISR}, \cite{CDF2}).

Apart from the advantages, discussed in Section 8  in connection with the background suppression 
(see also \cite{Ber}--\cite{Hu3}),
the ``$\gamma^{dir}+1~jet$'' final state may be easier for physical analysis
than inclusive photon production process ``$\gamma^{dir}+X$''  if we shall look at this problem from
the viewpoint of extraction of information on the gluon distribution in a proton.
Indeed, in the case of inclusive direct photon production the
cross section is given as an integral over the products of a fundamental $2\to2$ parton subprocess
cross sections and the corresponding parton distribution functions $f_a(x_a,Q^2)$ (a = quark or gluon), 
while in the case of $pp\rightarrow \gamma^{dir}+1~Jet+X$ for $\Pt^{Jet}\,
\geq \,30\, GeV/c$ (i.e. in the region where ``$k_t$ smearing effects''
\footnote{This terminology is different from ours, used in Sections 2 and 9, as we denote by ``$k_t$''
only the value of parton intrinsic transverse momentum.}
are not important, see \cite{Hu2}) the cross section is
expressed directly in terms of these distributions (see, for example,
\cite{Owe}): \\[-16pt]
\begin{eqnarray}
\frac{d\sigma}{d\eta_1d\eta_2d\Pt^2} = \sum\limits_{a,b}\,x_a\,f_a(x_a,Q^2)\,
x_b\,f_b(x_b,Q^2)\frac{d\sigma}{d\hat{t}}(a\,b\rightarrow c\,d),
\label{gl:4}
\end{eqnarray}
\vskip-3mm
\noindent
where \\[-9mm]
\begin{eqnarray}
x_{a,b} \,=\,\Pt/\sqrt{s}\cdot \,(exp(\pm \eta_{1})\,+\,exp(\pm \eta_{2})).
\label{eq:x_def}
\end{eqnarray}
\vskip-1mm

The designation used above are as the following:
$\eta_1=\eta^\gamma$, $\eta_2=\eta^{Jet}$; ~$\Pt=\Pt^\gamma$;~ $a,b=q,\bar{q},g$; 
$c,d=q,\bar{q},g,\gamma$. Formula (\ref{gl:4}) and the knowledge of 
$q, \,\bar{q}$ distributions 
allow the gluon  distribution $f_g(x,Q^2)$
to be determined after account of selection efficiencies for jets and  $\gamma^{dir}-$candi\-dates 
as well as after subtraction of the background contribution
(as it was discussed in Section 8 keeping in hand this physical application).

The earlier estimations of \gpj events suitable for jet energy
calibration, and thus for determination the gluon distribution
inside a proton \cite{9}, \cite{10}, 
showed that there would be many events with well-isolated
photons and suppressed cluster activity beyond the \gpj system. 
In the previous sections a lot of details connected with the
structure and topology of these events and the features of objects appearing
in them were discussed. Now with this information in mind we are
in position to discuss an application of the \gpj
event samples, selected with the previously proposed cuts, for estimating
the rates of gluon-based  subprocess (1a) in different $x$ and $Q^2$ intervals.
We shall use here the cuts $1-13$  of Table 13 
with the following values of parameters
\footnote{An application of cuts 14--17, as it is seen from Table 2, leads only
 to $20\%$-$30\%$  improvement of $S/B$ ratio and they are essential mostly for 
improvement of disbalance $(\Pt^{\gamma}\!-\!\Pt^J)/\Pt^{\gamma}$  value.}:

~\\[-13mm]
\begin{eqnarray}
\Ptg>40~ GeV/c,\quad |\eta^{\gamma}|<2.61,\quad \Pt^{jet}>30~ GeV/c,\quad |\eta^{jet}|<5.0,
\quad \Pt^{hadr}>5~ GeV/c,
\label{l1}
\nonumber
\end{eqnarray}
~\\[-15mm]
\begin{eqnarray}
\Pt^{isol}_{CUT}=5\;GeV/c, \;
{\epsilon}^{\gamma}_{CUT}=7\%, \;
\dphi<15^{\circ}, \; 
\Pt^{clust}_{CUT}=5\;GeV/c \;
\label{l2}
\end{eqnarray}
~\\[-11mm]

Table~\ref{tab:q/g-1} shows percentage of ``Compton-like" subprocess
(1a) (amounting to $100\%$ together with (1b)) in the samples of events
selected with cuts (17) -- (23) of Section 3.2 for $\Pt^{clust}_{CUT}=10~GeV/c$
for different $\Ptg$  and $\eta^{jet}$ intervals:
Barrel (HB) part ($|\eta^{jet}|<1.4$, see also tables of Appendix 1)
and  Endcap+Forward (HE+HF) part  ($1.4<|\eta^{jet}|<5.0$).
We see that the contribution of Compton-like subprocess drops with $|\eta^{jet}|$
enlarging and with growing $\Pt^{jet}(\approx\Ptg$ in the sample of the 
events collected with the cuts $1-13$  of Table 13).
~\\[-7mm]
\begin{table}[h]
\begin{center}
\caption{The percentage of ``Compton-like" process  $q~ g\rrr \gamma +q$.}
\normalsize
\vskip.1cm
\begin{tabular}{||c||c|c|c|}                  \hline \hline
\label{tab:q/g-1}
Calorimeter& \multicolumn{3}{c|}{$\Pt^{Jet}$ interval ($GeV/c$)} \\\cline{2-4}
    part   & 40--50 & 100--120 & 200--240   \\\hline \hline
HB         & 90     &  85   &  80  \\\hline
HE+HF      & 86     &  82   &  74  \\\hline
\end{tabular}
\end{center}
\end{table}
\normalsize
~\\[-11mm]


In Table~\ref{tab:q/g-2} we present the $Q^2 (\equiv(\Ptg)^2)$
\footnote{see \cite{PYT}}
and $x$ (defined according to (\ref{eq:x_def}))distribution of the number 
of events (divided by $10^3$) that are caused by the $q~ g\rrr \gamma +q$ 
subprocess and passed the following cuts ($\Pt^{out}$ was not limited).

~\\[-14mm]
\begin{table}[h]
\small
\begin{center}
\caption{ Number of~ $g\,q\to \gamma^{dir} \,+\,q$~
events (divided by $10^3$) at different $Q^2$ and $x$ values for $L_{int}=20~fb^{-1}$.}
\label{tab:q/g-2}
\vskip0.1cm
\begin{tabular}{|lc|c|c|c|c|c|c|c|}                  \hline
 & $Q^2$ &\multicolumn{4}{c|}{ \hspace{-0.9cm} $x$ values of a parton} &All $x$
&$\Pt^{\gamma}$   \\\cline{3-7}
 & $(GeV/c)^2$ & $10^{-4}$--$10^{-3}$ & $10^{-3}$--$10^{-2}$ &$10^{-2}$--
$10^{-1}$ & $10^{-1}$--$10^{0}$ & $10^{-4}$--$10^{0}$&$(GeV/c)$     \\\hline
&\hmm\hmm 1600-2500\hmm  & 735.7 &2319.2 &2229.0 & 236.9 &5521.0 & 40--50\\\hline
&\hmm\hmm 2500-5000\hmm  & 301.6 &1323.3 &1402.7 & 207.4 &3235.1 & 50--71\\\hline
&\hmm\hmm 5000-10000\hmm &  33.7 & 361.3 & 401.0 &  97.7 & 893.8 & 71--100\\\hline
&\hmm\hmm 10000-20000\hmm&   1.5 &  80.8 &  99.4 &  38.0 & 219.9 & 100--141\\\hline
&\hmm\hmm 20000-40000\hmm&     0 &  15.6 &  24.4 &  12.4 &  52.5 & 141--200\\\hline
&\hmm\hmm 40000-80000\hmm&     0 &   2.1 &   4.2 &   2.5 &   8.8 & 200--283\\\hline
\multicolumn{5}{c|}{}&\multicolumn{2}{c|}{~~~~Sum = ~~~\bf 9~931 $\times$ ~$10^3$}\\\cline{6-7}
\end{tabular}
\end{center}
\end{table}
\vskip-6mm

\begin{table}[h]
\small
\begin{center}
\vskip-0.5cm
\caption{Number of~ $g\,c\to \gamma^{dir} \,+\,c$~ events (divided by $10^3$) 
at different $Q^2$ and $x$ values for $L_{int}=20~fb^{-1}$.}
\label{tab:q/g-3}
\vskip0.1cm
\begin{tabular}{|lc|c|c|c|c|c|c|c|} \hline
& $Q^2$ &\multicolumn{4}{c|}{ \hspace{-1.2cm} $x$ values for $c$-quark} & All $x$
& $\Pt^{\gamma}$ \\\cline{3-7}
& $(GeV/c)^2$ & $10^{-4}$--$10^{-3}$ & $10^{-3}$--$10^{-2}$ &$10^{-2}$--
$10^{-1}$ & $10^{-1}$--$10^{0}$ & $10^{-4}$--$10^{0}$&$(GeV/c)$  \\\hline
&\hmm\hmm 1600-2500\hmm   &109.4 & 360.5 & 329.6 &  34.7 & 834.4 & 40--50   \\\hline
&\hmm\hmm 2500-5000\hmm   & 35.1 & 189.7 & 202.7 &  25.4 & 453.2 & 50--71  \\\hline
&\hmm\hmm 5000-10000\hmm  &  3.9 &  51.5 &  58.6 &  12.1 & 126.3 & 71--100\\\hline
&\hmm\hmm 10000-20000\hmm &  0.1 &   9.0 &  12.4 &   3.4 &  25.0 & 100--141      \\\hline
&\hmm\hmm 20000-40000\hmm &    0 &   1.4 &   3.2 &   1.0 &   5.6 & 141--200    \\\hline
&\hmm\hmm 40000-80000\hmm &    0 &   0.1 &   0.4 &   0.1 &   0.7 & 200--283   \\\hline
\multicolumn{5}{c|}{}&\multicolumn{2}{c|}{~~~~$Sum$ = ~~~\bf 1~446 $\times$ ~$10^3$}\\\cline{6-7}
\end{tabular}
\end{center}
~\\[-5mm]
\end{table}
\normalsize

The analogous information for events with the charmed quarks in the initial state
$g\,c\to \gamma^{dir} \,+\,c$ is presented in Table~\ref{tab:q/g-3} 
(see also tables of Appendix 1). The simulation of the process
$g\,b\to \gamma^{dir}\,+\,b$ $\;$ has shown that the rates for the 
$b$-quark are 8 -- 10 times smaller than for the $c$-quark (these event 
rates are also given in tables of Appendix 1 for different $\Ptg$ intervals).
\footnote{See also the estimations of heavy quarks production in \gpj events
at LHC energy that were done in \cite{MD1}, \cite{MD_}, \cite{MD2},
\cite{BKS_P1} and \cite{BKS_GLU}.}

  Thus one can expect  on total of about 10 millions events with clean
\gpj topology (in a sense of reduced $Pt$ cluster or mini-jet activity
in addition to a leading jet) at  $L_{int}=20~fb^{-1}$ and among them of
about 1.5 million of events with c-quark jets.
Fig.~33 includes the widely used $(x,Q^2)$ kinematic plot (see \cite{Sti}
and  also \cite{Hu2}) to show what area can be covered by  
$q~ g\rrr \gamma +q$ 
\normalsize
\begin{flushleft}
\begin{figure}[h]
   \vskip-10.5mm
   \hspace{-1mm} \includegraphics[width=.50\linewidth,height=9.0cm,angle=0]{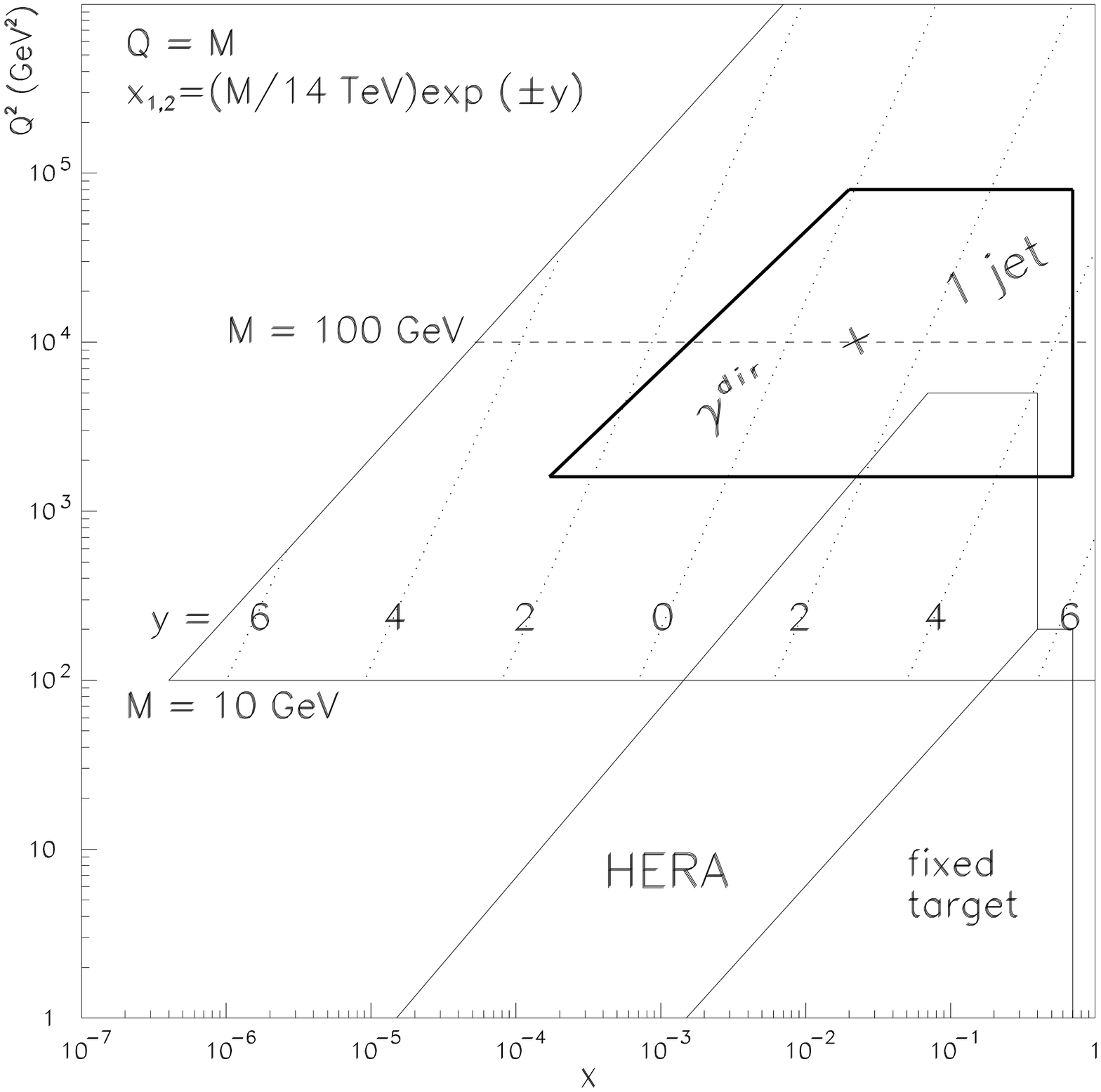}
\label{fig:q/g}
\end{figure}
\end{flushleft}
\begin{flushright}
\vskip-98mm
\parbox[r]{.48\linewidth}
{ \vspace*{-3mm}
events selected with the discussed above cuts.
The number of events in this area is given in Table~\ref{tab:q/g-2}.
From this figure and Table~\ref{tab:q/g-2} it becomes clear that even at 
low LHC luminosity  it would be possible (after gaining of $L_{int}=20~fb^{-1}$)
to study with a good statistics of \gpj events the gluon distribution of
in the region of small $x$ , attainable now at HERA, but in a new interval
of $Q^2$ that may be about 2--3 orders of magnitude higher than now reached
at HERA. It is worth emphasising that extension of the experimentally reachable
region at the LHC to the region of lower $Q^2$, overlapping with the area
covered by HERA, would also be of great interest.}
\end{flushright}

~\\[3mm]
{
\hspace*{-.1cm} {Figure~33: \footnotesize { $(x,Q^2)$ kinematic region
that can be covered at LHC in first few months of low luminosity operation by
$pp\to \gamma+Jet$ events selected for gluon distribution measurement.}}
}

~\\[-7mm]

%
\section{SUMMARY.}

We have done an attempt here to consider, following 
\cite{9}--\cite{BKS_GLU}, the physics of high $\Pt$ direct photon and jet 
associative production in proton-proton collisions
basing on the predictions of PYTHIA generator and the models implemented there.
This work may be useful for two practical goals: for absolute jet energy scale
determination and for gluon distribution measurement at LHC energy.

The detailed information provided in the PYTHIA event listings allows to track the
origin of different particles (like photons) and of objects (like clusters and jets) 
that appear in the final state. 
So, the aim of this work was to explore at the particle level of simulation 
this information as much as possible  for finding out what effect may be produced
by new variables, proposed in \cite{9}--\cite{BKS_P5} for describing \gpj events,  
and the cuts on them for solution of the mentioned above practical tasks.

For the first problem of the jet energy scale determination an important task 
is find a way to select the events
that may be caused (with a high probability) by the $q\bar{q}\to g+\gamma$ and 
$qg\to q+\gamma$~ fundamental parton subprocesses of direct photon production. 
To take into account a possible effect
of initial state radiation (which spectra are presented in different \ptg ~intervals 
in Tables 2--7 of Section 5) we used here the $\Pt$-balance equation, see (16),
written for an event as a whole. It allows to express \ptgj fractional disbalance
(see (28)) through new variables \cite{9}--\cite{BKS_P5} that describe the $\Pt$ 
activity {\it out} of \gpj system. They are $\Pt^{out}$ and $\Pt^{clust}$, i.e. $\Pt$  
of mini-jets or clusters that are additional to the main jet in event.
The latter are the most ``visible'' part of $\Pt^{out}$.


 It is shown that the limitation of $\Pt$
of clusters, i.e. $\Pt^{clust}$,  can help to decrease this disbalance
(see Figs.~15--21 and Tables 4--12 of Appendices 2--5). 

Analogously, the limitation of $\Pt$ activity of all detectable particles 
($|\eta_i|\lt5.0$) beyond the \gpj system, i.e. $\Pt^{out}$, also leads 
to a noticeable \ptgj disbalance reduction (see Figs.~22 and 23).

It is demonstrated that in the events, selected by means of simultaneous
restriction  from above of the $\Pt^{clust}$ and $\Pt^{out}$ activity, the 
values of \Ptgj are well balanced with each other while considering the PYTHIA 
particle level of simulation. The samples of these \gpj events, gained in this 
way,  are expected to be of a large enough volume for jet energy scale  determination in 
the interval $40\lt\Ptg\lt360~GeV/c$ (see Tables 1--12 of Appendix 6). 

It is worth mentioning that the most effect of improvement of \Ptgj balance 
can be reached by applying additionally the jet isolation criterion defined 
in \cite{9}--\cite{BKS_P5}. As it can be seen from Figs.~18, 19 
(Selection 2) as well as from Figs.~20, 21 (Selection 3) and also from
Tables 13--18 of Appendices 2--5 and Tables 13--24 of Appendix 6, the 
application of this criterion allows to select the events having the 
$\Fptgj$ disbalance at the particle level less than $1\%$
\footnote{The achieved disbalance value at the particle level of simulation 
shows the most optimistic value of\\
$\Fptgj$.}.

Definitely, the detector effects may worsen the balance determination
due to the limited accuracy of the experimental measurement.
We are planning to present the results of full GEANT simulation with 
the following digitization and reconstruction of signals by using the 
corresponding CMS packages (like CMSIM) in the forthcoming papers.

We present also PYTHIA predictions for the dependence of the distributions 
of the number of  selected \gpj events on \ptg~ and $\eta^{jet}$
(see Tables 8--12 of Section 5 and also tables of Appendices 2--5 with account 
of $\Pt^{clust}$ variation). The features of \gpj events in the barrel region 
of the CMS detector ($|\eta^{jet}|<1.4$) are exposed (see Figs.~8, 10). 
The $\Pt$ structure of the region in the $\eta-\phi$ space inside and beyond 
a jet is established (see Figs.~11--14).

The corrections to the measurable values of $\Pt^{jet}$ that  take into 
account the contribution from neutrinos belonging to a jet are presented for 
different $\Pt^{Jet}(\approx\Ptg$ for  the selected events$)$ intervals
in the tables of Appendix 1. It is shown in Section 4 that a cut on 
$\Pt^{miss}\lt10~GeV/c$ allows 
to reduce this contribution down to the value of
$\Delta_\nu=\la\Pt^{Jet}_{(\nu)}\ra_{all\; events}=0.2~GeV/c$. 
At the same time, as it is shown in  Section 8 (see also \cite{BKS_P5}), 
this cut noticeably decreases the number of the background $e^\pm$-events 
in which $e^\pm$ (produced in the $W^\pm\to e^\pm\nu$ weak decay) may be 
registered as direct photon.

The study of the fractional disbalance $\Fptgj$ dependence on an intrinsic 
parton transverse momentum $\la k_{\,t}\ra$, 
performed in Section 9, has shown its weak impact on the disbalance
in the case of initial state radiation account.

The possibility of the background events (caused by QCD subprocesses of
$qg, gg, qq$ scattering) 
suppression  was studied in Section 8. Basing on the introduced selection
criteria that include 17 cuts (see Table 13 of Section 8),
the background suppression relative factors and the values of signal event selection
efficiencies  are estimated (see Table 14). 

It is shown that after applying the first 5 ``photonic'' cuts 
(that may be used, for example, for selecting events with inclusive photon 
production and lead to signal-to-background 
ratio $S/B=1.9$ in the interval $\Ptg>100 ~GeV/c$, see Table 14) 
the use of the next 12 ``hadronic'' cuts of Table 13 may lead to further 
essential improvement  of $S/B$ ratio (by a factor of 12 for the same 
$\Ptg>100 ~GeV/c$ thus  $S/B$ becomes equal to $24.5$, see Table 14).

It is important to underline that this improvement is achieved by applying 
``hadronic'' cuts that select the events having clear \gpj topology at the 
particle level and also having rather ``clean'' area
(in a sense of limited $\Pt$ activity) beyond a \gpj system.
The  consideration of the cuts, connected with detector effects 
(e.g., based on an electromagnetic shower profile),
may lead to  further improvement of $S/B$ ratio (see \cite{GMS} and \cite{GMS_NN}).
In this sense and taking into account also the fact that these ``hadronic'' cuts 
lead to an essential improvement of \ptgj balance,
we conclude  that the cuts on $\Pt^{clust}$ and $\Pt^{out}$, considered here,
are  quite effective  for selection of  the events caused by leading order 
diagrams (see Fig.~1) and they  do suppress the contribution of NLO corrections
shown by diagrams of Figs.~2, 4.

Another interesting predictions of PYTHIA is about the dominant contribution of
``$\gamma$-brem'' events into the total background at LHC energy, as in was 
already mentioned in Section 8 (see also \cite{BKS_P5} ). As the ``$\gamma$-brem'' 
background has an irreducible nature its careful estimation is an important task and 
we plan to make the analogous estimation with other generators (HERWIG, for instance). 

To finish the discussion of the jet calibration study let us mention that the 
main results on this subject are summed up in tables 1--12 (Selection 1) and
in tables 13--24 (Selection 2 with jet isolation criterion)
of Appendix 6 and in  Fig.~\ref{fig:mu-sig}.

It should be emphasized that numbers presented in all mentioned tables and figures 
were found within the PYTHIA particle level of simulation. They may depend on the 
used generator and on the particular choice of a long set of its parameters
\footnote{The comparison of predictions of different generators (PYTHIA, HERWIG,
etc.) with the experimental results is a part of a work in any experiment.}
as well as they may change after account of the results of the full GEANT-based simulation.


It is shown that the samples of the \gpj events, gained with the cuts used for the jet energy calibration,
can provide an information suitable also for determining the
gluon distribution inside a proton in the kinematic region (see Fig.~33) that includes
$x$ values as small as accessible at HERA  \cite{H1}, \cite{ZEUS}, but
at much higher $Q^2$ values (by about two orders of magnitude):
$2\cdot10^{-4}\leq x \leq 1.0$ with $1.6\cdot10^3\leq Q^2\leq8\cdot10^4 ~(GeV/c)^2$.
The number of  events, based on the gluonic process (1a), that may be collected 
with $L_{int}=20~fb^{-1}$ in different $x$- and $Q^2$- intervals of this new kinematic region
for extraction of the information about gluon distribution  are presented in Table 25
(all quarks included) and in Table 26 (only for charm quarks) 
\footnote{see also tables of Appendix 1.}

~\\[5mm]
\bf{ACKNOWLEDGMENTS.}  \\[5pt]
\normalsize
\rm
We are greatly thankful to D.~Denegri who initiated our interest to study
the physics of \gpj processes, for his permanent support and fruitful
suggestions. It is a pleasure for us to express our deep recognition for 
helpful discussions to P.~Aurenche,
M.~Dittmar, M.~Fontannaz, J.Ph.~Guillet, M.L.~Mangano, E.~Pilon,
H.~Rohringer, S.~Tapprogge, H.~Weerts and J.~Womersley. Our special gratitude is
to J.~Womersley also for supplying us with the preliminary version of paper [1],
interest in the work and encouragements.

\newpage

\setcounter{page}{103}

\normalsize

~\\[-5mm]

\end{document}